\documentclass[12pt]{article}

 \usepackage{graphicx}
 \usepackage{cite}
 \usepackage[english]{babel}

\begin{document}


\begin{center}
{\LARGE\bf On the suitability of the Brillouin action\\[2mm]
           as a kernel to the overlap procedure}
\end{center}

\vspace{10pt}

\begin{center}
{\large\bf Stephan D\"urr$\,^{a,b}$}
\,\,\,and\,\,\,
{\large\bf Giannis Koutsou$\,^{c}$}
\\[10pt]
${}^a${\sl University of Wuppertal, Gau{\ss}stra{\ss}e 20, 42119 Wuppertal, Germany}\\
${}^b${\sl IAS/JSC, Forschungszentrum J\"ulich GmbH, 52425 J\"ulich, Germany}\\
${}^c${\sl Cyprus Institute, CaSToRC, 20 Kavafi Street, Nicosia 2121, Cyprus}
\end{center}

\vspace{10pt}

\begin{abstract}
\noindent
We investigate the Brillouin action in terms of its suitability as a kernel to the overlap procedure, with a view on both heavy and light quark physics.
We use the diagonal elements of the Kenney-Laub family of iterations for the sparse matrix sign function, since they grow monotonically and facilitate cascaded preconditioning strategies with different rational approximations to the sign function.
We find that the overlap action with the Brillouin kernel is significantly better localized than the version with the Wilson kernel.
\end{abstract}

\vspace{10pt}


\newcommand{\pad}{\partial}
\newcommand{\hqu}{\hbar}
\newcommand{\til}{\tilde}
\newcommand{\pri}{^\prime}
\renewcommand{\dag}{^\dagger}
\newcommand{\<}{\langle}
\renewcommand{\>}{\rangle}
\newcommand{\gaf}{\gamma_5}
\newcommand{\nab}{\nabla}
\newcommand{\lap}{\triangle}
\newcommand{\dal}{{\sqcap\!\!\!\!\sqcup}}
\newcommand{\trc}{\mathrm{tr}}
\newcommand{\Trc}{\mathrm{Tr}}
\newcommand{\Mpi}{M_\pi}
\newcommand{\Fpi}{F_\pi}
\newcommand{\Mka}{M_K}
\newcommand{\Fka}{F_K}
\newcommand{\Met}{M_\et}
\newcommand{\Fet}{F_\et}
\newcommand{\Mss}{M_{\bar{s}s}}
\newcommand{\Fss}{F_{\bar{s}s}}
\newcommand{\Mcc}{M_{\bar{c}c}}
\newcommand{\Fcc}{F_{\bar{c}c}}

\newcommand{\al}{\alpha}
\newcommand{\be}{\beta}
\newcommand{\ga}{\gamma}
\newcommand{\de}{\delta}
\newcommand{\ep}{\epsilon}
\newcommand{\ve}{\varepsilon}
\newcommand{\ze}{\zeta}
\newcommand{\et}{\eta}
\renewcommand{\th}{\theta}
\newcommand{\vt}{\vartheta}
\newcommand{\io}{\iota}
\newcommand{\ka}{\kappa}
\newcommand{\la}{\lambda}
\newcommand{\rh}{\rho}
\newcommand{\vr}{\varrho}
\newcommand{\si}{\sigma}
\newcommand{\ta}{\tau}
\newcommand{\ph}{\phi}
\newcommand{\vp}{\varphi}
\newcommand{\ch}{\chi}
\newcommand{\ps}{\psi}
\newcommand{\om}{\omega}

\newcommand{\psb}{\bar{\psi}}
\newcommand{\etb}{\bar{\eta}}
\newcommand{\psh}{\hat{\psi}}
\newcommand{\eth}{\hat{\eta}}
\newcommand{\psd}{\psi^{\dagger}}
\newcommand{\etd}{\eta^{\dagger}}
\newcommand{\qh}{\hat{q}}
\newcommand{\kh}{\hat{k}}

\newcommand{\bdm}{\begin{displaymath}}
\newcommand{\edm}{\end{displaymath}}
\newcommand{\bea}{\begin{eqnarray}}
\newcommand{\eea}{\end{eqnarray}}
\newcommand{\beq}{\begin{equation}}
\newcommand{\eeq}{\end{equation}}

\newcommand{\mr}{\mathrm}
\newcommand{\mb}{\mathbf}
\newcommand{\ri}{\mr{i}}
\newcommand{\Nf}{N_{\!f}}
\newcommand{\Nc}{N_{ c }}
\newcommand{\Nt}{N_{ t }}
\newcommand{\Dst}{D^\mr{st}}
\newcommand{\Dov}{D^\mr{ov}}
\newcommand{\Dit}{D^\mr{it}}
\newcommand{\Dke}{D^\mr{ke}}
\newcommand{\Dwi}{D^\mr{wils}}
\newcommand{\Dbr}{D^\mr{bril}}
\newcommand{\Dstm}{D_m^\mr{st}}
\newcommand{\Dovm}{D_m^\mr{ov}}
\newcommand{\Ditm}{D_m^\mr{it}}
\newcommand{\Dkem}{D_m^\mr{ke}}
\newcommand{\Dker}{D_{-\rh/a}^\mr{ke}}
\newcommand{\Hker}{H_{-\rh/a}^\mr{ke}}
\newcommand{\Ake}{A^\mr{ke}}
\newcommand{\Bke}{B^\mr{ke}}
\newcommand{\Drh}{D_{-\rh}}
\newcommand{\Hrh}{H_{-\rh}}
\newcommand{\Arh}{A_{-\rh}}
\newcommand{\Brh}{B_{-\rh}}
\newcommand{\MeV}{\,\mr{MeV}}
\newcommand{\GeV}{\,\mr{GeV}}
\newcommand{\fm}{\,\mr{fm}}
\newcommand{\MSbar}{\overline{\mr{MS}}}

\hyphenation{topo-lo-gi-cal simu-la-tion theo-re-ti-cal mini-mum con-tinu-um}


\section{Introduction}


One of the key issues in a numerical study of lattice QCD is a suitable choice of the lattice Dirac operator, as this choice has a major impact on the overall cost, in terms of CPU time, of the computation.
Whenever processes with non-zero momentum transfer are considered (e.g.\ in meson and baryon form factors which are relevant for
semileptonic decays) the lattice dispersion relation is of interest, i.e.\ how much the continuum relation $(aE)^2-(a\mb{p})^2=(am)^2$ is violated, where we use the lattice spacing $a$ to build dimensionless quantities.

The Wilson Dirac operator \cite{Wilson:1974sk,Wilson:1975id} and the Brillouin Dirac operator \cite{Durr:2010ch}
\beq
D_\mr{wil}(x,y)=\sum_\mu \ga_\mu \nab_\mu^\mr{std}(x,y)
-\frac{a}{2}\lap^\mr{std}(x,y)+m_0\de_{x,y}
-\frac{c_\mr{SW}}{2}\sum_{\mu<\nu}\si_{\mu\nu}F_{\mu\nu}\de_{x,y}
\label{def_wils}
\eeq
\beq
D_\mr{bri}(x,y)=\sum_\mu \ga_\mu \nab_\mu^\mr{iso}(x,y)
-\frac{a}{2}\lap^\mr{bri}(x,y)+m_0\de_{x,y}
-\frac{c_\mr{SW}}{2}\sum_{\mu<\nu}\si_{\mu\nu}F_{\mu\nu}\de_{x,y}
\label{def_bril}
\eeq
both show cut-off effects $\propto a$ which can be reduced to $\propto a^2$ by proper tuning of the coefficient $c_\mr{SW}$ \cite{Sheikholeslami:1985ij,Heatlie:1990kg,Luscher:1996sc,Luscher:1996ug}.
The only difference is the discretization used for the covariant derivative $\nab_\mu$ and the gauged Laplacian $\lap$; the former operator uses a 9-point stencil for the Laplacian, while the latter operator uses a 81-point stencil (the Nabla operator always uses a subset of that stencil).
The larger stencil allows for an improved dispersion relation (see Refs.\,\cite{Durr:2010ch,Durr:2012dw,Cho:2015ffa} and below), but obviously the numerical cost is increased.

Regardless whether $c_\mr{SW}$ is zero or tuned to remove the $O(a)$ on-shell cut-off effects, the operators (\ref{def_wils}, \ref{def_bril}) are subject to limitations concerning the (renormalized) quark mass (which derives from the bare quark mass $m_0$) that can be used in a simulation.
For light quarks there is an algorithmic bound for such non-chiral actions \cite{DelDebbio:2005qa,Durr:2010aw}, and for heavy quark masses cut-off effects tend to proliferate (unless special measures are taken, see e.g.\ Refs.\,\cite{ElKhadra:1996mp,Oktay:2008ex,Cho:2015ffa}).

The algorithmic limitation how light a quark mass may be taken at a given value of the gauge coupling $\be=6/g_0^2$ is absent for chiral actions, i.e.\ for actions which satisfy the Ginsparg-Wilson relation \cite{Ginsparg:1981bj,Hasenfratz:1997ft,Hasenfratz:1998jp,Luscher:1998pqa}.
The overlap construction (here and below this term is meant to include both the ``domain-wall'' \cite{Kaplan:1992bt,Shamir:1993zy,Furman:1994ky} and the ``overlap'' \cite{Neuberger:1997fp,Neuberger:1998wv} emanation of this idea) manages to upgrade a non-chiral into a chiral action.
This is a highly practical procedure, though it is somewhat expensive in terms of CPU time (see below).

As a side effect, the overlap construction leads to automatic $O(a)$ improvement.
In other words no tuning of a coefficient like $c_\mr{SW}$ in (\ref{def_wils}, \ref{def_bril}) is needed; the requirement of chiral symmetry automatically kills odd powers of $a$ in on-shell quantities \cite{Niedermayer:1998bi}.
This is the reason why the overlap action with the Wilson kernel has proven very useful in heavy quark physics, see for instance the charm physics programs by the Kentucky group, JLQCD, and RBC/UKQCD \cite{Yang:2014sea,Fahy:2015xka,Boyle:2016lzk}.

\bigskip

In this paper we wish to explore whether there is any relevant improvement if one replaces, in the overlap construction, the Wilson kernel by the Brillouin kernel.
Ideally, such an action would enable one to use a uniform relativistic formulation to simulate all hadronizing quarks ($d,u,s,c,b$) at their physical mass values, on accessible lattices.
The first technical question is whether the improved dispersion relation of the Brillouin operator for light quark masses (both at the quark-level \cite{Durr:2010ch,Cho:2015ffa} and for hadronic quantities \cite{Durr:2012dw}) would persist after the overlap procedure has been applied.
The second question is whether the CPU requirements of the Brillouin-overlap action are roughly comparable to those of the Wilson-overlap formulation or whether they proliferate.
The third question is whether there is any notable technical difference between the two overlap formulations, e.g.\ in terms of operator locality.

\bigskip

The remainder of this paper addresses these questions in due turn, intertwined with a few reminders on the overlap formulation and its technical implementation in a sparse matrix setup to make it self-contained.
Sec.\,\ref{sec:DR} presents an investigation of the free-field dispersion relations of both the Wilson and Brillouin kernel, along with their overlap descendents.
Sec.\,\ref{sec:KL} summarizes some knowledge about the Kenney-Laub family of iterations for the matrix sign function, since the diagonal members of this family show properties which we consider particularly convenient for the implementation of an overlap-times-vector application.
Sec.\,\ref{sec:overlap} gives a quick review of the overlap construction and discusses a way of introducing the mass in the overlap operator which avoids any ``extra prescription'' if the Green's function is used in the computation of a decay constant or matrix element.
Sec.\,\ref{sec:eigs} illustrates the eigenvalues of some low-order Kenney-Laub iterates of the Wilson and Brillouin kernels on small lattices where all eigenvalues can be calculated.
Sec.\,\ref{sec:flow} presents the spectral flow, i.e.\ eigenvalues of the shifted hermitean kernels (for both formulations) on some selected gauge backgrounds.
Sec.\,\ref{sec:tests} addresses the aforementioned technical issues, such as the operator locality, and reports on a pilot spectroscopy calculation on $40^3\times64$ lattices generated by QCDSF.
Sec.\,\ref{sec:precondition} is a reminder that the Kenney-Laub family of matrix iterations offers many possibilities for cascaded preconditioning strategies where very-low-order polynomial approximations to the sign function are used to speed-up computations with not-so-low-order approximations.
Sec.\,\ref{sec:outlook} gives reasons why we feel optimistic about the use of the framework portrayed in this article in future studies of full QCD with exact (i.e.\ arbitrarily good) chiral symmetry.
Sec.\,\ref{sec:summary} contains our summary, and some technical material is arranged in three appendices.
A preliminary account of this work appeared in Ref.\,\cite{Durr:2016xoc}.


\section{Quark-level dispersion relations \label{sec:DR}}


In this section we discuss the free-field dispersion relations of the Wilson and Brillouin operators, as well as those of their overlap descendents.

For the Wilson operator the dispersion relation reads (see App.\,\ref{app:dispersion} for details)
\beq
2\cosh(aE)\Big[4+am-\sum_i\cos(ap_i)\Big]=1+\sum_i\sin^2(ap_i)+\Big[4+am-\sum_i\cos(ap_i)\Big]^2
\eeq
and an expansion in powers of $a$ yields \cite{Cho:2015ffa}
\bea
(aE)^2-(a\mb{p})^2
&=&\Big[(am)^2-(am)^3+\frac{11}{12}(am)^4-\frac{5}{6}(am)^5\Big]
\nonumber
\\
&+&\Big[-\frac{2}{3}(am)^2+\frac{7}{6}(am)^3\Big](a\mb{p})^2
\nonumber
\\
&+&\Big[-\frac{2}{3}+\frac{am}{2}\Big]\Big(\sum_{i<j}a^4p_i^2p_j^2+\sum_{i}(ap_i)^4\Big)+O(a^6)
\;.
\label{free_wils}
\eea
For the Brillouin operator the dispersion relation reads (see App.\,\ref{app:dispersion} for details)
\bea
\textstyle
\frac{1}{729}\sum_i s_i^2 \prod_{j\neq i}\{c_j+2\}^2\{\cosh^2+4\cosh+4\}+
\frac{1}{729}\{1-\cosh^2\} \prod_{i}\{c_i+2\}^2&&
\nonumber
\\[2mm]
\textstyle
+\frac{1}{64}\prod_i\{c_i+1\}^2\{\cosh^2+2\cosh+1\}
-\frac{1}{4}\prod_i\{c_i+1\}\{\cosh+1\}[2+am]
+[2+am]^2&=&0
\eea
with $s_i=\sin(ap_i), c_i=\cos(ap_i)$, and an expansion of the physical solution yields \cite{Cho:2015ffa}
\bea
(aE)^2-(a\mb{p})^2
&=&\Big[(am)^2-(am)^3+\frac{11}{12}(am)^4-\frac{5}{6}(am)^5\Big]
\nonumber
\\
&+&\Big[0+\frac{1}{12}(am)^3\Big](a\mb{p})^2
\nonumber
\\
&+&\Big[0+\frac{am}{12}\Big]\Big(\sum_{i<j}a^4p_i^2p_j^2+\sum_{i}(ap_i)^4\Big)+O(a^6)
\;.
\label{free_bril}
\eea

\begin{figure}[!tb]
\centering
\includegraphics[width=0.5\textwidth]{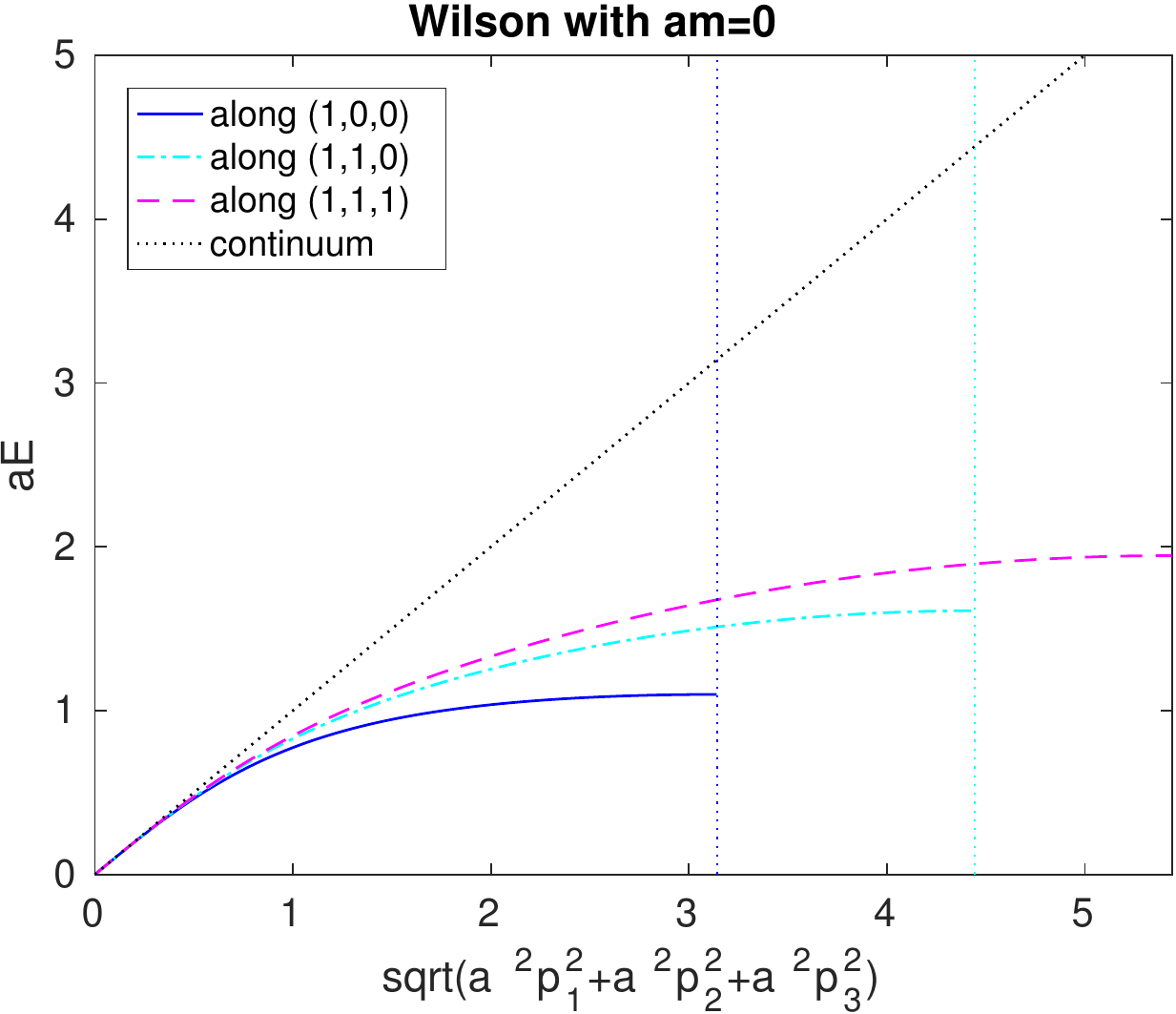}%
\includegraphics[width=0.5\textwidth]{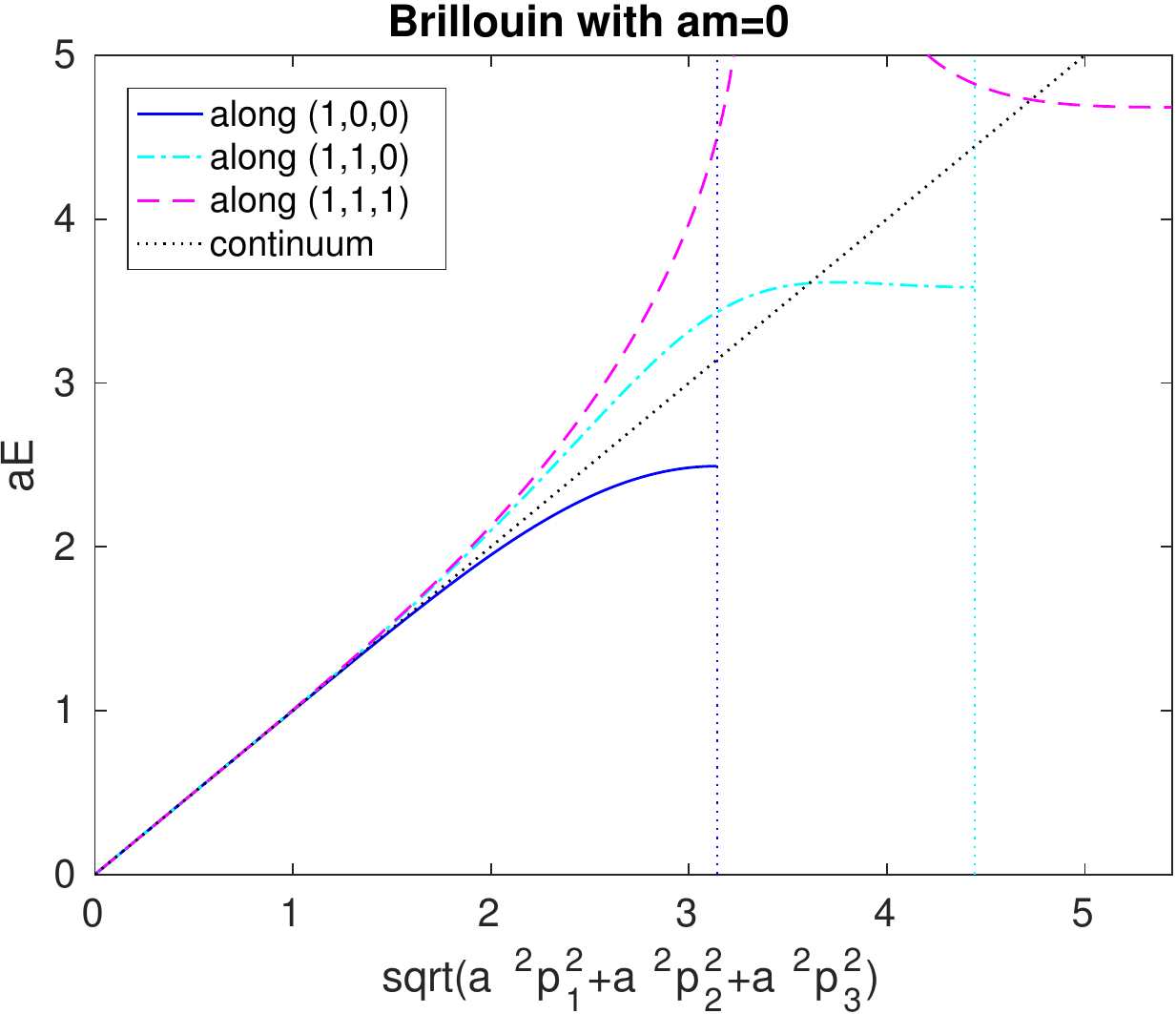}%
\\[2mm]
\includegraphics[width=0.5\textwidth]{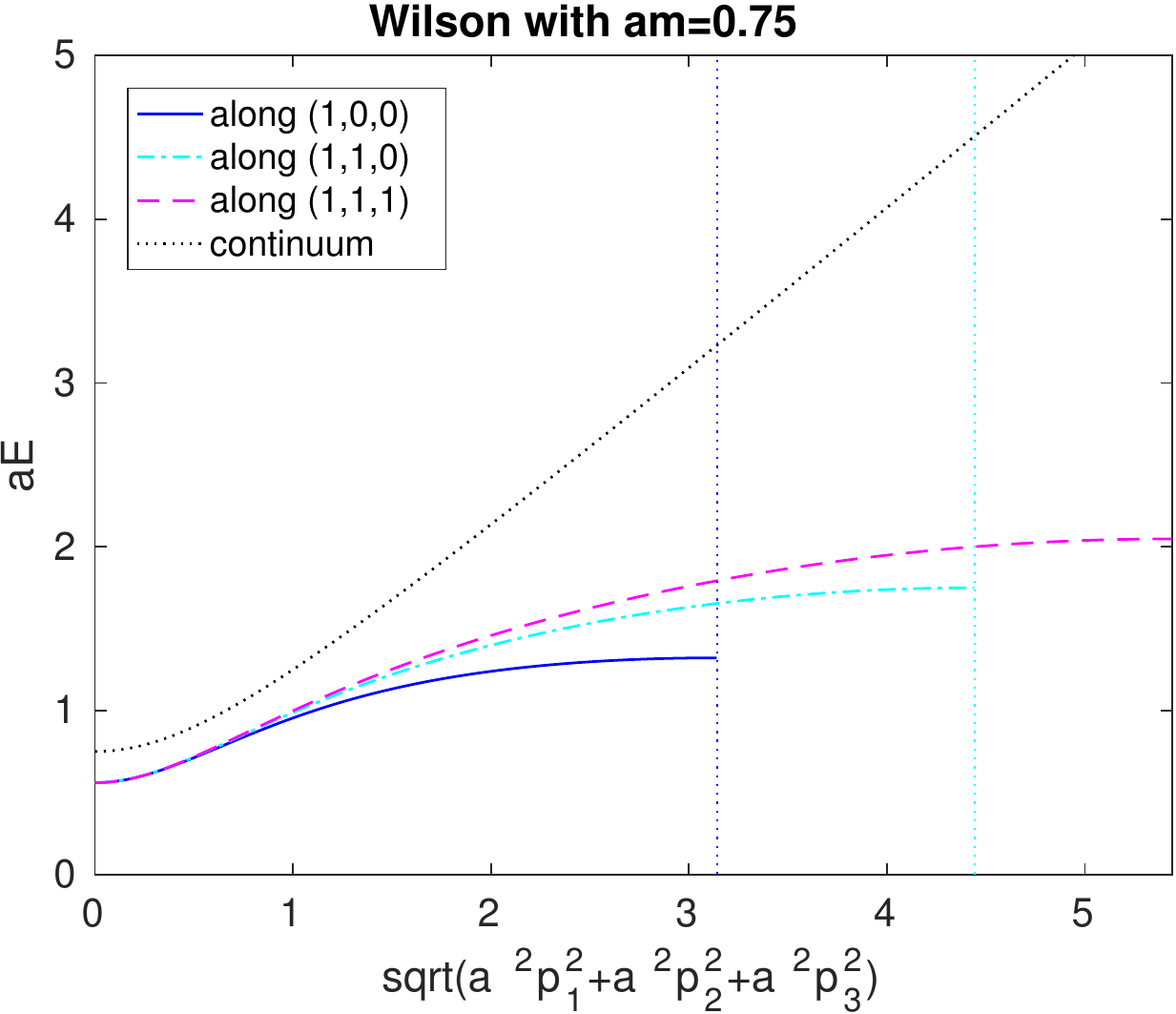}%
\includegraphics[width=0.5\textwidth]{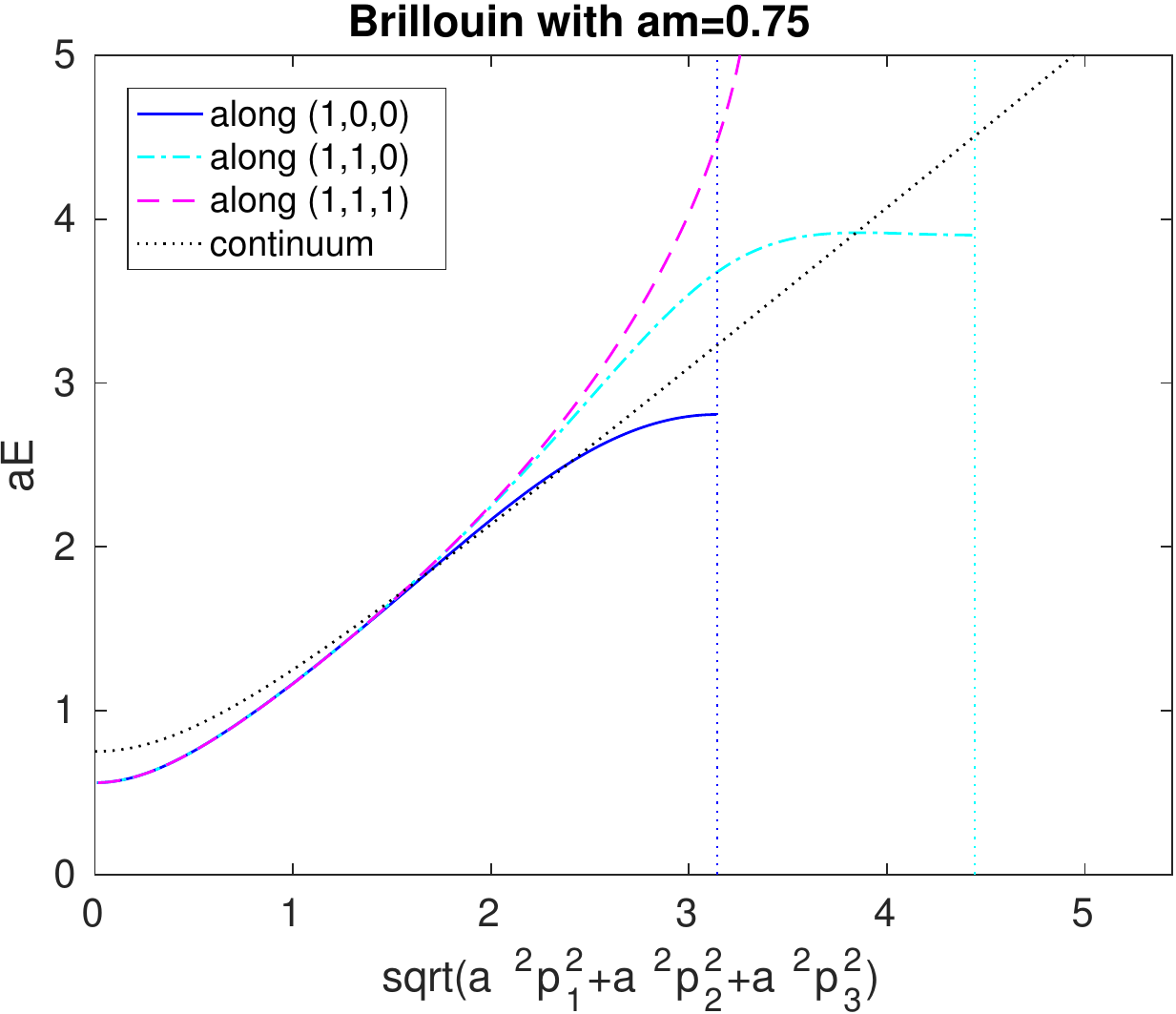}%
\caption{\label{fig:MDR_kern}\sl
Free field dispersion relations of $D_\mr{W}$ (left) and $D_\mr{B}$ (right) for the bare quark masses $am=0$ (top) and $am=0.75$ (bottom).
We plot the spatial directions $(1,0,0)$, $(1,1,0)$, and $(1,1,1)$, where the Brillouin zone ends at $\pi/a$, $\sqrt{2}\pi/a$, and $\sqrt{3}\pi/a$, respectively.}
\end{figure}

As was already noted in Ref.\,\cite{Cho:2015ffa}, a comparison of (\ref{free_wils}) and (\ref{free_bril}) shows that the Brillouin construction manages to reduce the amount of isotropy breaking (the term $\propto a^4$ in the last line vanishes, and the term $\propto a^5$ receives a factor $1/6$).
However, the momentum independent part in the first line, which is an expansion of $\log^2(1+am)$, is unchanged from the Wilson case \cite{Cho:2015ffa}.
This suggests that the Brillouin construction brings an advantage for heavy quark spectroscopy only in case non-zero spatial momenta are involved.

This conclusion is supported by the plots shown in Fig.\,\ref{fig:MDR_kern}.
The Wilson operator at $am=0$ shows significant deviations from the continuum dispersion relation and strong isotropy violations (differences  between the momentum directions).
Furthermore, at the heavy quark mass $am=0.75$ strong cut-off effects even at $a\mb{p}=\mb{0}$ become visible.
The Brillouin operator at $am=0$ features a significantly improved dispersion relation with much smaller isotropy violations, but at $am=0.75$ the cut-off effects are equally large as in the Wilson case.

\begin{figure}[!tb]
\centering
\includegraphics[width=0.5\textwidth]{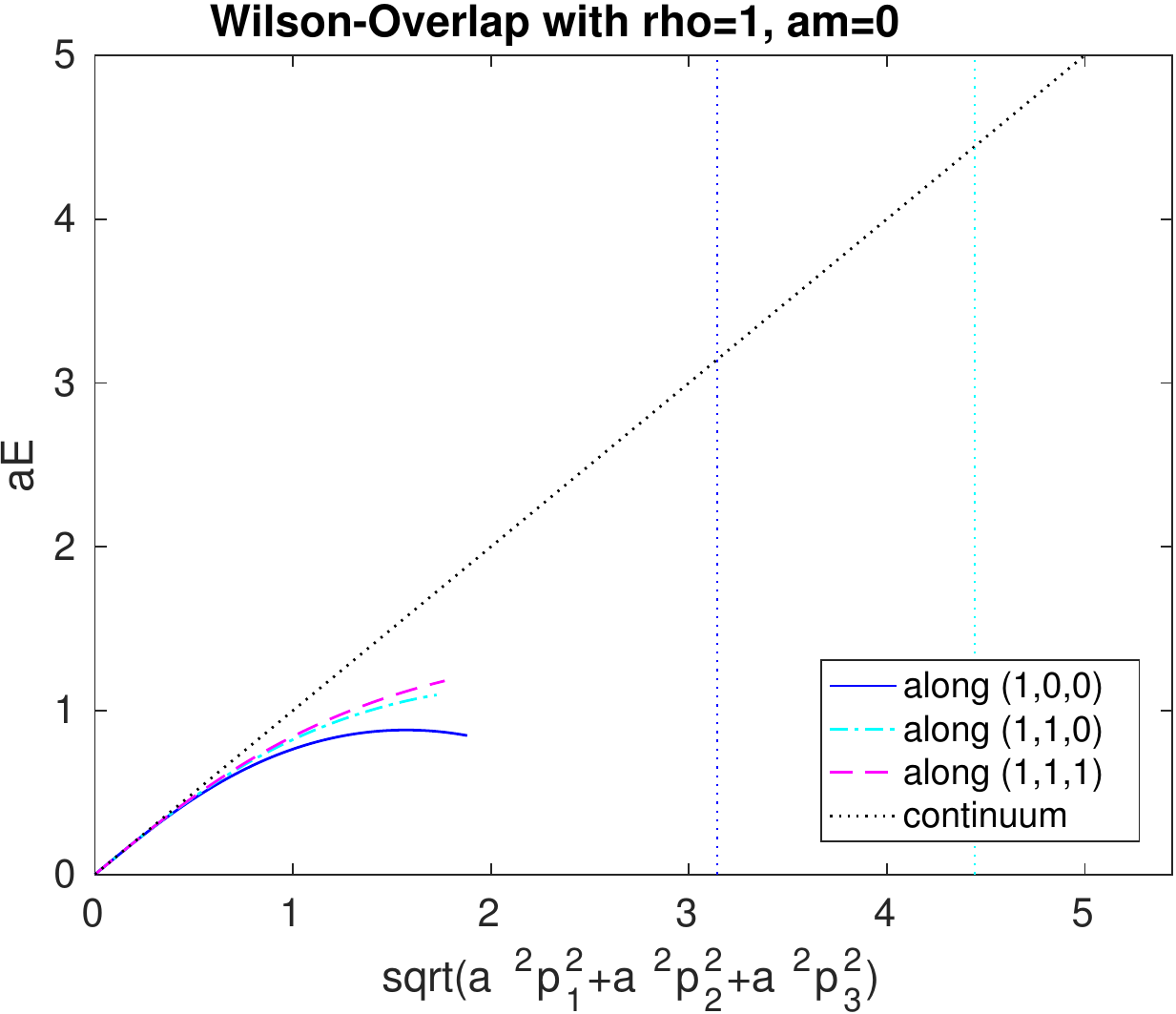}%
\includegraphics[width=0.5\textwidth]{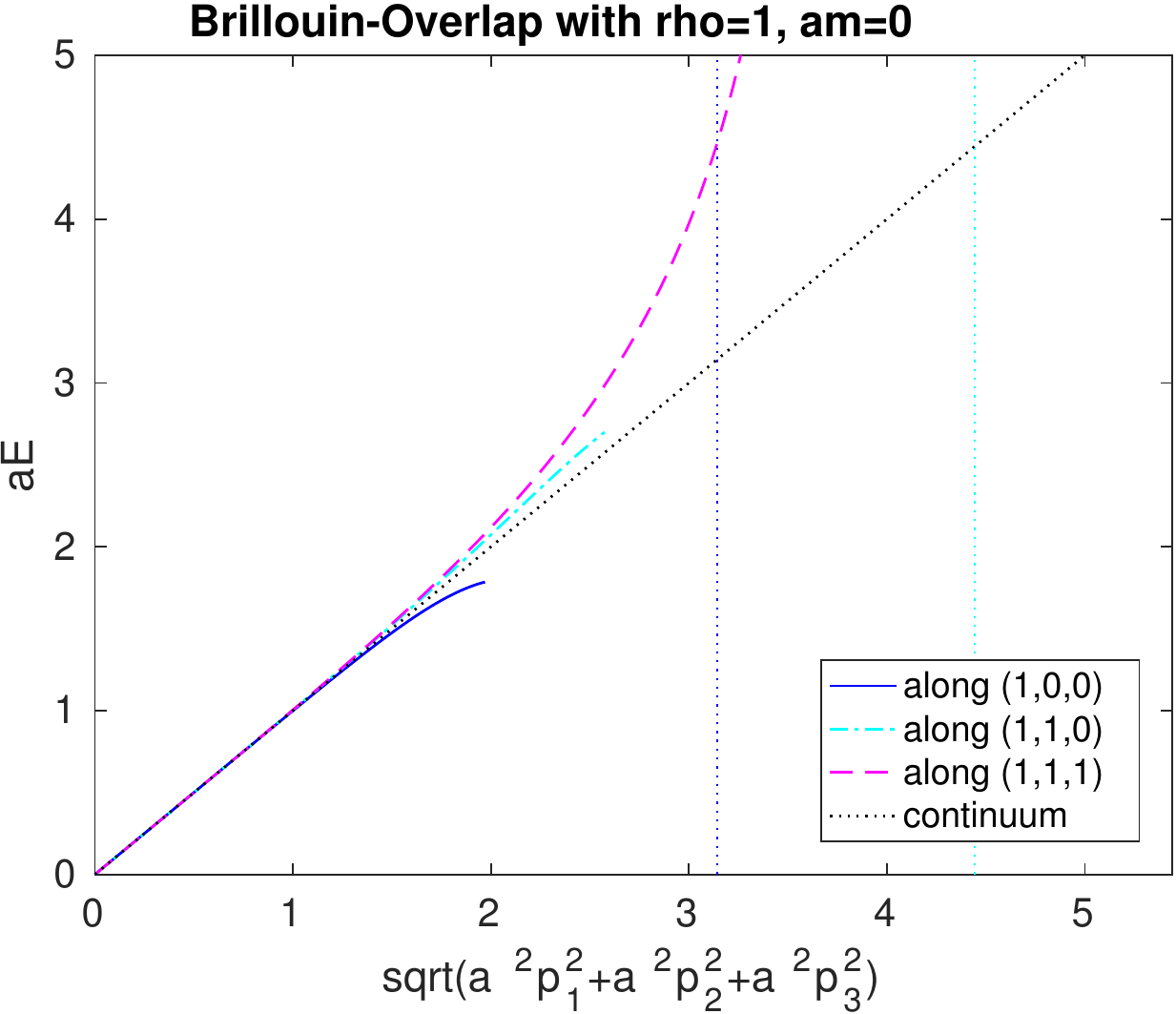}%
\\[2mm]
\includegraphics[width=0.5\textwidth]{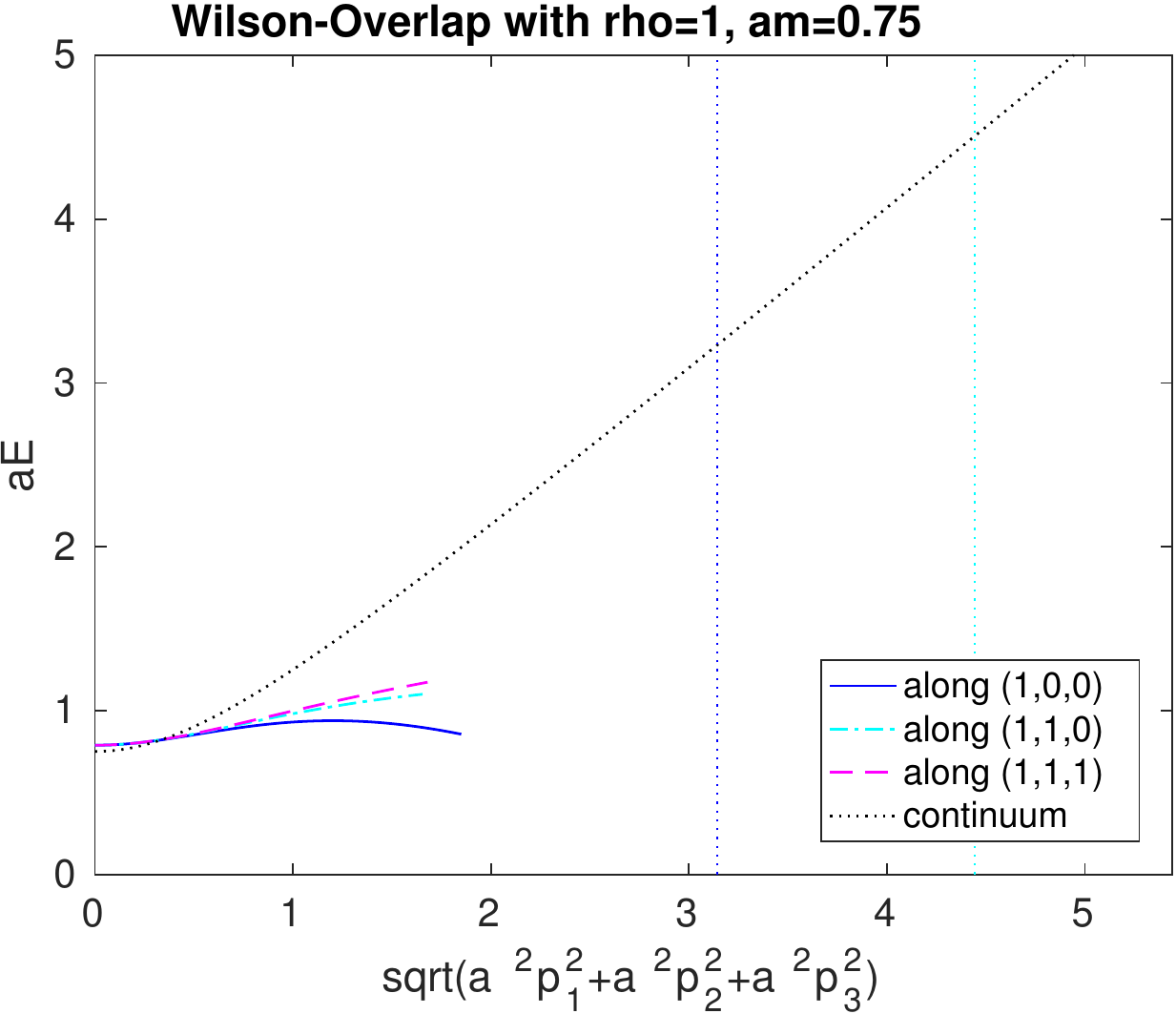}%
\includegraphics[width=0.5\textwidth]{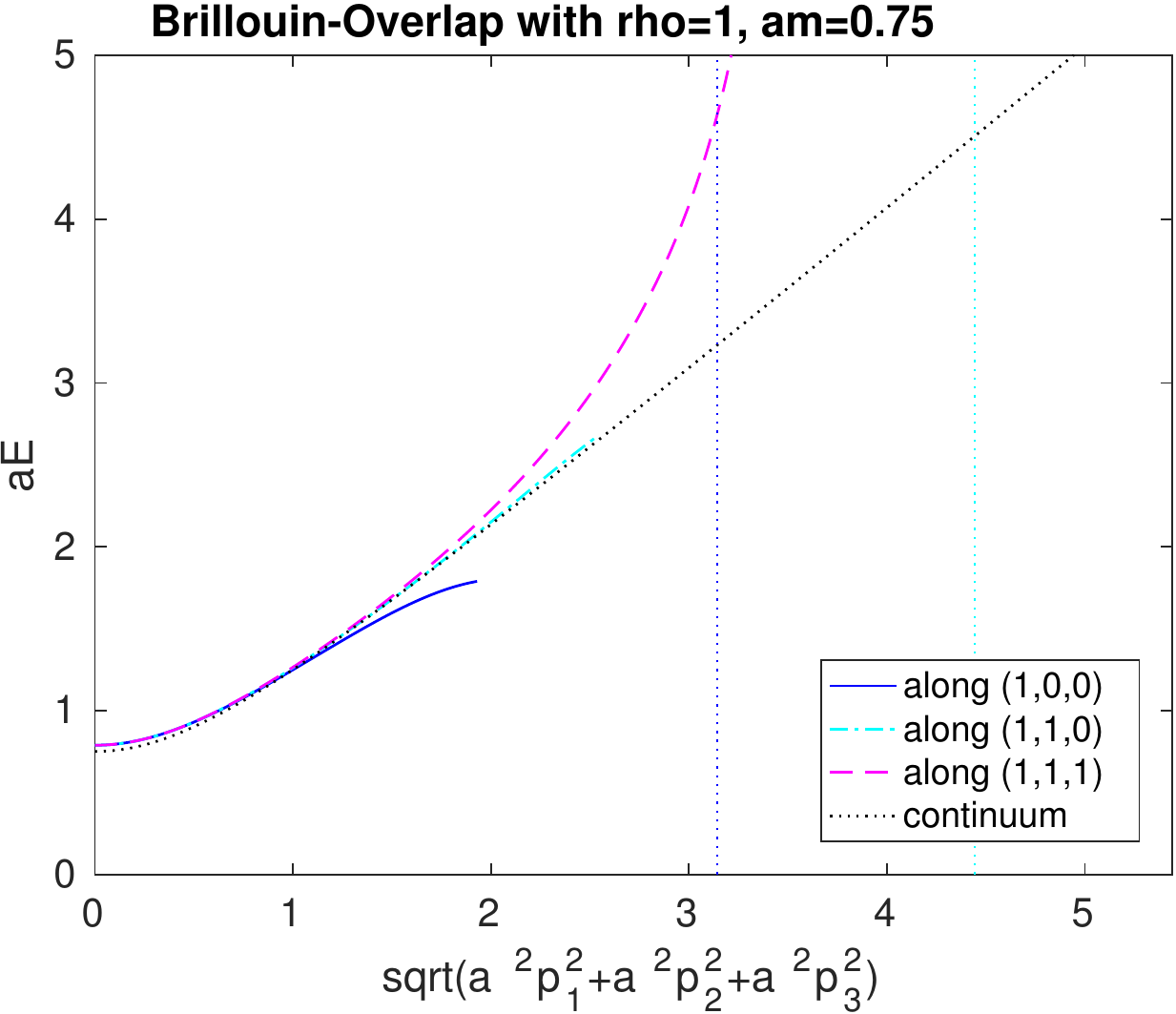}%
\caption{\label{fig:MDR_over}\sl
Same as Fig.\,\ref{fig:MDR_kern} but for the overlap actions based on the Wilson (left) and Brillouin (right) kernels at $\rh=1$.
The overlap mass is again $am=0$ (top) and $am=0.75$ (bottom).}
\end{figure}

For the overlap operator with the Wilson kernel the dispersion relation follows from searching for zeros of $c^2+2cd(\frac{a}{2}\hat{p}^2-\frac{\rh}{a})[\bar{p}^2+(\frac{a}{2}\hat{p}^2-\frac{\rh}{a})^2]^{-1/2}+d^2=0$ with $c,d,\hat{p}^2,\bar{p}^2$ given in App,\,\ref{app:dispersion}, and an expansion in powers of $a$ yields
\bea
(aE)^2-(a\mb{p})^2
&=&\Big[(am)^2-\frac{2\rh^2-6\rh+3}{6\rh^2}(am)^4\Big]
\nonumber
\\
&+&\Big[-\frac{2}{3}(am)^2+0\Big](a\mb{p})^2
\nonumber
\\
&+&\Big[-\frac{2}{3}+0\Big]\Big(\sum_{i<j}a^4p_i^2p_j^2+\sum_{i}(ap_i)^4\Big)+O(a^6)
\;.
\label{over_wils}
\eea
For the overlap operator with the Brillouin kernel the dispersion relation follows from searching for zeros of $c^2+2cd(\frac{a}{2}\check{p}^2-\frac{\rh}{a})[\til{p}^2+(\frac{a}{2}\check{p}^2-\frac{\rh}{a})^2]^{-1/2}+d^2=0$ with $c,d,\check{p}^2,\til{p}^2$ given in App,\,\ref{app:dispersion}, and an expansion in powers of $a$ yields
\bea
(aE)^2-(a\mb{p})^2
&=&\Big[(am)^2-\frac{2\rh^2-6\rh+3}{6\rh^2}(am)^4\Big]
\nonumber
\\
&+&\Big[0+0\Big](a\mb{p})^2
\nonumber
\\
&+&\Big[0+0\Big]\Big(\sum_{i<j}a^4p_i^2p_j^2+\sum_{i}(ap_i)^4\Big)+O(a^6)
\;.
\label{over_bril}
\eea

Comparing (\ref{over_wils}) and (\ref{over_bril}) suggests that the Brillouin overlap inherits the reduced isotropy breaking from its kernel action.
The cut-off effects at $a\mb{p}=\mb{0}$ are still identical for the two overlap actions, and the removal of odd powers of $am$ clearly reduces the momentum-dependent cut-off effects in comparison to the non-chiral predecessors.
Note that in (\ref{over_wils}, \ref{over_bril}) the coefficient of $(am)^4$ can be made zero by choosing $\rh=(3-\sqrt{3})/2\simeq0.634$; in this case the free-field Brillouin overlap dispersion relation  is free of cut-off effects through $O(a^5)$.

This conclusion is supported by the plots shown in Fig.\,\ref{fig:MDR_over}.
The $\rh=1$ Wilson overlap operator at $am=0$ shows similar (or even worse) deviations from the continuum dispersion relation as its predecessor, but at $a\mb{p}=\mb{0}$ the cut-off effects for a heavy quark mass are much mitigated.
The $\rh=1$ Brillouin overlap operator at $am=0$ still enjoys a rather good dispersion relation, and at $a\mb{p}=\mb{0}$ the cut-off effects are equally small as for the Wilson overlap operator.

Evidently, the nice behavior of the free field dispersion relation of the Brillouin overlap operator at arbitrary quark mass and generic $\rh$ is a necessary (and not a sufficient) condition for this formulation to be useful in real physics applications.
However, given this property, we think it is worth while to investigate the Brillouin overlap action in more detail.


\section{Kenney-Laub iterates for the matrix sign function \label{sec:KL}}


\subsection{Definition}

Kenney and Laub proposed a family of iterations to compute the matrix sign function (equivalently to compute the unitary factor in the polar decomposition) that have some remarkable properties \cite{KenneyLaub}.
The $(m,n)$ iteration for a matrix $A$ with no purely imaginary eigenvalue is
\beq
X_{k+1}=X_k\;p_{mn}(I\!-\!X_k^2)\;[q_{mn}(I\!-\!X_k^2)]^{-1}
\equiv
f_{mn}(X_k)
\;,\qquad X_0=A
\label{pade_def}
\eeq
where $r_{mn}(t)=p_{mn}(t)/q_{mn}(t)$ is the $(m,n)$ Pad\'e approximant to $h(t)=(1-t)^{-1/2}$.
Here $I$ is the identity, $p_{mn}$ is a polynomial of order $m$ in $t=1-x^2$ (or of order $2m$ in $x$), and $q_{mn}$ is a polynomial of order $n$ in $t=1-x^2$ (or $2n$ in $x$).
To compute the polar decomposition $A=UP$ with unitary $U$ and positive semi-definite $P$ one simply replaces $X_k^2\to X_k\dag X_k$.

\begin{table}[!b]
\centering\large
\setlength{\tabcolsep}{4pt}
\begin{tabular}{|c|cccc|}
\hline
 & \small $n=0$ & \small $n=1$ & \small $n=2$ & \small $n=3$
\\[0mm]
\hline
{\small $m=0$} &
 $\frac{\phantom{X^X}}{\phantom{X^X}}$ &
 $\frac{2x}{1+x^2}$ &
 $\frac{8x}{3+6x^2-x^4}$ &
 $\frac{16x}{5+15x^2-5x^4+x^6}$
\\[2mm]
{\small $m=1$} &
 $\frac{x(3-x^2)}{2}$ &
 $\frac{x(3+x^2)}{1+3x^2}$ &
 $\frac{4x(1+x^2)}{1+6x^2+x^4}$ &
 $\frac{8x(3+5x^2)}{5+45x^2+15x^4-x^6}$
\\[2mm]
{\small $m=2$} &
 $\frac{x(15-10x^2+3x^4)}{8}$ &
 $\frac{x(15+10x^2-x^4)}{4(1+5x^2)}$ &
 $\frac{x(5+10x^2+x^4)}{1+10x^2+5x^4}$ &
 $\frac{2x(3+10x^2+3x^4)}{1+15x^2+15x^4+x^6}$
\\[2mm]
{\small $m=3$} &
 $\frac{x(35-35x^2+21x^4-5x^6)}{16}$ &
 $\frac{x(35+35x^2-7x^4+x^6)}{8(1+7x^2)}$ &
 $\frac{x(35+105x^2+21x^4-x^6)}{2(3+42x^2+35x^4)}$ &
 $\frac{x(7+35x^2+21x^4+x^6)}{1+21x^2+35x^4+7x^6}$
\\[2mm]
\hline
\end{tabular}
\caption{\label{tab1}\sl
Iteration functions $f_{mn}(x)=x\,p_{mn}(x^2)/q_{mn}(x^2)$ for $m,n=0,...,3$ from the Kenney-Laub family (\ref{pade_def}) for the matrix sign function.
The element $f_{00}(x)=x$ is not useful.}
\end{table}

\begin{table}[!tb]
\centering\large
\begin{tabular}{|c|c|}
\hline
{\small $m=n=4$} &
$\frac{x(9+84x^2+126x^4+36x^6+x^8)}{1+36x^2+126x^4+84x^6+9x^8}$
\\[2mm]
{\small $m=n=5$} &
$\frac{x(11+165x^2+462x^4+330x^6+55x^8+x^{10})}{1+55x^2+330x^4+462x^6+165x^8+11x^{10}}$
\\[2mm]
{\small $m=n=6$} &
$\frac{x(13+286x^2+1287x^4+1716x^6+715x^8+78x^{10}+x^{12})}{1+78x^2+715x^4+1716x^6+1287x^8+286x^{10}+13x^{12}}$
\\[2mm]
{\small $m=n=7$} &
$\frac{x(15+455x^2+3003x^4+6435x^6+5005x^8+1365x^{10}+105x^{12}+x^{14})}{1+105x^2+1365x^4+5005x^6+6435x^8+3003x^{10}+455x^{12}+15x^{14}}$
\\[2mm]
{\small $m=n=8$} &
$\frac{x(17+680x^2+6188x^4+19448x^6+24310x^8+12376x^{10}+2380x^{12}+136x^{14}+x^{16})}{1+136x^2+2380x^4+12376x^6+24310x^8+19448x^{10}+6188x^{12}+680x^{14}+17x^{16}}$
\\[2mm]
\hline
\end{tabular}
\caption{\label{tab2}\sl
Diagonal iteration functions $f_{nn}(x)$ for $n=4,...,8$ from the Kenney-Laub family (\ref{pade_def}).}
\end{table}

In Tab.\,\ref{tab1} the first few members $f_{m,n}$ of this family are listed (which one also finds in the literature \cite{KenneyLaub}) and in Tab.\,\ref{tab2} the elements $f_{n,n}$ with $n=4,...,8$ are given.
The convergence order [in $k$] of the $(m,n)$ element is $m+n+1$.
Two subsets of this family have special properties.
First of all, the elements on the diagonal ($m=n$) and first upper diagonal ($n-m=1$) are globally convergent, i.e.\ they work with any argument \cite{KenneyLaub}.
Second, the elements in the first column ($n=0$) do not require an inverse, but they tend to be numerically unstable.
In fact, the element $(m=1,n=0)$ is the Newton-Schulz iteration for the matrix sign function, which derives from the Newton method $X_{k+1}=\frac{1}{2}(X_k+X_k^{-1})$ through an additional expansion of the inverse.
The $(m=0,n=1)$ element generates the inverses of the Newton-Schulz series, and the $(m=1,n=1)$ element is sometimes named after Halley.

It seems to us that the diagonal mappings $f_{n,n}$ with $n\geq1$ are most convenient for practical use.
They have the special algebraic property that the polynomial $q_{nn}(x)$ in the denominator is the \emph{mirror polynomial} of $p_{nn}(x)$ in the numerator, i.e.\ the coefficients show up in reverse order (e.g.\ $5+10x^2+x^4$ versus $1+10x^2+5x^4$ in $f_{2,2}$).


\subsection{Principal Pad\'e iteration functions}


The diagonal ($m=n$) and first upper diagonal ($m=n-1$) elements of the family (\ref{pade_def}) are singled out as the ``principal Pad\'e iteration functions''.
For these $m$ and $n$ one defines
\beq
g_\ell(x)\equiv g_{m+n+1}(x) \equiv f_{m,n}(x)
\eeq
which means that the index $\ell$ counts them in Tab.\,\ref{tab1} in a zig-zag fashion, i.e.\ $g_1(x)=x$, $g_2(x)=2x/(1+x^2)$, $g_3(x)=x(3+x^2)/(1+3x^2)$, $g_4(x)=4x(1+x^2)/(1+6x^2+x^4)$, and so on.
These functions have a number of important properties \cite{Higham}:
\begin{itemize}
\item[$(i)$]
The coefficients of the numerator and the denominator follow from the binomial theorem
\beq
g_\ell(x)=\frac{(1+x)^\ell-(1-x)^\ell}{(1+x)^\ell+(1-x)^\ell}
\label{def_gell}
\eeq
and the numerator/denominator are thus the odd/even parts of $(1+x)^\ell$.
\item[$(ii)$]
This implies the following symmetry properties (for $x>0$) about $x=1$
\beq
g_{2n}(1/x)=g_{2n}(x)\quad[\mbox{upper diagonal}]
\;,\quad
g_{2n+1}(1/x)=1/g_{2n+1}(x)\quad[\mbox{diagonal}]
\label{pade_symmetry}
\eeq
and ditto (for $x<0$) about $x=-1$, since the overall functions are all odd in $x$.
\item[$(iii)$]
All principal Pad\'e iteration functions allow a tanh(.) representation, viz.\
\beq
g_\ell(x)=\tanh(\ell\,\mr{artanh}(x))
\;.
\eeq
\item[$(iv)$]
Nesting two principal Pad\'e iteration functions yields another one, viz.\
\beq
g_{\ell'}(g_{\ell''}(x))=g_{\ell'\ell''}(x)
\;.
\label{nesting}
\eeq
Since the product of two odd numbers is odd, and the product of two even numbers is even, it follows that both the diagonal and the first upper diagonal Kenney-Laub mappings satisfy this ``semigroup property'' separately.
Obviously, this implies that nestings commute for principal Pad\'e iteration functions, while this does not hold in general, i.e.\ $f_{mn}(f_{pq}(x)) \neq f_{pq}(f_{mn}(x))$ for arbitrary $m,n,p,q$.
\end{itemize}
In addition, $g_\ell(x)$ admits a partial fraction form which we will discuss in the next subsection.

\begin{figure}[!tb]
\centering
\includegraphics[width=5.5cm]{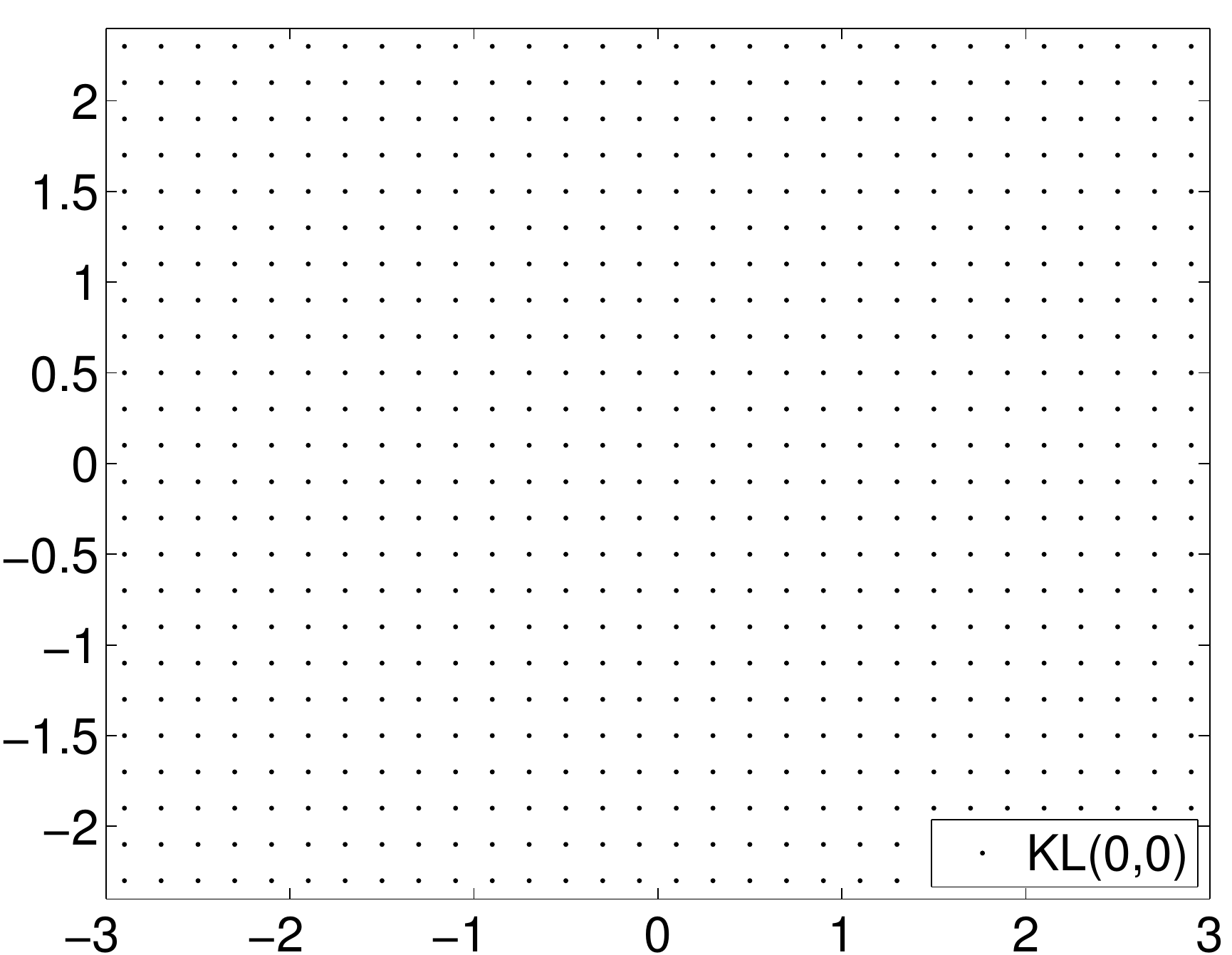}%
\includegraphics[width=5.5cm]{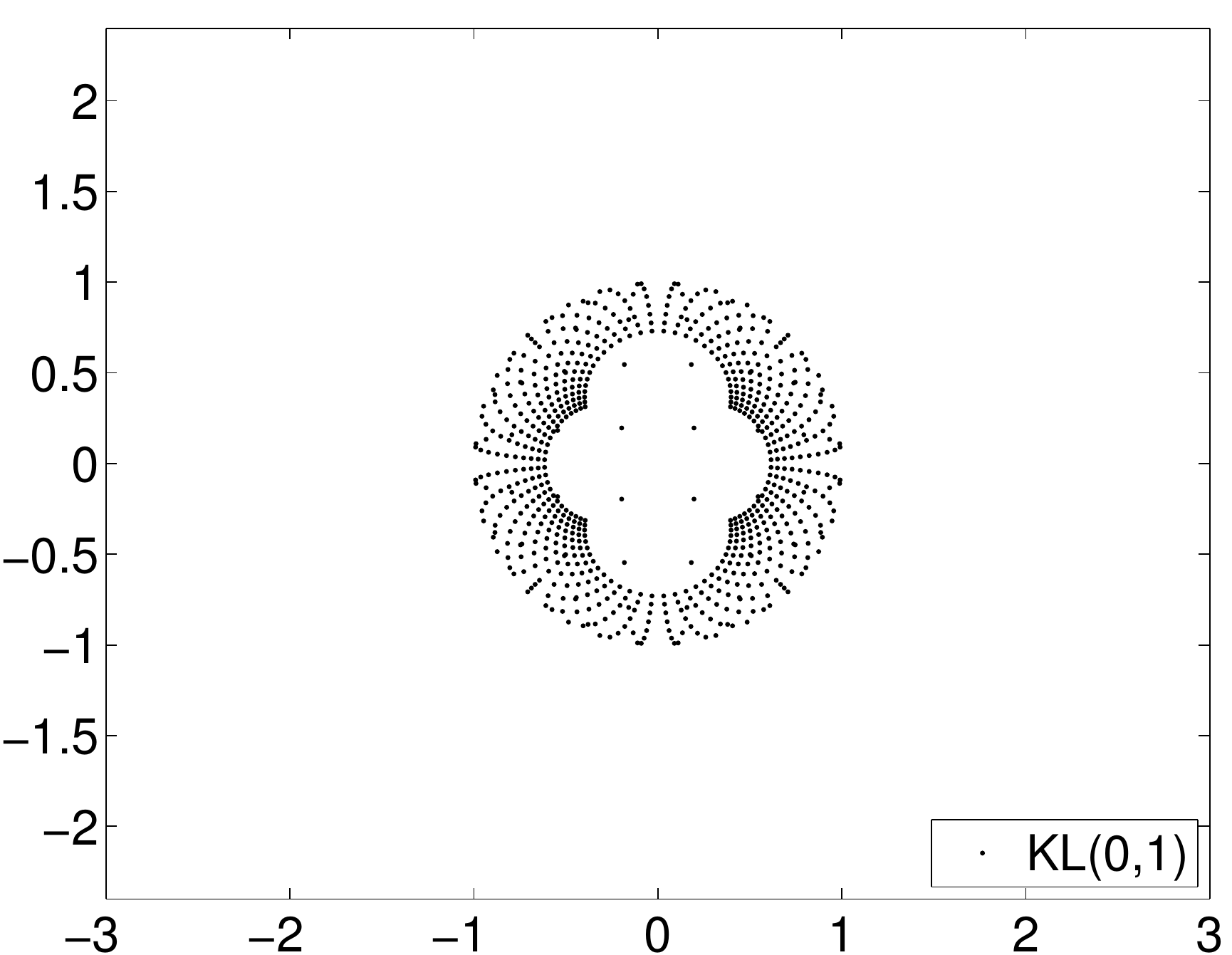}%
\includegraphics[width=5.5cm]{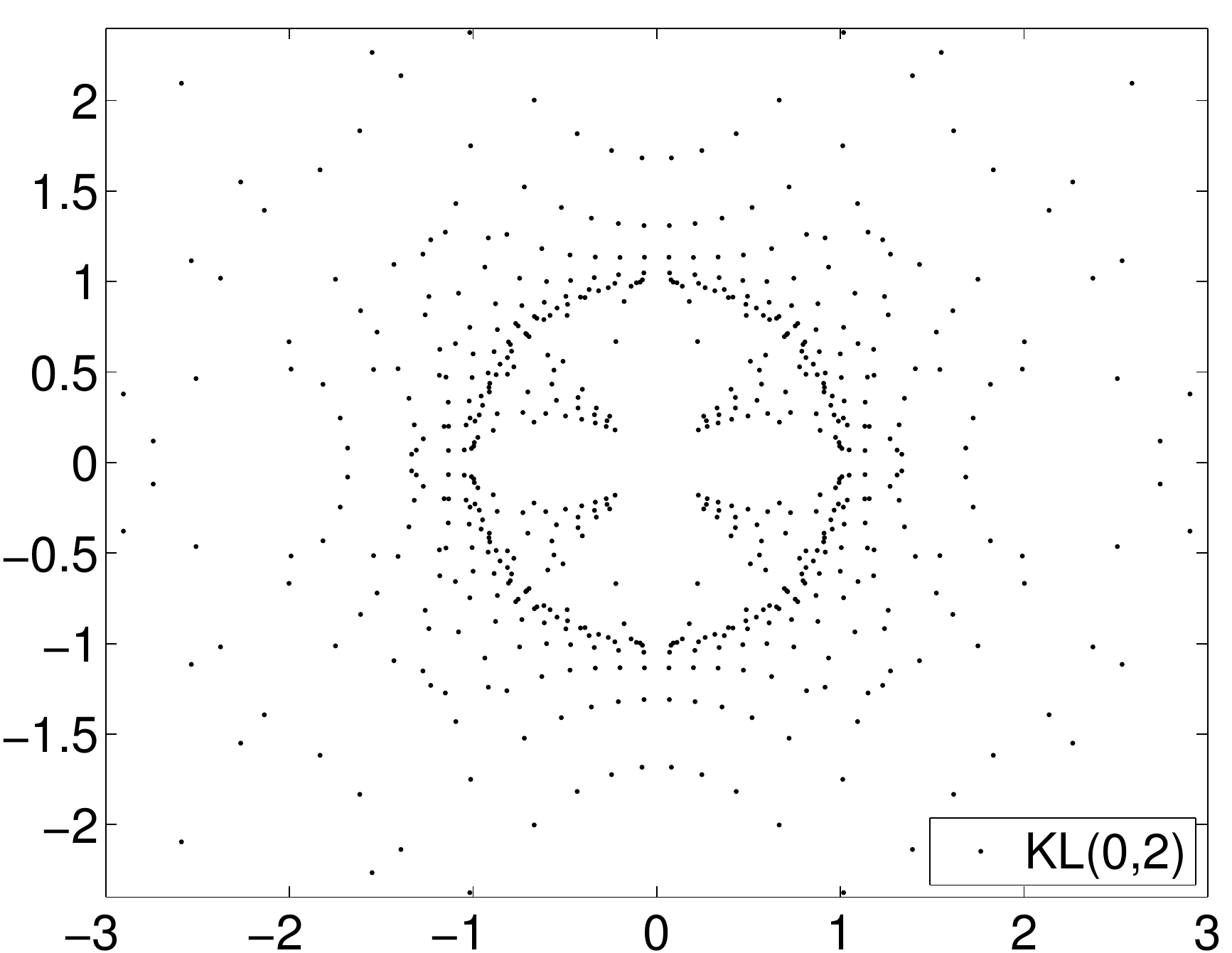}%
\\
\includegraphics[width=5.5cm]{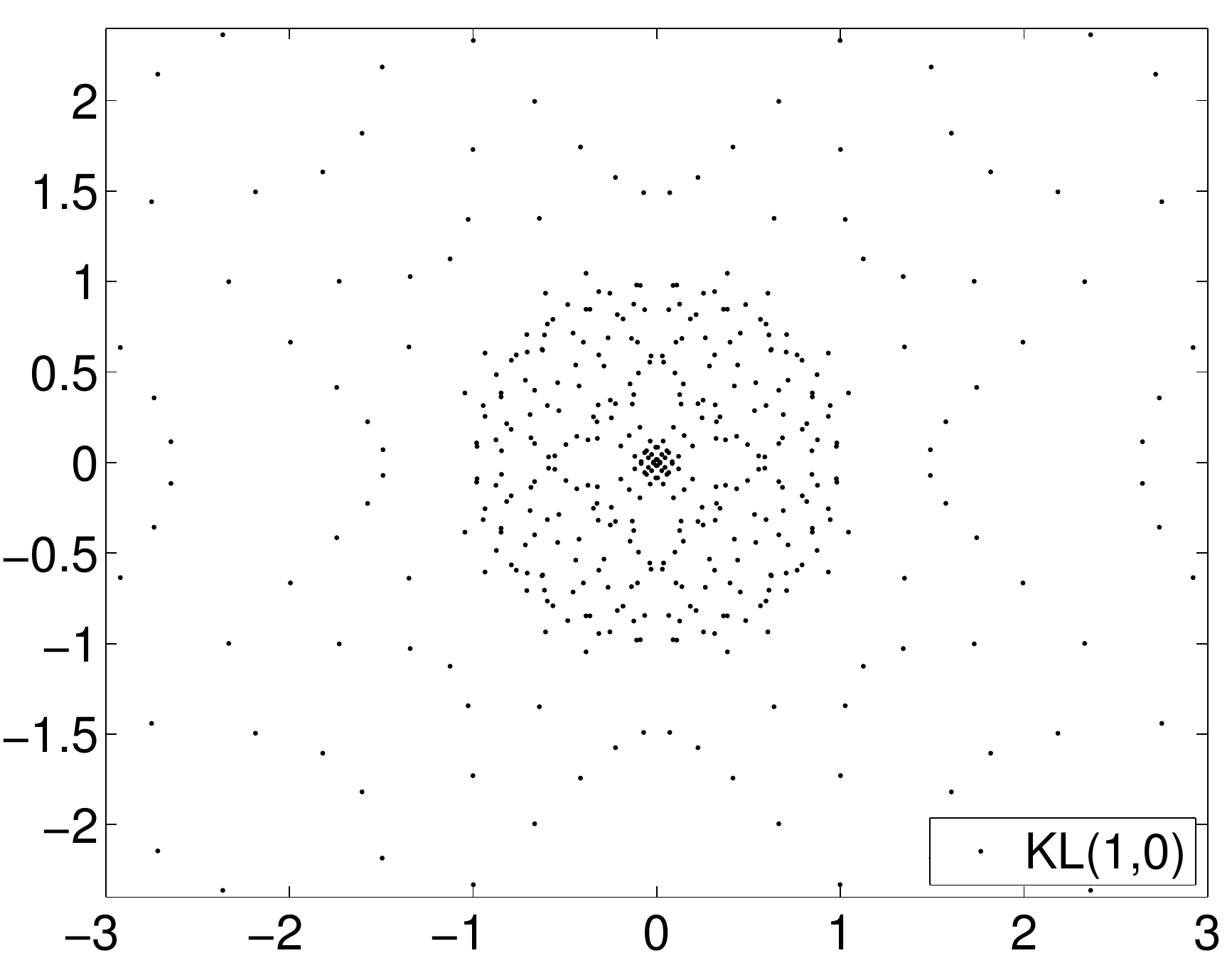}%
\includegraphics[width=5.5cm]{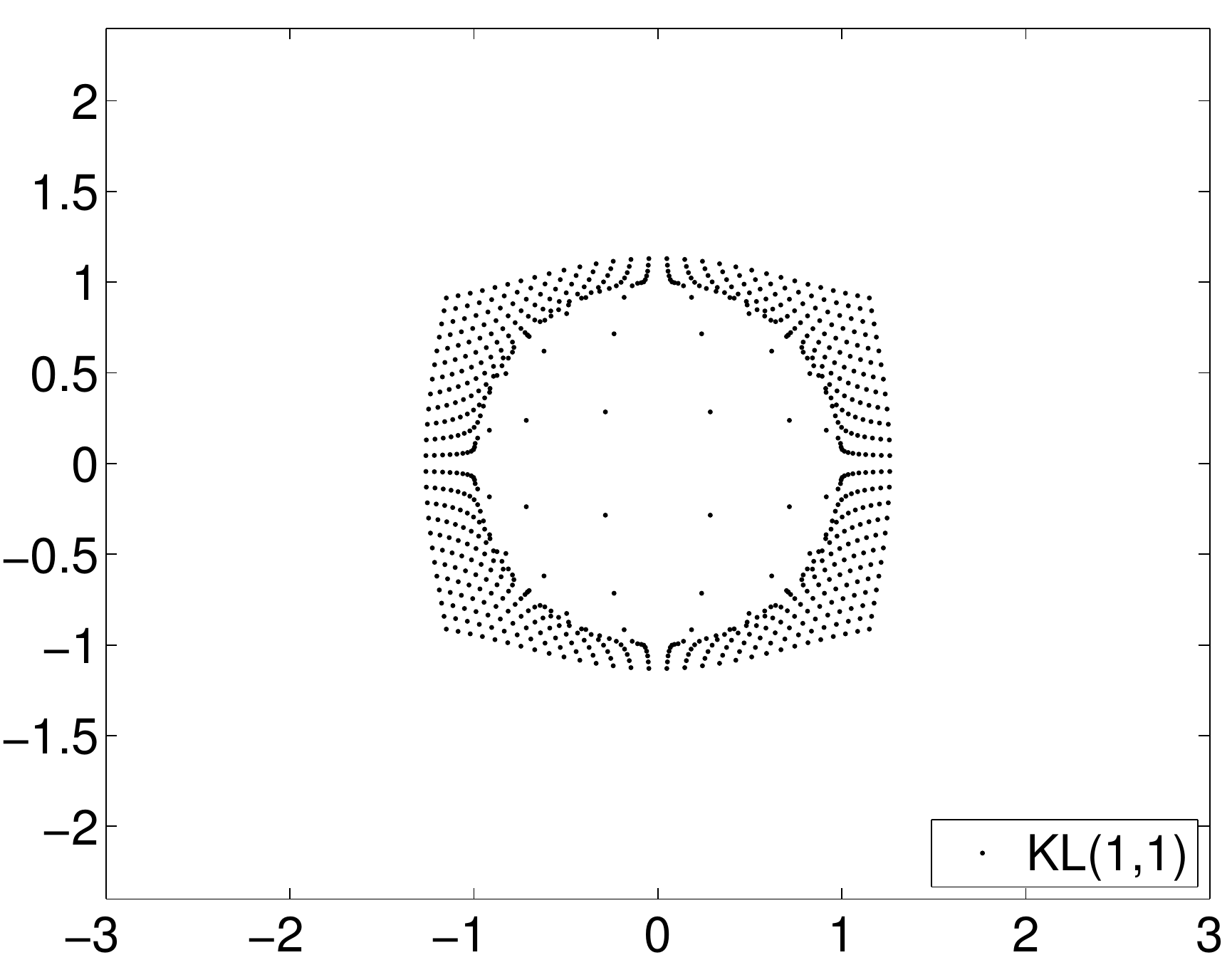}%
\includegraphics[width=5.5cm]{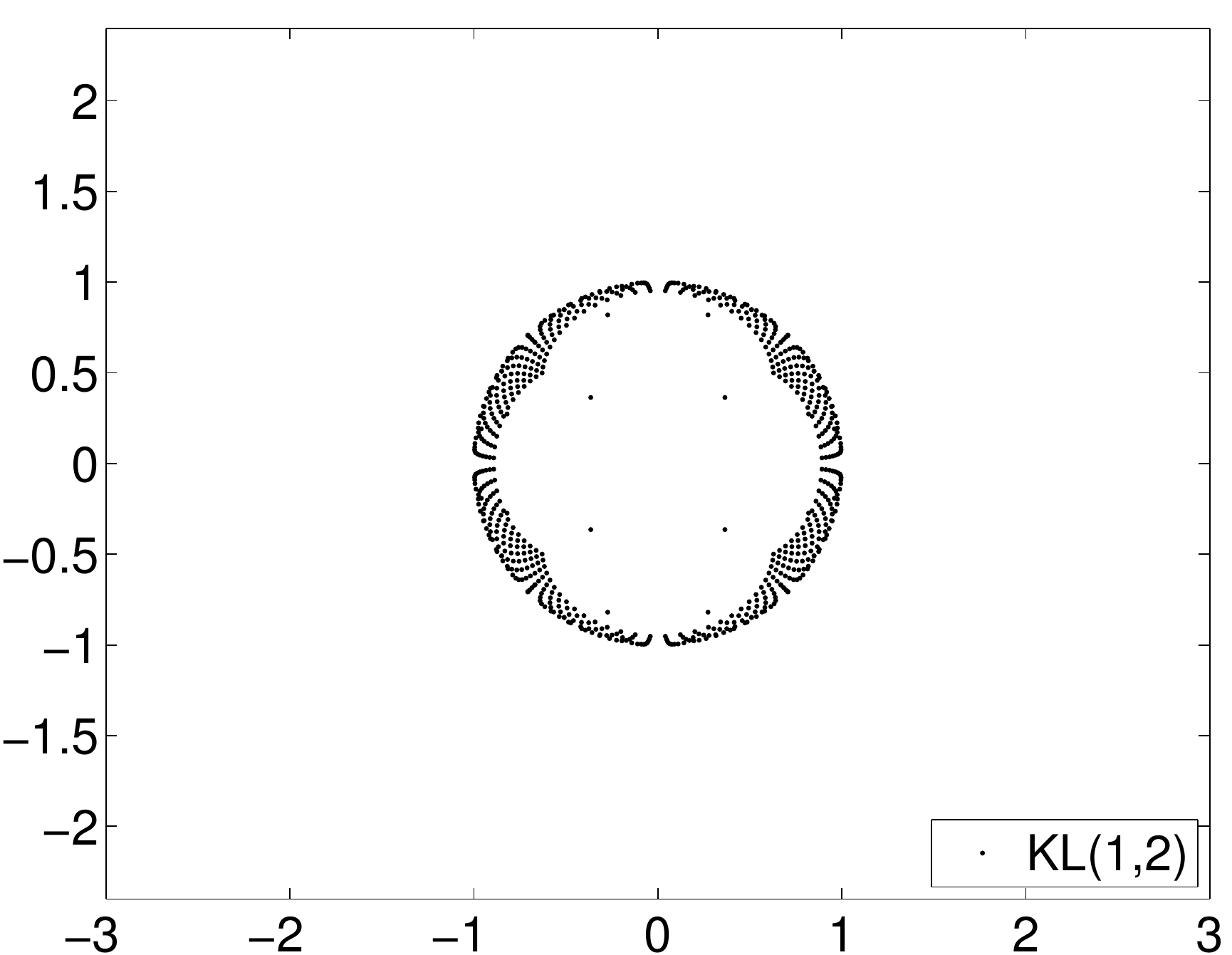}%
\\
\includegraphics[width=5.5cm]{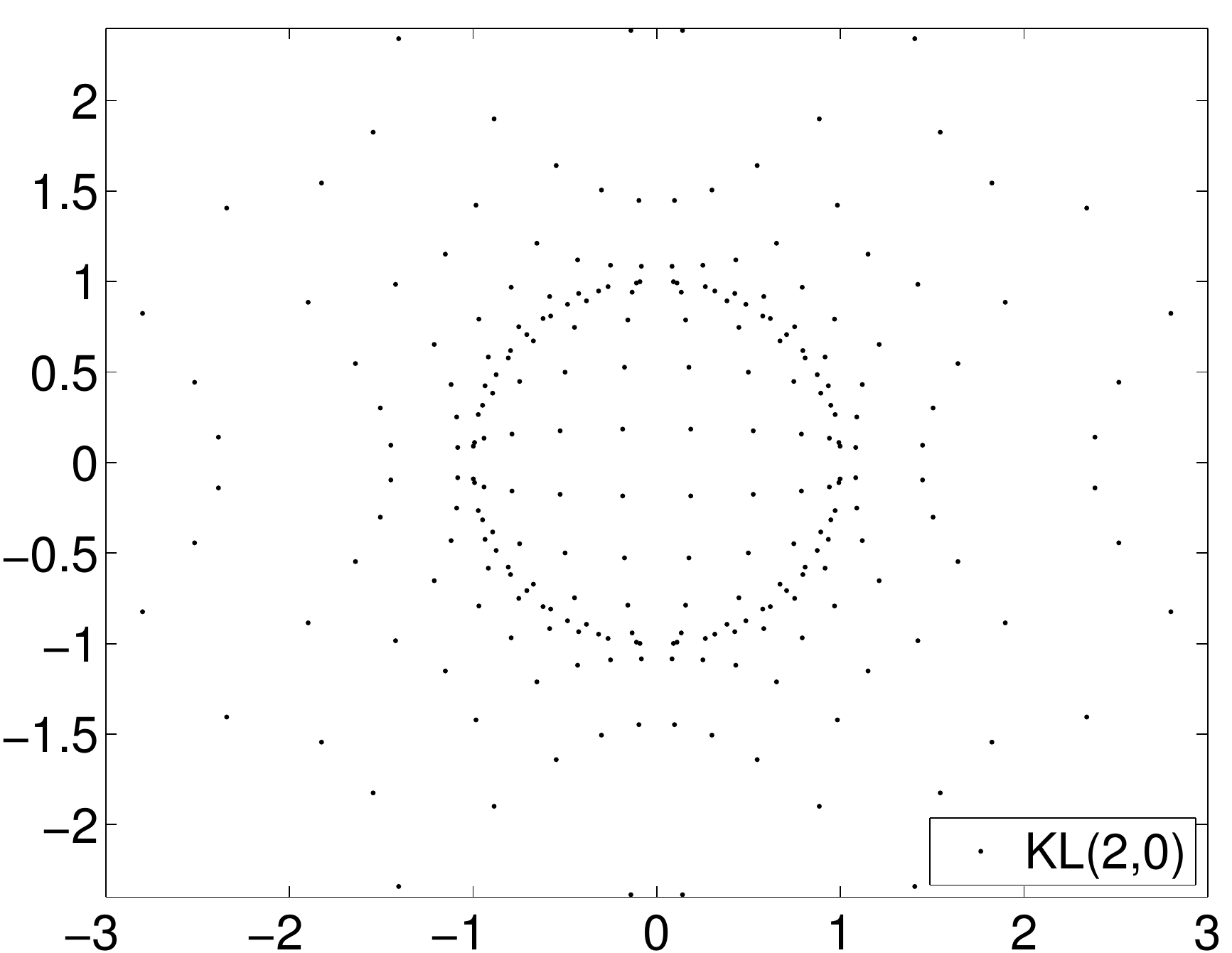}%
\includegraphics[width=5.5cm]{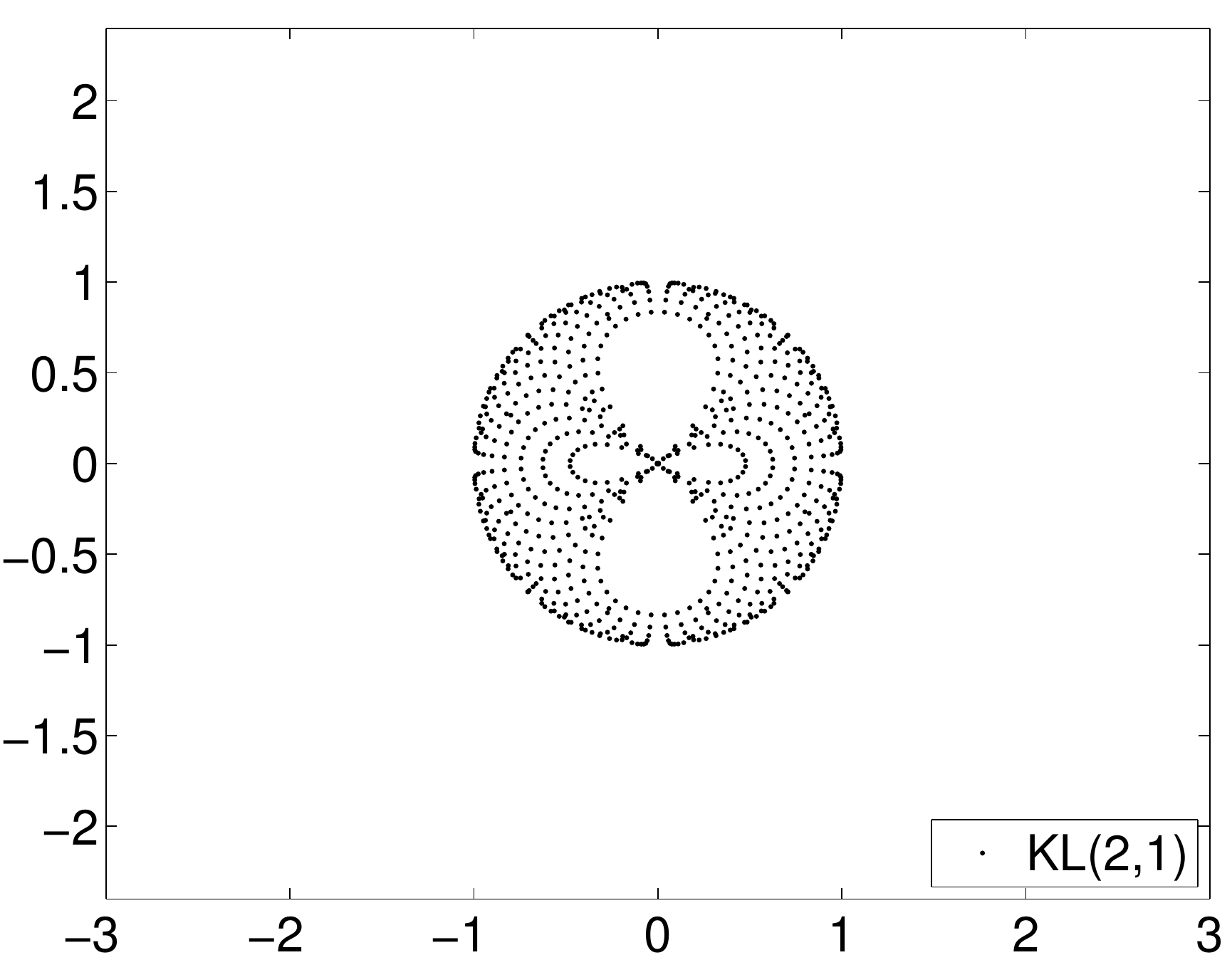}%
\includegraphics[width=5.5cm]{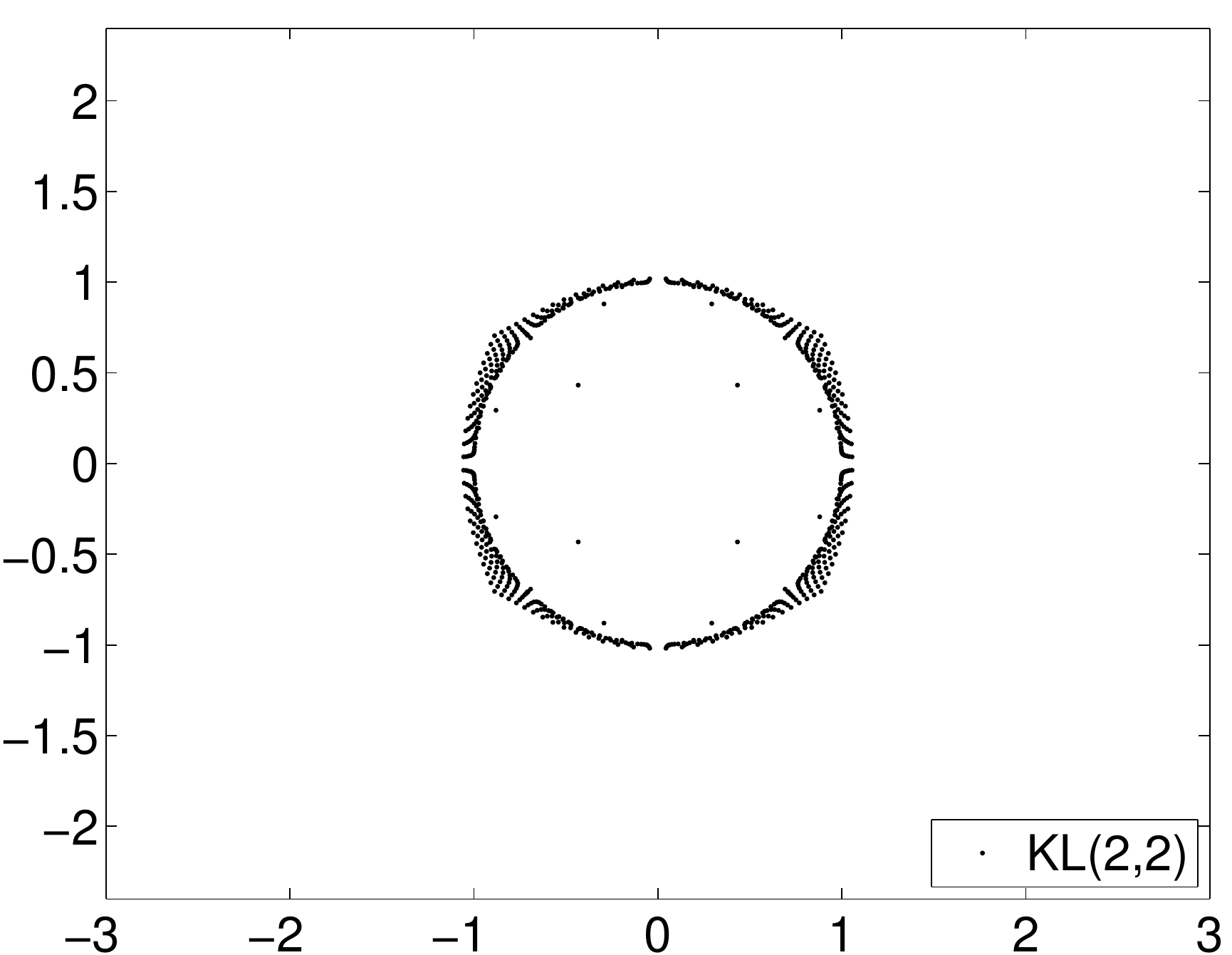}%
\caption{\label{fig:KL_unitarity}\sl
Image of regularly spaced points $x\!\in\!\mb{C}$ with $-3\!\leq\!\mr{Re}(x)\!\leq\!3$ and $-2.4\!\leq\!\mr{Im}(x)\!\leq\!2.4$ under one iteration of $f_{0,0}(.)$ [identity, top left] to $f_{2,2}(.)$ [bottom right] as defined in Tab.\,\ref{tab1}.}
\end{figure}

The effect of the Kenney-Laub mappings for the unitary projection [i.e.\ for (\ref{pade_def}) with $X_k^2\to X_k\dag X_k$] may be visualized in the complex plane.
The nine panels in Fig.\,\ref{fig:KL_unitarity} illustrate the effect of one operation of $f_{m,n}$ with $0\leq m,n \leq2$, as given in Tab.\,\ref{tab1}.
It seems plausible that the mappings on the diagonal and first upper diagonal are globally convergent, while mappings far away from the diagonal work only for near-unitary arguments $X\dag X\simeq I$.

\begin{figure}[!tbp]
\centering
\includegraphics[width=8.4cm]{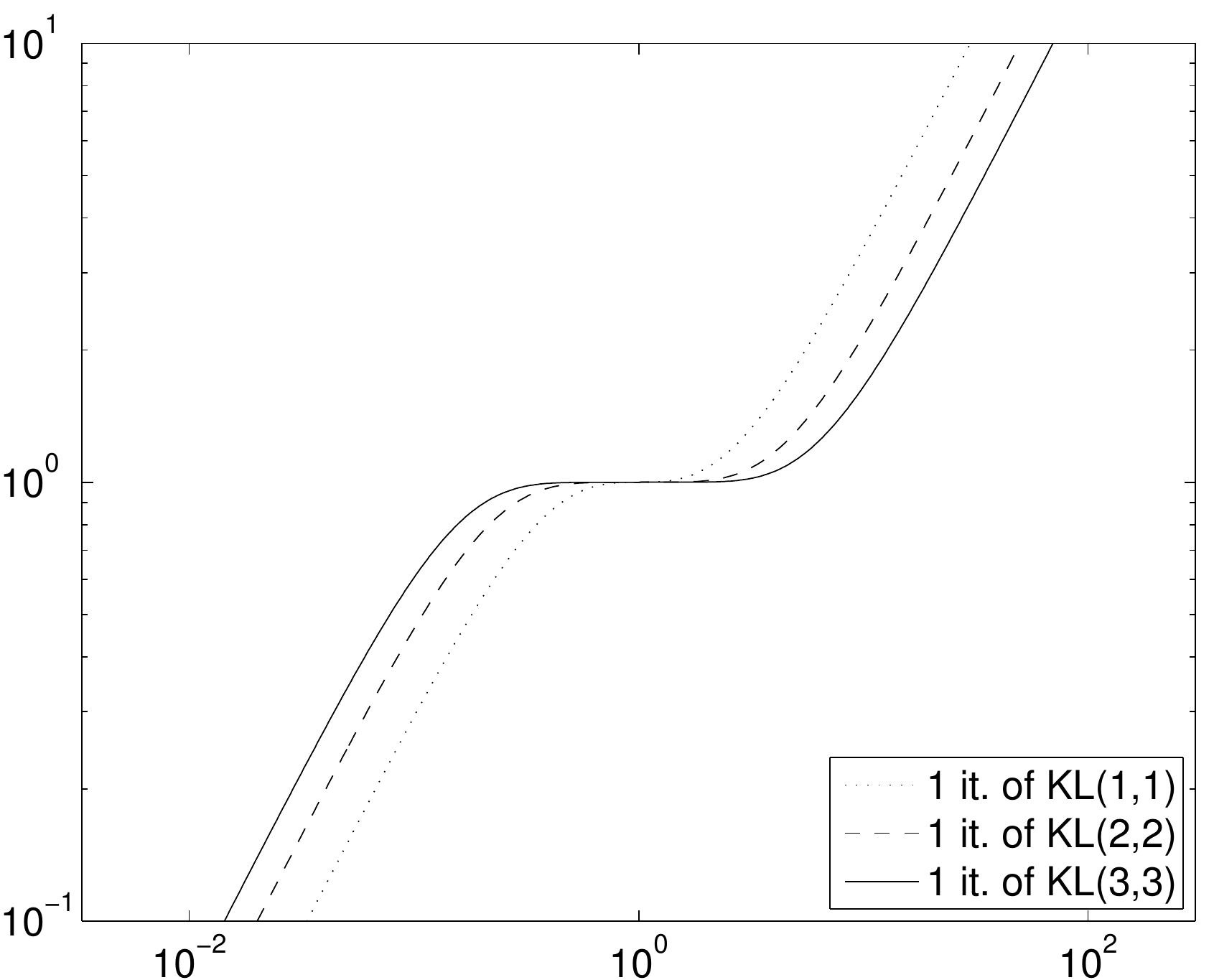}%
\includegraphics[width=8.4cm]{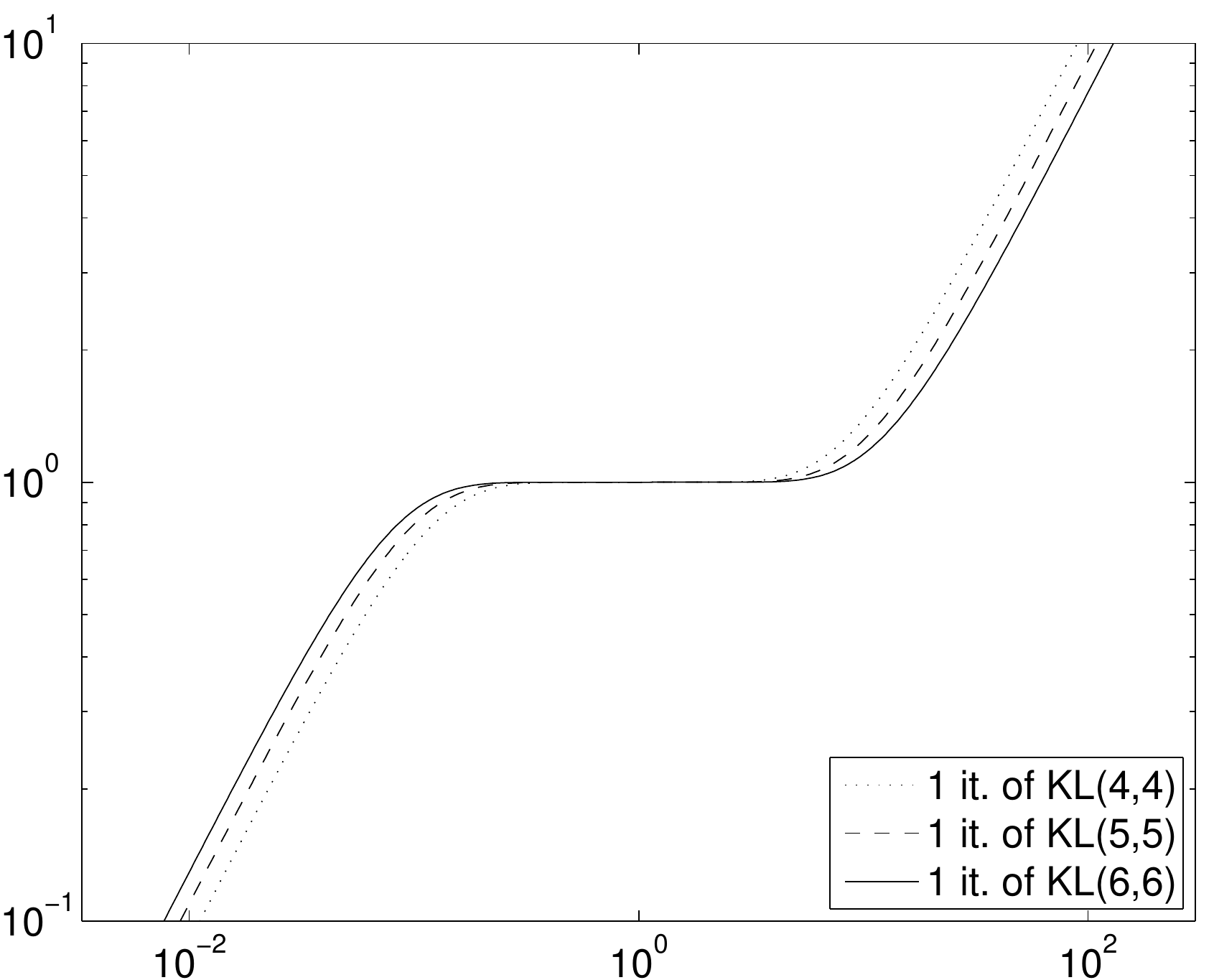}%
\\
\includegraphics[width=8.4cm]{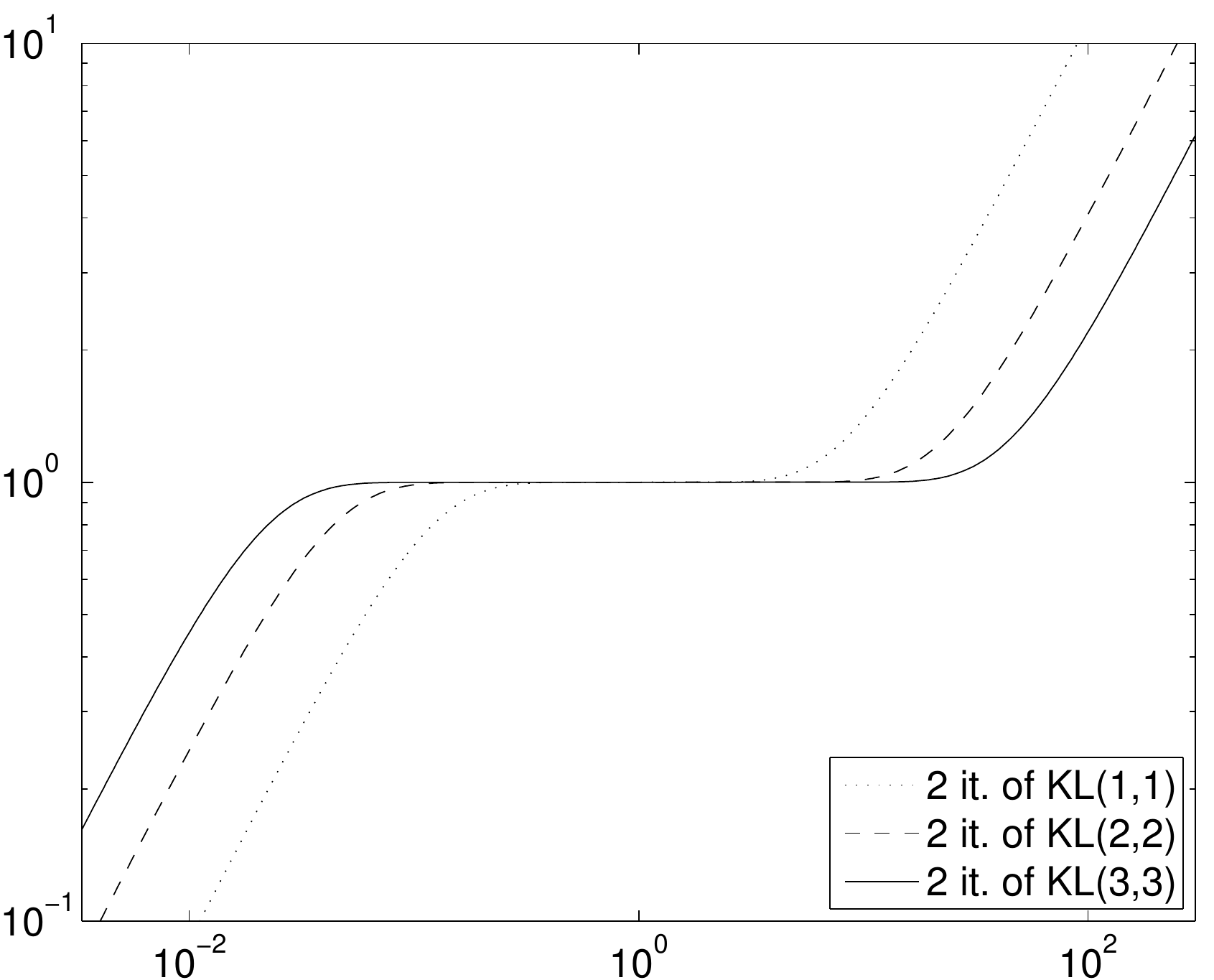}%
\includegraphics[width=8.4cm]{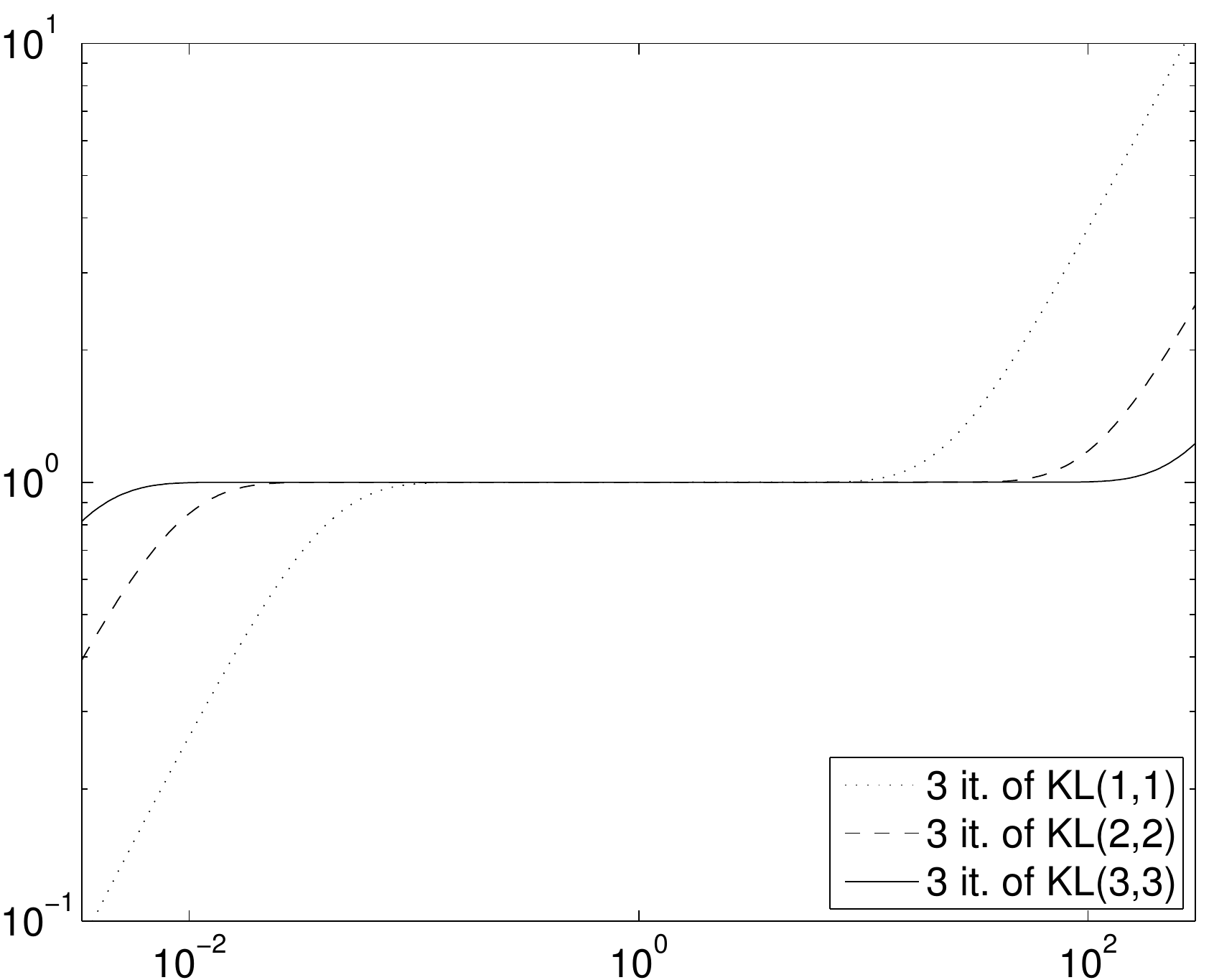}%
\\
\includegraphics[width=8.4cm]{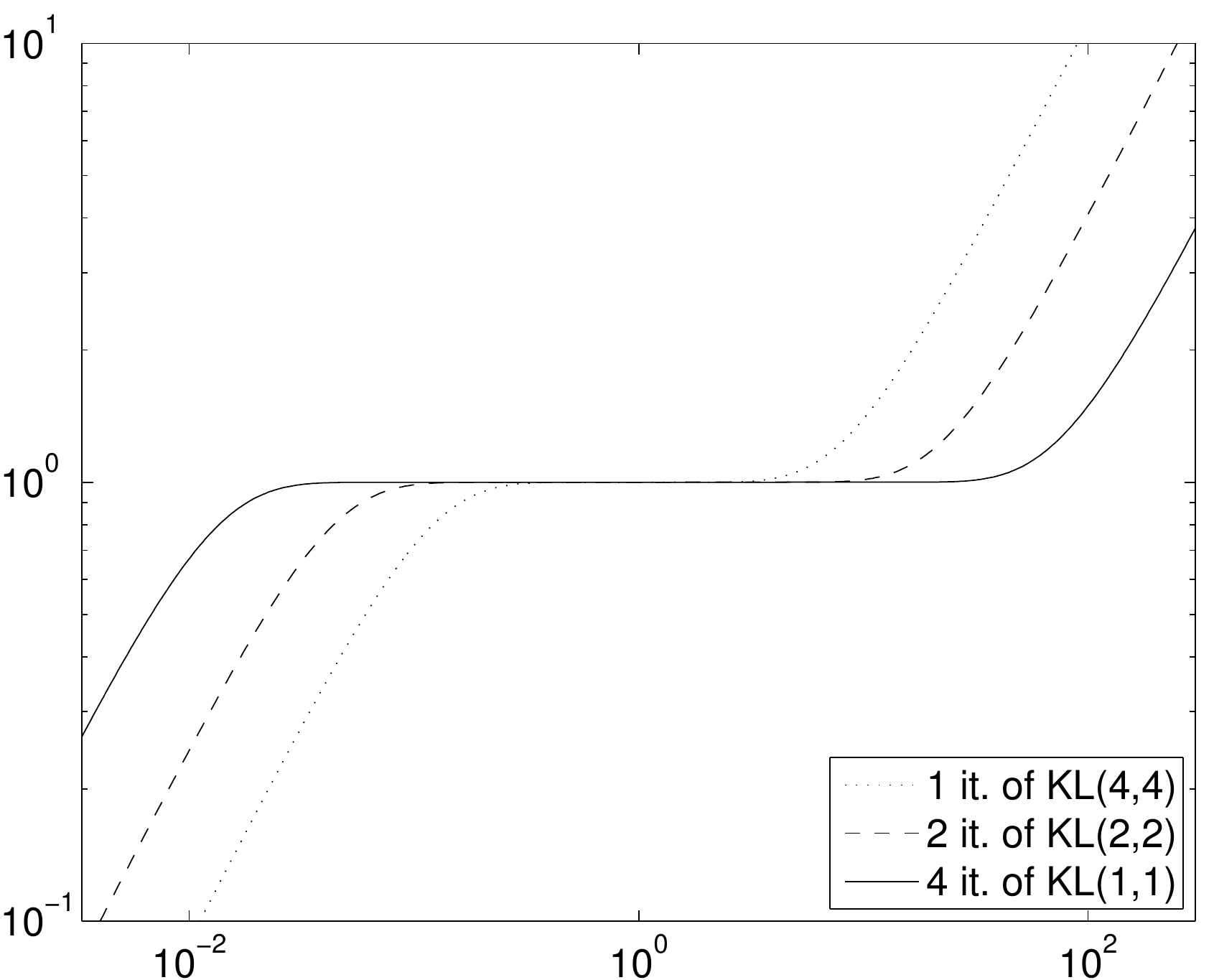}%
\includegraphics[width=8.4cm]{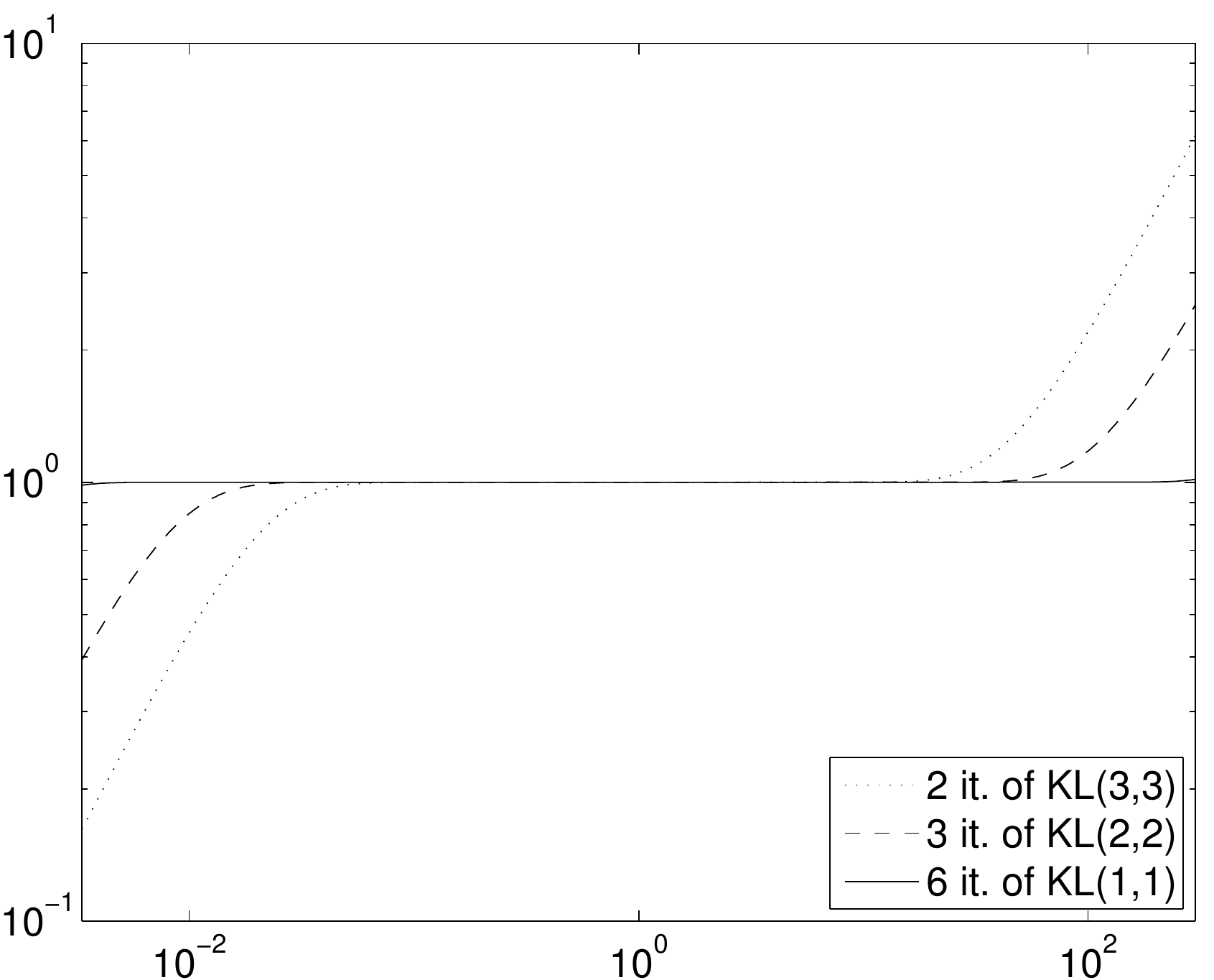}%
\caption{\label{fig:KL_signfunction}\sl
Image of the interval $x\in[0.003,300]$ under one or several iterations of the diagonal Kenney-Laub mappings $x\to f_{nn}(x)=x\,p_{nn}(x^2)/q_{nn}(x^2)$ in log-log representation.}
\end{figure}

To visualize the approximations to the sign function that derive from the Kenney-Laub mappings it suffices to restrict the discussion to $x>0$, since each $f_{m,n}(x)$ is an odd function of $x$.
The panels of Fig.\,\ref{fig:KL_signfunction}) illustrate various combinations of diagonal ($m=n$) mappings on the interval $x\in[0.003,300]$.
Evidently, these approximations work best for $x\simeq1$, with monotonically decreasing quality for small ($x\ll1$) and large ($x\gg1$) arguments.
That this decrease in quality is symmetric about $1$ is a direct consequence of (\ref{pade_symmetry}).

In lattice QCD the elements of the first upper diagonal have been used before \cite{Neuberger:1998my}.
The most obvious difference to the diagonal mappings which we advocate is that the former set of functions assumes a maximum/minimum at $x=\pm1$, respectively, while the diagonal functions $f_{n,n}(x)=g_{2n+1}(x)$ increase without any bound.
In a similar vein we emphasize that -- unlike optimal rational approximations \cite{Edwards:1998yw,vandenEshof:2002ms,Kennedy:2006ax} -- diagonal Kenney-Laub functions show \emph{no wiggles}; the value $\pm1$ at $x=\pm1$ is approached monotonically, both from the origin and from $\pm\infty$.


\subsection{Partial fraction and continued fraction representations}

In view of numerical applications let us rewrite the diagonal Kenney-Laub mappings in partial fraction form.
The first two diagonal mappings can be brought into the form
\bea
g_3(X)=f_{1,1}(X)&=&\frac{X}{3}
\Big(1+\frac{8/3}{X\dag X+1/3}\Big)
\label{KL11_parfrac}
\\
g_5(X)=f_{2,2}(X)&=&\frac{X}{5}
\Big(1+\frac{4(1-1/\sqrt{5})}{X\dag X+1-2/\sqrt{5}}
      +\frac{4(1+1/\sqrt{5})}{X\dag X+1+2/\sqrt{5}}\Big)
\label{KL22_parfrac}
\eea
while for higher $n$ the roots of the denominator polynomial can only be given over the field of complex numbers (though they happen to be real).
The general formula reads \cite{Higham}
\beq
g_{2n+1}(x)=\frac{x}{2n+1}
\sum_{i=0}^{n}\frac{2-\de_{i,n}}{\sin^2((2i+1)\pi/(4n+2))+\cos^2((2i+1)\pi/(4n+2))x^2}
\label{KLnn_parfrac}
\eeq
and from the explicit form provided in Tab.\,\ref{tab5} of the appendix it is easy to see that the smallest shift gets progressively smaller with increasing $n$.
Moreover, the coefficients in the numerator are all positive, and they grow synchronously with the shift in the denominator.
This formula is reminiscent of the one for the first upper diagonal \cite{Higham,Neuberger:1998my}
\beq
g_{2n}(x)=\frac{x}{n}
\sum_{i=0}^{n-1}\frac{1}{\sin^2((2i+1)\pi/(4n))+\cos^2((2i+1)\pi/(4n))x^2}
\eeq
except that the former expression has a constant contribution ($i=n$), while the latter one has not.
The bottom line is that one can use a multi-shift conjugate gradient (CG) solver to evaluate $f_{n,n}(X)v$ on a given vector $v$ \cite{Frommer:1995ik,Jegerlehner:1996pm}.
In our view it is convenient that the coefficients can be worked out beforehand, i.e.\ independent of the spectral properties of $A\equiv X\dag X$.

For $f_{n,n}$ with $n\geq2$ also a continued fraction representation can be given, for instance
\bea
f_{2,2}(X)&=&\frac{X}{5}
\bigg(1+\frac{8}{X\dag X+7/5-\frac{16/25}{X\dag X+3/5}}\bigg)
\label{KL22_confrac}
\\
f_{3,3}(X)&=&\frac{X}{7}
\bigg(1+\frac{16}{X\dag X+3-\frac{24/7}{X\dag X+5/3-\frac{8/63}{X\dag X+1/3}}}\bigg)
\\
f_{4,4}(X)&=&\frac{X}{9}
\bigg(1+\frac{80/3}{X\dag X+77/15-\frac{264/25}{X\dag X+139/45-\frac{520/891}{X\dag X+103/117-\frac{96/1859}{X\dag X+3/13}}}}\bigg)
\\
f_{5,5}(X)&=&\frac{X}{11}
\bigg(1+\frac{40}{X\dag X+39/5-\frac{624/25}{X\dag X+73/15-\frac{160/99}{X\dag X+61/39-\frac{408/1859}{X\dag X+131/221-\frac{8/289}{X\dag X+3/17}}}}}\bigg)
\\
f_{6,6}(X)&=&\frac{X}{13}
\bigg(1+\frac{56}{X\dag X+11-\frac{352/7}{X\dag X+7-\frac{272/77}{X\dag X+31/13-\frac{1064/1859}{X\dag X+227/221-\frac{616/5491}{X\dag X+53/119-\frac{16/931}{X\dag X+1/7}}}}}}\bigg)
\eea
where all coefficients are found to be given by (small-over-small) rational numbers.


\section{Overlap operator construction \label{sec:overlap}}


Given any undoubled (or doubled but with one chirality in the physical branch) ``kernel'' Dirac operator $\Dke_m$ at a quark mass $m$, the massless overlap operator $\Dov$ is defined as a backshifted version of the (unique) unitary part of the kernel at negative mass $-\rh/a$ \cite{Neuberger:1997fp,Neuberger:1998wv}
\beq
a\Dov \equiv a\Dov_0=
\left\{
\begin{array}{l}
\displaystyle
\rh\,
\Big[a\Dke_{-\rh/a}(a^2D_{-\rh/a}^{\mr{ke}\,\dagger}\Dke_{-\rh/a})^{-1/2}+1\Big]=
\rh\,
\Big[\gaf\,\mr{sign}(\gaf a\Dke_{-\rh/a})+1\Big]
\\[4mm]
\displaystyle
\rh\,
\Big[(a^2\Dke_{-\rh/a}D_{-\rh/a}^{\mr{ke}\,\dagger})^{-1/2}a\Dke_{-\rh/a}+1\Big]=
\rh\,
\Big[\mr{sign}(a\Dke_{-\rh/a} \gaf)\,\gaf+1\Big]
\end{array}
\right.
\label{def_over}
\eeq
where $0\!<\!\rh\!<\!2$ is an arbitrary parameter (its canonical value is $1$).
The equivalence of the two lines follows from the singular value decomposition $a\Dke_{-\rh/a}=USV\dag$ with unitary $U,V$ and $S>0$, by means of which $a^2D_{-\rh/a}^{\mr{ke}\;\dagger}\Dke_{-\rh/a}=VS^2V\dag$ and $a^2\Dke_{-\rh/a}D_{-\rh/a}^{\mr{ke}\;\dagger}=US^2U\dag$.
This implies $a\Dov=\rh[USV\dag VS^{-1}V\dag+1]=\rh[UV\dag+1]$ and $a\Dov=\rh[US^{-1}U\dag USV\dag+1]=\rh[UV\dag+1]$, respectively, which completes the proof.
Note that the reformulation in terms of the matrix sign function in eqn.\,(\ref{def_over}) holds only if the kernel is $\gaf$-hermitean, i.e.\ $\gaf\Dke\gaf=D^\mr{ke}{}\dag$.

The massless overlap operator (\ref{def_over}) fulfills the Ginsparg-Wilson (GW) relation \cite{Ginsparg:1981bj}
\beq
D \gaf+\gaf D = \frac{a}{\rh} D \gaf D
\quad\Longleftrightarrow\quad
D\gaf(1-\frac{aD}{2\rh})+(1-\frac{aD}{2\rh})\gaf D=0
\label{ginswils_one}
\eeq
and $\Dov$ is thus said to be ``chirally symmetric'', regardless the details of the kernel.
In practice there is a choice to be made regarding the type of kernel (e.g.\ Wilson or Brillouin), how much link smearing one wants to apply, and whether the kernel shall be equipped with a clover term.
Whenever $\Dke$ is $\gaf$-hermitean, this property extends to $\Dov$, and in this case multiplying (\ref{ginswils_one}) with $\gaf$ from the left or the right yields (note that $[D,D\dag]=0$ is implied)
\beq
D\dag + D = \frac{a}{\rh} D\dag D = \frac{a}{\rh} D D\dag
\quad\Longleftrightarrow\quad
D\dag(1-\frac{aD}{2\rh})+(1-\frac{aD}{2\rh})\dag D=0
\label{ginswils_two}
\;.
\eeq


\subsection{Kenney-Laub iterates of shifted Dirac kernels}

For the sake of clarity let us consider the use of a diagonal Kenney-Laub mapping $f_{n,n}$ to define, for a given kernel $a\Dke$, an approximation to the overlap operator (\ref{def_over}).
With $X=a\Dke-\rh$ the relation $Y=X(X\dag X+3)(3X\dag X+1)^{-1}$ defines a Dirac operator $a\Dit=\rh[Y+1]$ with improved chiral symmetry.
After another iteration, which may involve a different mapping, for instance $Z=Y(Y\dag YY\dag Y+10Y\dag Y+5)(5Y\dag YY\dag Y+10Y\dag Y+1)^{-1}$, the redefinition $a\Dit=\rh[Z+1]$ yields a Dirac operator with an even smaller violation of the relation (\ref{ginswils_one}).

In usual applications one cannot hold any of these matrices in memory.
The challenge is thus to implement the forward application $\Dit x$ on a given vector $x$ in such a form that everything boils down to repeated matrix-vector multiplications of the form $D^{\mr{ke}\dagger}\Dke y$ and $\Dke z$.

\begin{table}[!tb]
\centering
\begin{tabular}{|c|cccccc|}
\hline
 & $f_{1,1}^{(k)}$ & $f_{2,2}^{(k)}$ & $f_{3,3}^{(k)}$ & $f_{4,4}^{(k)}$ & $f_{5,5}^{(k)}$ & $f_{6,6}^{(k)}$
\\
\hline
$k=1$ & $2.7\,10^{-  3}$ & $3.4\,10^{-  5}$ & $4.2\,10^{-  7}$ & $5.2\,10^{-  9}$ & $6.3\,10^{-  11}$ & $7.9\,10^{-  13}$ \\
$k=2$ & $5.2\,10^{-  9}$ & $2.8\,10^{- 24}$ & $3.5\,10^{- 47}$ & $1.0\,10^{- 77}$ & $6.9\,10^{- 116}$ & $1.1\,10^{- 161}$ \\
$k=3$ & $3.4\,10^{- 26}$ & $1.0\,10^{-119}$ & $1.0\,10^{-327}$ & $4.6\,10^{-696}$ & $1.6\,10^{-1270}$ & $6.8\,10^{-2097}$ \\
\hline
\end{tabular}
\caption{\label{tab:wouldbe}\sl
Image of the would-be zero-mode $\la=0.2$ under 1 to 3 iterations of the diagonal Kenney-Laub mappings $\la \to f_{n,n}(\la\!-\!1)+1$ as defined in Tabs.\,\ref{tab1},\,\ref{tab2}.
The order of convergence (in $k$) of the columns is $3,5,7,9,11,13$, respectively.}
\end{table}

For the diagonal iterations $f_{n,n}$ the image of the ``would-be zero-mode'' $\la=0.2$ of $\Dke$ under $1$ to $3$ iterations is summarized in Tab.\,\ref{tab:wouldbe}.
For such a mode and in double-precision arithmetics the mappings $f_{1,1}$ and $f_{2,2}$ achieve exact chiral symmetry after $3$ and $2$ iterations, respectively.
Note that the nesting formula (\ref{nesting}) says $f_{1,1}^{(3)}=g_3^{(3)}=g_{27}^{(1)}=f_{13,13}^{(1)}$ and $f_{2,2}^{(2)}=g_5^{(2)}=g_{25}^{(1)}=f_{12,12}^{(1)}$.


\subsection{Massive overlap action -- traditional version}

Let $\la$ be an eigenvalue of the massless overlap operator $a\Dov$ with parameter $\rh$, i.e.\ $\la=\rh(1+e^{\ri\vp})$ with $\vp\in]0,2\pi[$.
This circular eigenvalue spectrum is mapped onto the imaginary axis through the stereographic projection $\la\to\til\la\equiv\la/(1-\la/[2\rh])=2\ri\rh/\tan(\vp/2)$.
The massive overlap operator follows by shifting this line by $am$ to the right, and inverting the mapping.

The traditional way of doing this is to multiply $\til\la+am$ with the factor $(1-\la/[2\rh])$ which then leads to $\la+am(1-\la/[2\rh])$.
In operator language this means that the massive overlap operator $\Dov_m$ follows by adding a ``chirally rotated'' scalar term
\cite{Neuberger:1997fp,Neuberger:1998wv}
\bea
D_m^\mr{tra}&\equiv&
\Dov+m\Big(1-\frac{a}{2\rh}\Dov\Big)=
\Big(1-\frac{am}{2\rh}\Big)\Dov+m
\nonumber
\\
&=&
\Big(\frac{\rh}{a}-\frac{m}{2}\Big)
\gaf\,\mr{sign}(\gaf a\Dke_{-\rh/a})+
\Big(\frac{\rh}{a}+\frac{m}{2}\Big)
\label{def_mass_traditional}
\eea
which yields an operator with a circular eigenvalue spectrum of radius $\rh-am/2$ around the point $(\rh+am/2,0)$ in the complex plane.
Obviously this implies the constraint $am<2\rh$.
An ad hoc way of removing this constraint would be to replace $m$ in (\ref{def_mass_traditional}) by $\bar{m}\equiv1/(1/m\!+\!a/[2\rh])=m/(1+am/[2\rh])$, so that $a\bar{m}\leq2\rh$ for all $am$.
With the traditional definition (\ref{def_mass_traditional}) solving the massive Dirac equation $D_m^\mr{tra}x=b$ for $x$ with a given right-hand side $b$ is equivalent to solving
\beq
\Big(\Dov+\til{m}\Big)\,x=\til{b}
\quad\mbox{with}\quad
\til{m}=\frac{m}{1-am/[2\rh]}
\quad\mbox{and}\quad
\til{b}=\frac{b}{1-am/[2\rh]}=
\frac{\til{m}}{m}\,b
\label{upshot_traditional}
\eeq
for $x$, with the massless $\Dov$ defined in (\ref{def_over}).


\subsection{Massive overlap action -- complete version}


Alternatively, one might start from the proper inversion of the stereographic mapping, which is $\til\la\to\til\la/(1+\til\la/[2\rh])=\la$, and by adding the mass to $\til\la$ one ends up with the massive spectrum
\bdm
\frac{\til\la+am}{1+\til\la/[2\rh]+am/[2\rh]}=
\frac{\la/(1-\la/[2\rh])+am}{1+\la/(1-\la/[2\rh])/[2\rh]+am/[2\rh]}=
\frac{\la+am(1-\la/[2\rh])}{1+am(1-\la/[2\rh])/[2\rh]}
\edm
which does not entail any constraint on $am$ (with $am\to\infty$ the eigenvalue spectrum shrinks to a point at $2\rh$).
In operator language this means that the complete definition
\bea
D_m^\mr{com}&\equiv&
\frac{\Dov+m\Big(1-\frac{a}{2\rh}\Dov\Big)}{1+\frac{am}{2\rh}\Big(1-\frac{a}{2\rh}\Dov\Big)}=
\frac{(1-\frac{am}{2\rh})\Dov+m}{(1+\frac{am}{2\rh})-\frac{a^2m}{2\rh^2}\Dov}
\nonumber
\\
&=&
\frac{(\frac{\rh}{a}-\frac{m}{2})\gaf\,\mr{sign}(\gaf a\Dke_{-\rh/a})+(\frac{\rh}{a}+\frac{m}{2})}
     {1-\frac{am}{2\rh}\gaf\,\mr{sign}(\gaf a\Dke_{-\rh/a})}
\label{def_mass_complete}
\eea
looks superficially similar to the Moebius kernel that was proposed for the massless case \cite{Brower:2004xi}.
Note that the fractional notation in (\ref{def_mass_complete}) is well-defined, since the normality of $\Dov$ ensures that the numerator and the inverse of the denominator would commute.
Hence with the definition (\ref{def_mass_complete}) solving the massive Dirac equation $D_m^\mr{com}x=b$ for $x$ with a given $b$ amounts to solving
\beq
\Big(\Dov+\til{m}\Big)\,x=\til{b}
\quad\mbox{with}\quad
\til{m}=\frac{m}{1-am/[2\rh]}
\quad\mbox{and}\quad
\til{b}=\frac{1+am/[2\rh]\,(1-a\Dov/[2\rh])}{1-am/[2\rh]}\;b
\label{upshot_complete}
\eeq
for the vector $x$, with the massless $\Dov$ defined in (\ref{def_over}).
A comparison with (\ref{upshot_traditional}) shows that the procedure is the same, except that the right-hand side $\til{b}$ is now defined in a different manner.


\subsection{Massive overlap action -- proof of equivalence}

It is well known that the traditional form (\ref{upshot_traditional}) of the massive overlap action is to be used in conjunction with a ``chiral symmetry ensuring factor'' $(1-aD/[2\rh])$ to be attached to the external densities $S$, $P$ and currents $V_\mu$, $A_\mu$.
This leads to an effective Green's function (propagator)
\beq
S_m^\mr{tra}=
\frac{1-\frac{a}{2\rh}\Dov}{\Dov+m\Big(1-\frac{a}{2\rh}\Dov\Big)}=
\frac{1}{\frac{\Dov}{1-\frac{a}{2\rh}\Dov}+m}
\label{inv_traditional}
\eeq
which, thanks to the ``extra prescription'', has the same form as in the continuum \cite{Chiu:1998gp,Kikukawa:1999sy,Capitani:1999uz,Liu:2002qu}.

In the complete approach, when inverting (\ref{def_mass_complete}) without any extra factors one arrives at
\beq
S_m^\mr{com}=
\frac{1+\frac{am}{2\rh}\Big(1-\frac{a}{2\rh}\Dov\Big)}{\Dov+m\Big(1-\frac{a}{2\rh}\Dov\Big)}=
\frac{\frac{1}{1-\frac{a}{2\rh}\Dov}+\frac{am}{2\rh}}{\frac{\Dov}{1-\frac{a}{2\rh}\Dov}+m}
\label{inv_complete}
\eeq
which differs from (\ref{inv_traditional}) by just a contact term ($I$ is the identity)
\beq
S_m^\mr{com}-S_m^\mr{tra}=\frac{a}{2\rh}I
\;.
\label{contact}
\eeq

In short, we recommend to abandon the definition (\ref{def_mass_traditional}, \ref{upshot_traditional}) and to use (\ref{def_mass_complete}, \ref{upshot_complete}) instead.
In this complete form chiral symmetry is genuinely built in, and there is no need for invoking any ``extra prescription'' if decay constants and other matrix elements are to be determined.


\section{Eigenvalue spectra with Wilson and Brillouin kernel \label{sec:eigs}}


To gain an understanding of the difference between an approximate overlap operator with Wilson kernel and the same fixed-order approximant with the Brillouin kernel it is useful to take a look at the eigenvalue spectrum of either kernel on a given background.

\begin{figure}[!tb]
\centering
\includegraphics[width=8.4cm]{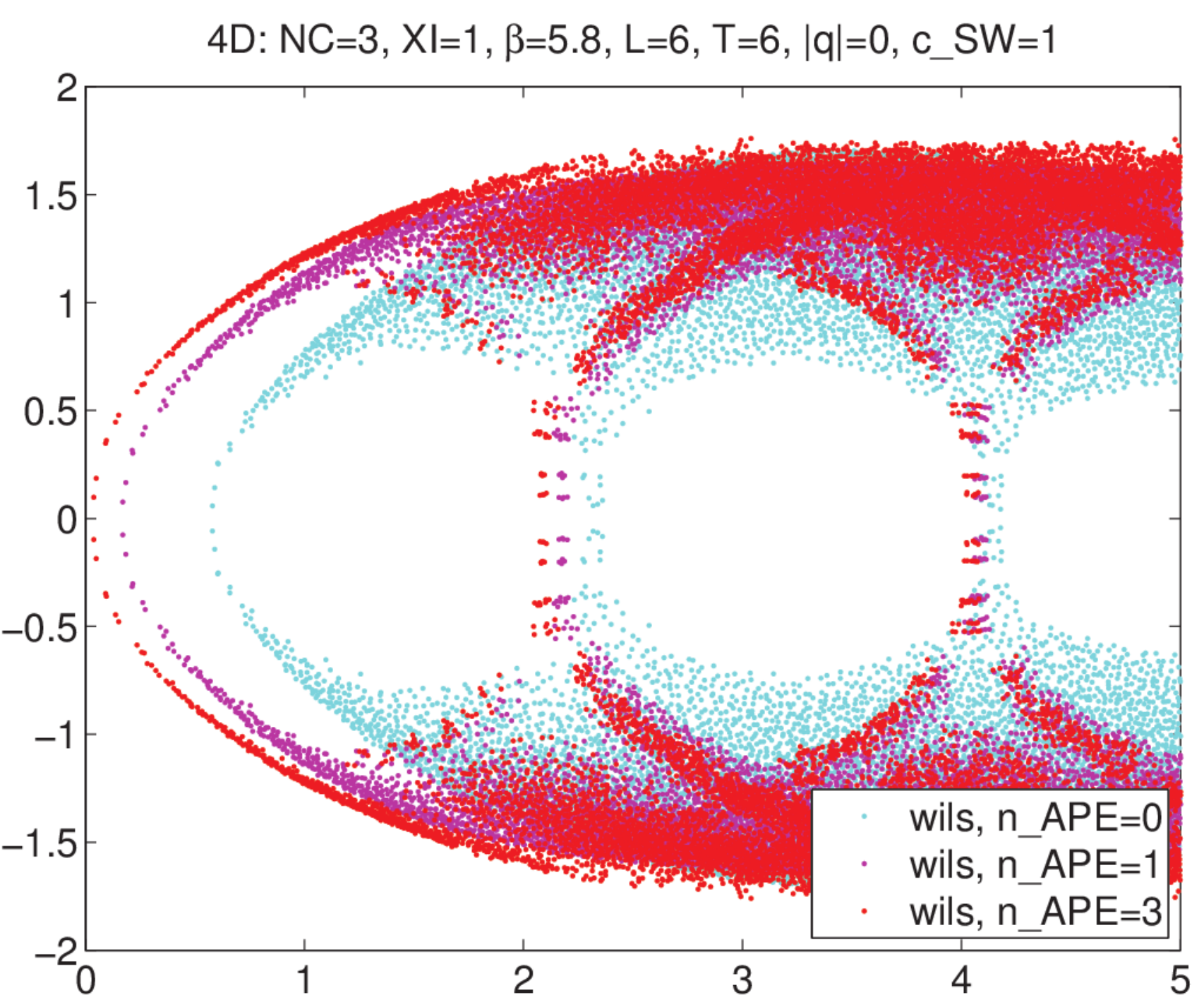}%
\includegraphics[width=8.4cm]{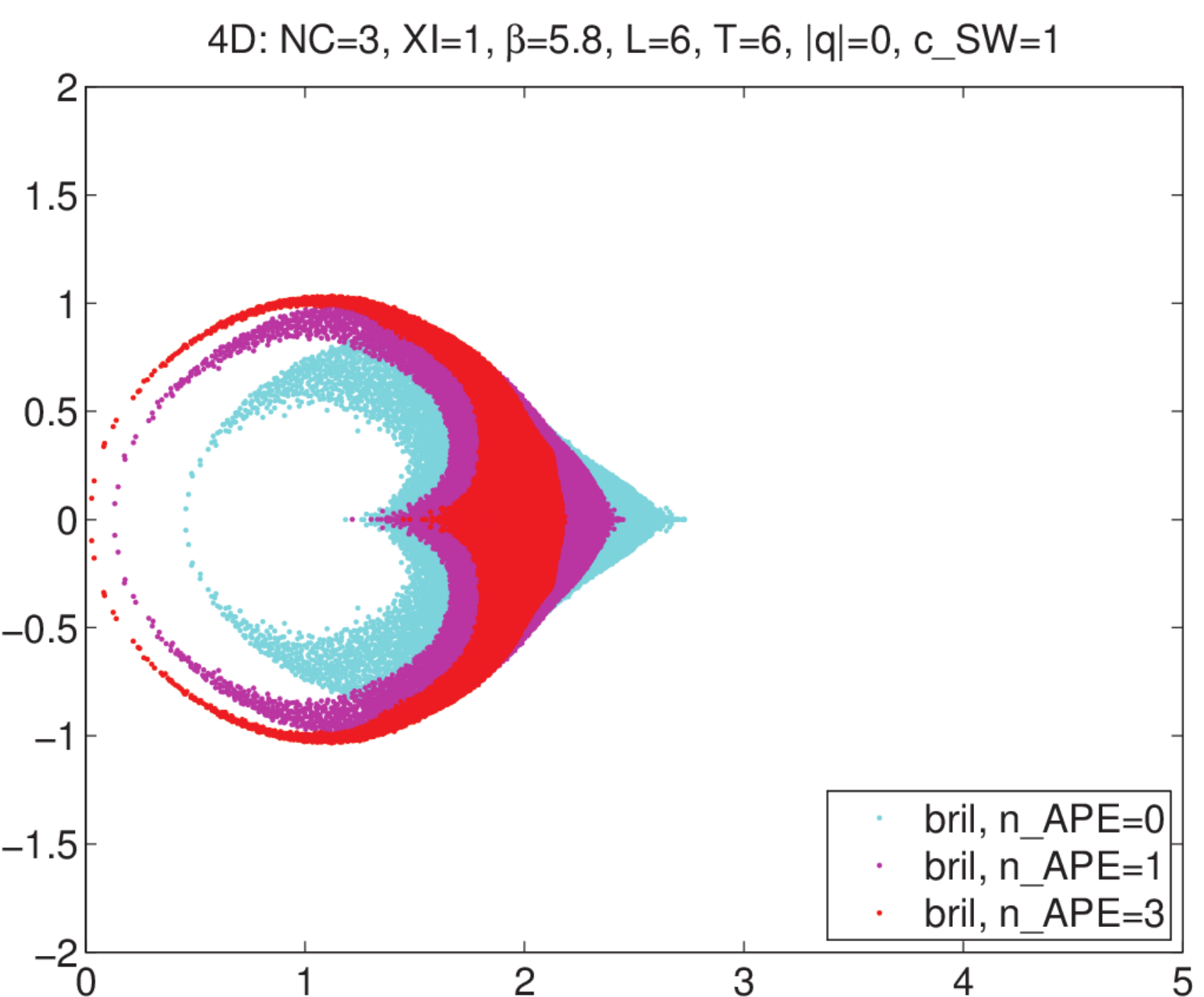}%
\\
\includegraphics[width=8.4cm]{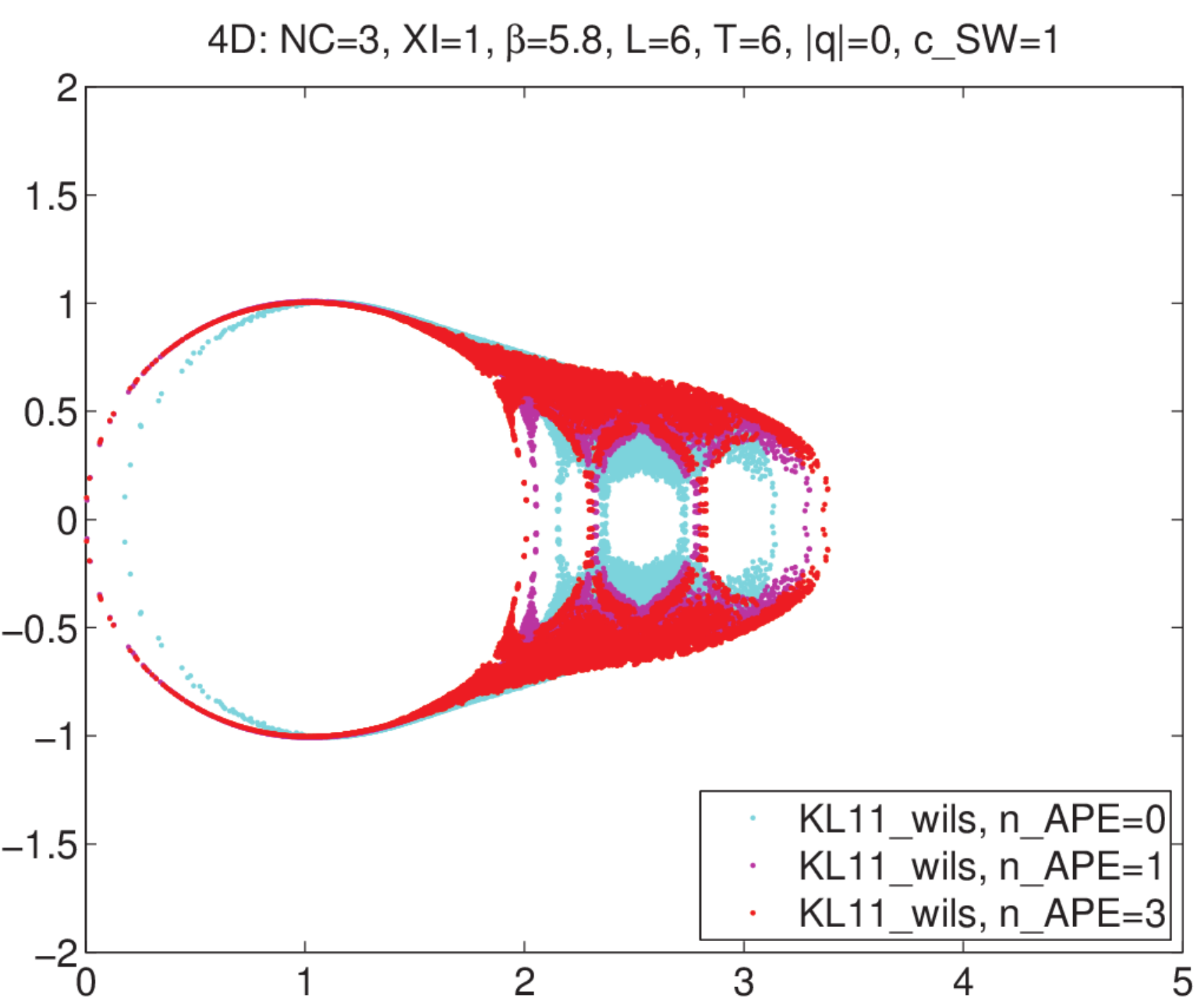}%
\includegraphics[width=8.4cm]{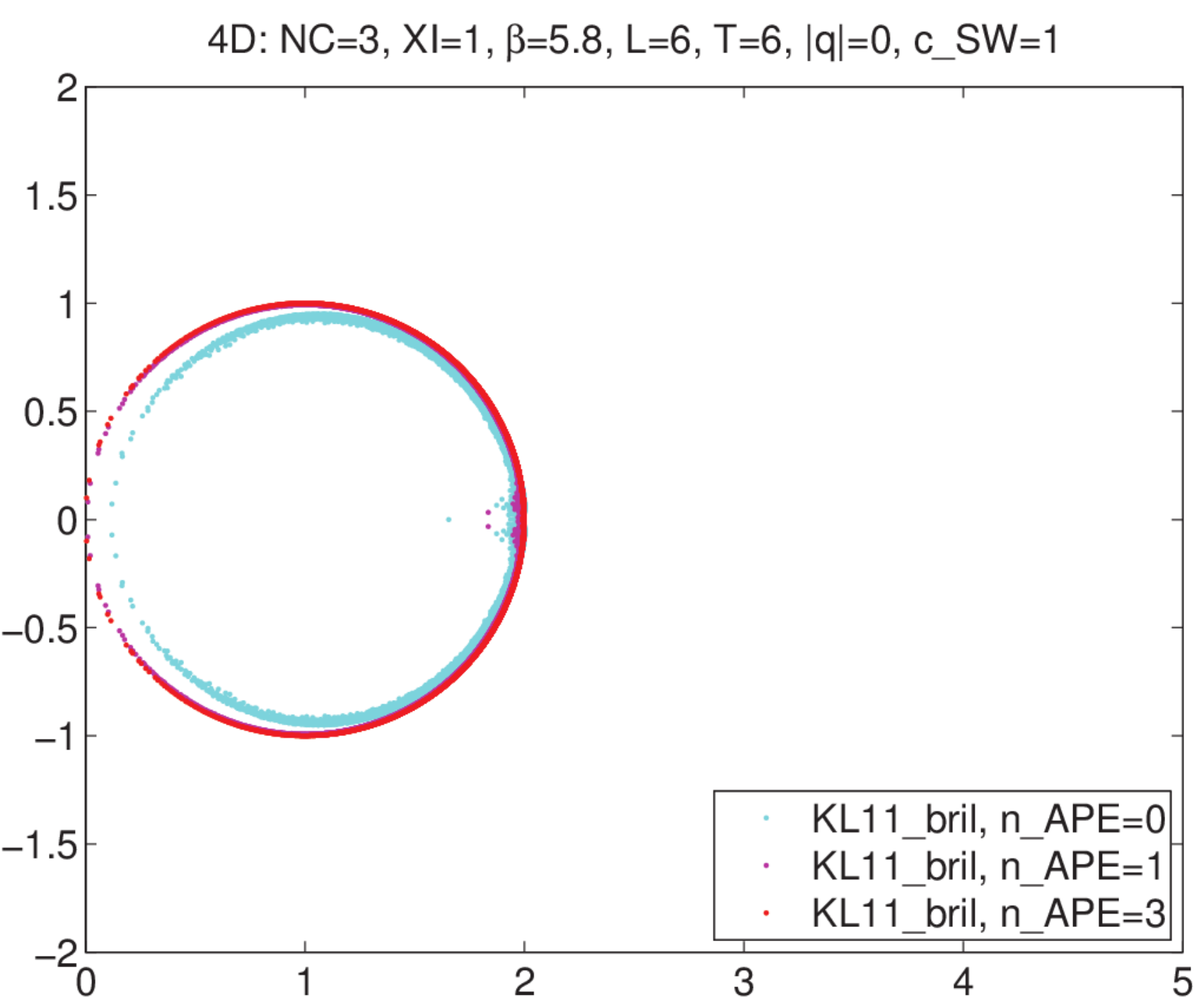}%
\caption{\label{fig:eigenvalues}\sl
Eigenvalue spectra of the Wilson (left) and Brillouin (right) operators with 0,1,3 APE smearings before (top) and after (bottom) one iteration of $f_{1,1}$.
The titles specify the number of colors ($\Nc=3$), the anisotropy coefficient ($\xi=1$ means $a_s=a_t$), the box size ($N_s/a=N_t/a=6$), the absence of topological charge ($q_\mr{top}=0$), and tree-level clover improvement ($c_\mr{SW}=1$).}
\end{figure}

The first row of Fig.\,\ref{fig:eigenvalues} displays such eigenvalue spectra on a thermalized SU(3) gauge configuration.
The Wilson operator has 5 branches with multiplicities 1,4,6,4,1 (from left to right, the last two are cut off), respectively.
Only the first (leftmost) branch contributes to continuum physics.
The Brillouin operator has only two branches, with multiplicities 1,15, respectively.
Again, only the first branch contributes in the continuum, but the advantage is that the unphysical species are more condensed; they all sit near $a\la=2$ (which proves useful in the overlap projection, see below).
The figures show the effect of the link smearing combined with tree-level (that is $c_\mr{SW}=1$) clover improvement.
In the Wilson case the horizontal ``jitter'' in the physical branch gets ameliorated by the smearing; after 3 smearings the segment of the physical branch close to the origin is fairly close to a GW circle.
Also in the Brillouin case both the additive mass shift and the ``jitter'' in the physical branch get reduced by the smearing; after 3 steps the Brillouin eigenvalue spectrum looks similar to that of a ``parameterized fixed point action'' (which is the practical implementation of the ``perfect action'') \cite{Hasenfratz:1993sp,DeGrand:1995ji,Bietenholz:1995cy,Hasenfratz:1997ft,Hasenfratz:1998jp,Hasenfratz:2000xz,Hasenfratz:2002rp}.

The second row of Fig.\,\ref{fig:eigenvalues} displays the eigenvalues of the Kenney-Laub iterate $f_{1,1}$ of the two kernels at $\rh=1$.
With either kernel the eigenvalue spectrum gets attracted (compared to the first row) towards the unit circle, but the effect is more stringent with the Brillouin kernel.
With the Wilson kernel (left) there is a significant left-over from the 15 unphysical branches, now at $1.9<\mr{Re}(z)<3.4$.
With the Brillouin kernel (right) the eigenvalue spectrum is essentially a GW circle, at least if the version with smearing is considered.
Clearly, further iterations of the Kenney-Laub mapping (tantamount to higher $n$ in $f_{n,n}$) will bring the eigenvalue spectrum of the resulting operator arbitrarily close to a GW circle, and a higher value of $n$ is needed with the Wilson kernel to reach a certain level of proximity than with the Brillouin kernel.

\begin{figure}[!tb]
\centering
\includegraphics[width=8.4cm]{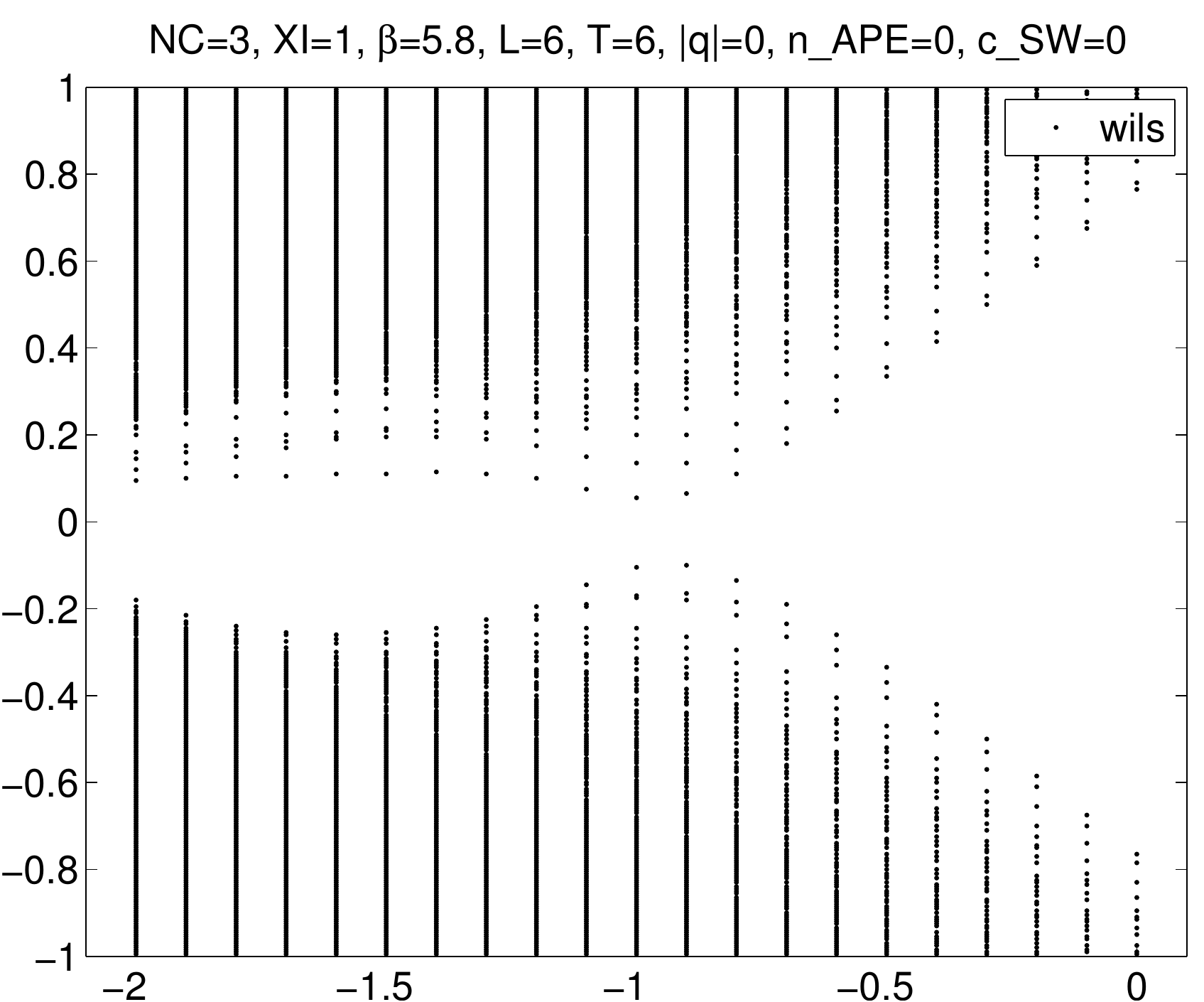}%
\includegraphics[width=8.4cm]{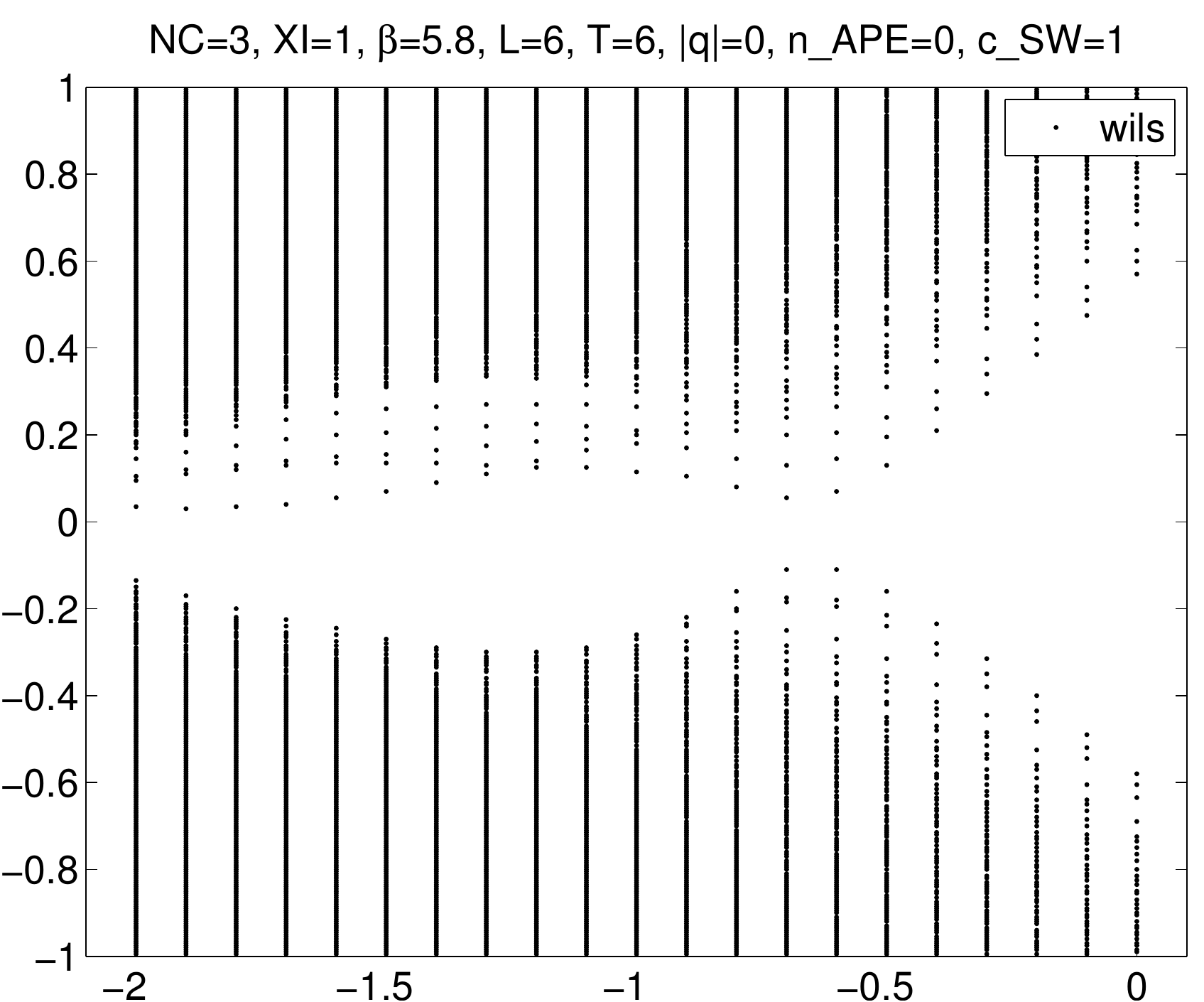}%
\\
\includegraphics[width=8.4cm]{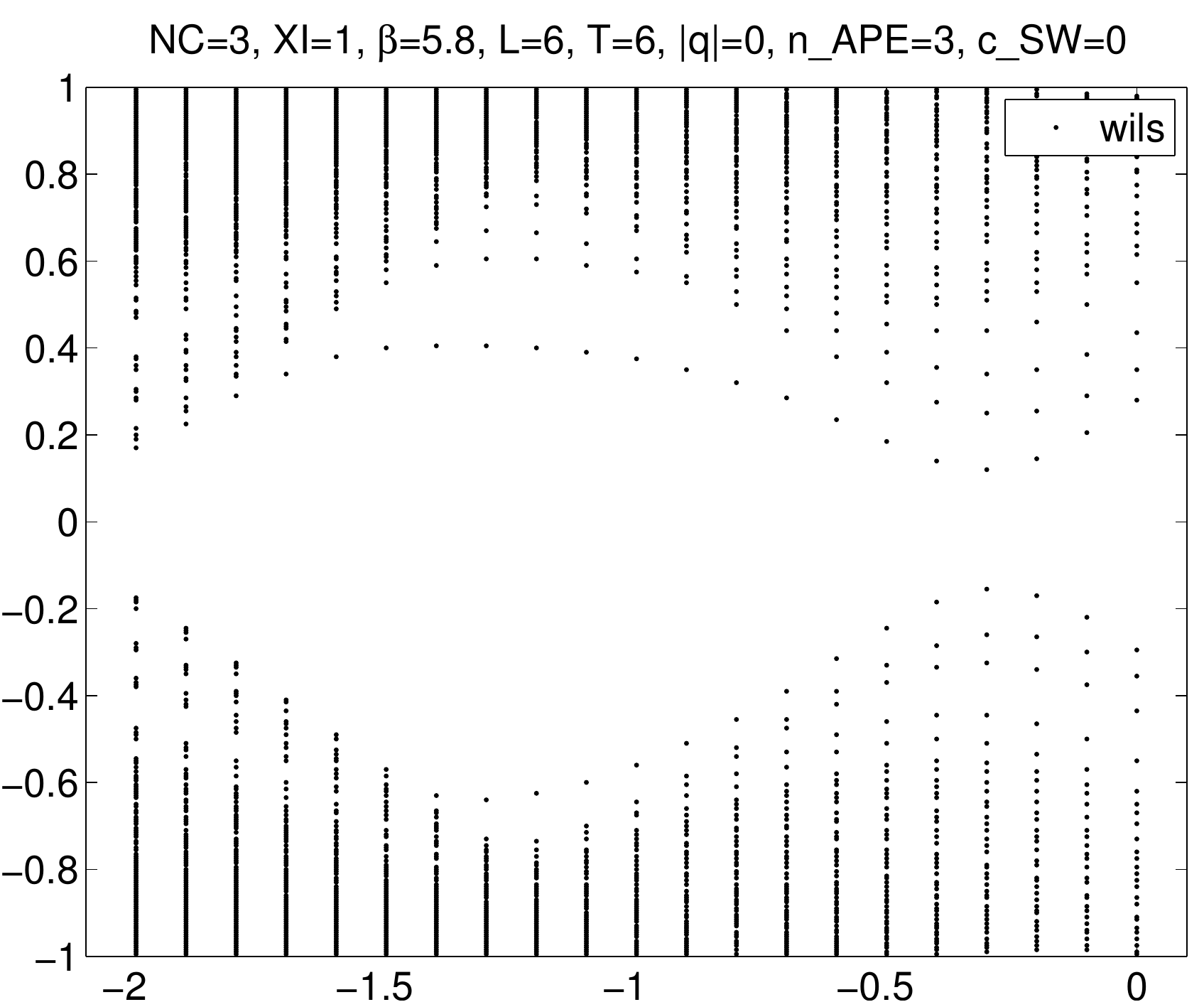}%
\includegraphics[width=8.4cm]{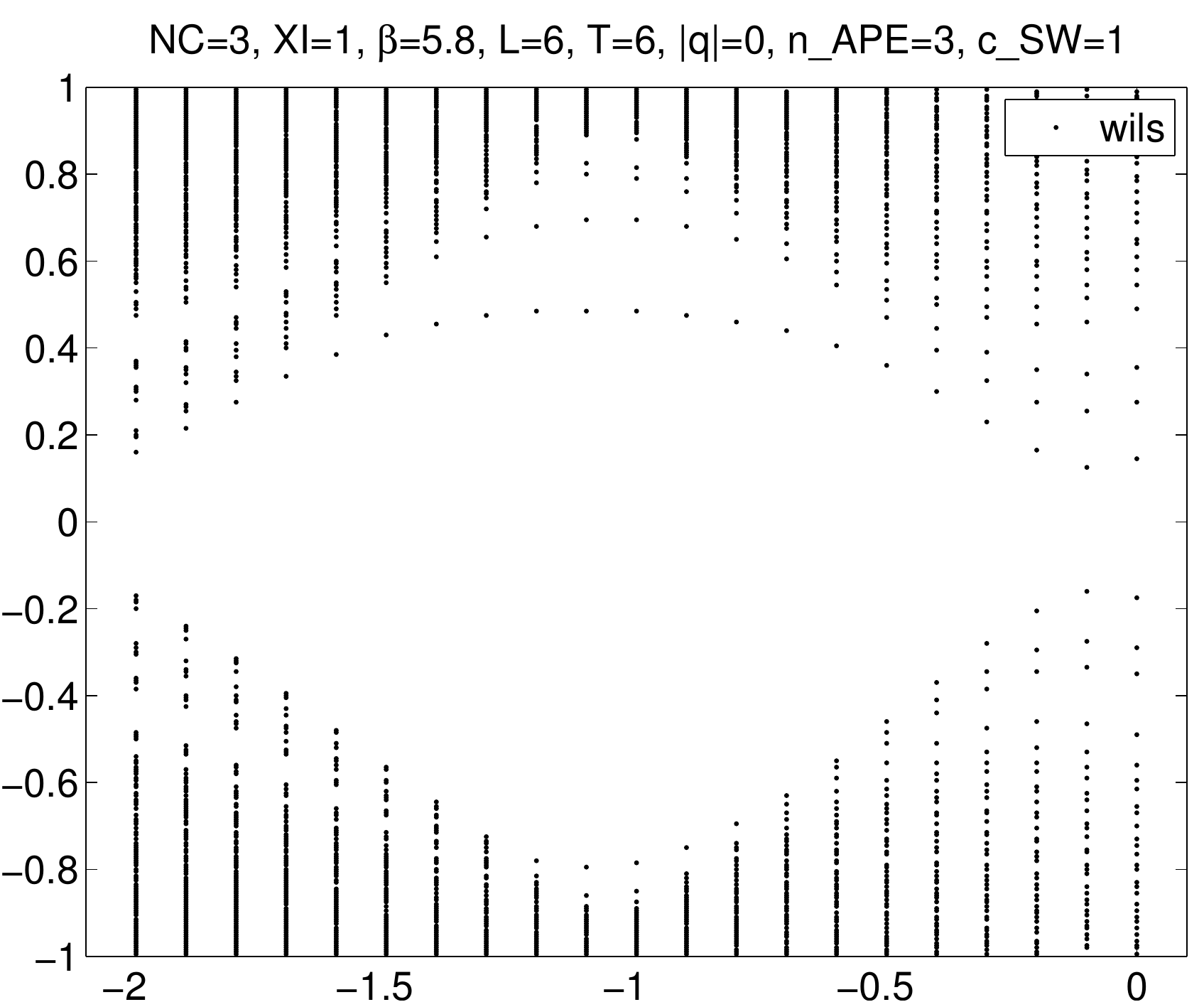}%
\caption{\label{fig:belly_wils}\sl
Spectrum of $\gaf D_{-\rh}^\mr{wils}$ on one gauge configuration as a function of $m_0=-\rh$ with $c_\mr{SW}=0$ (left) or $c_\mr{SW}=1$ (right) and 0 (top) or 3 (bottom) APE smearings.
Compare to Fig.\,\ref{fig:belly_bril}.}
\end{figure}

\begin{figure}[!tb]
\centering
\includegraphics[width=8.4cm]{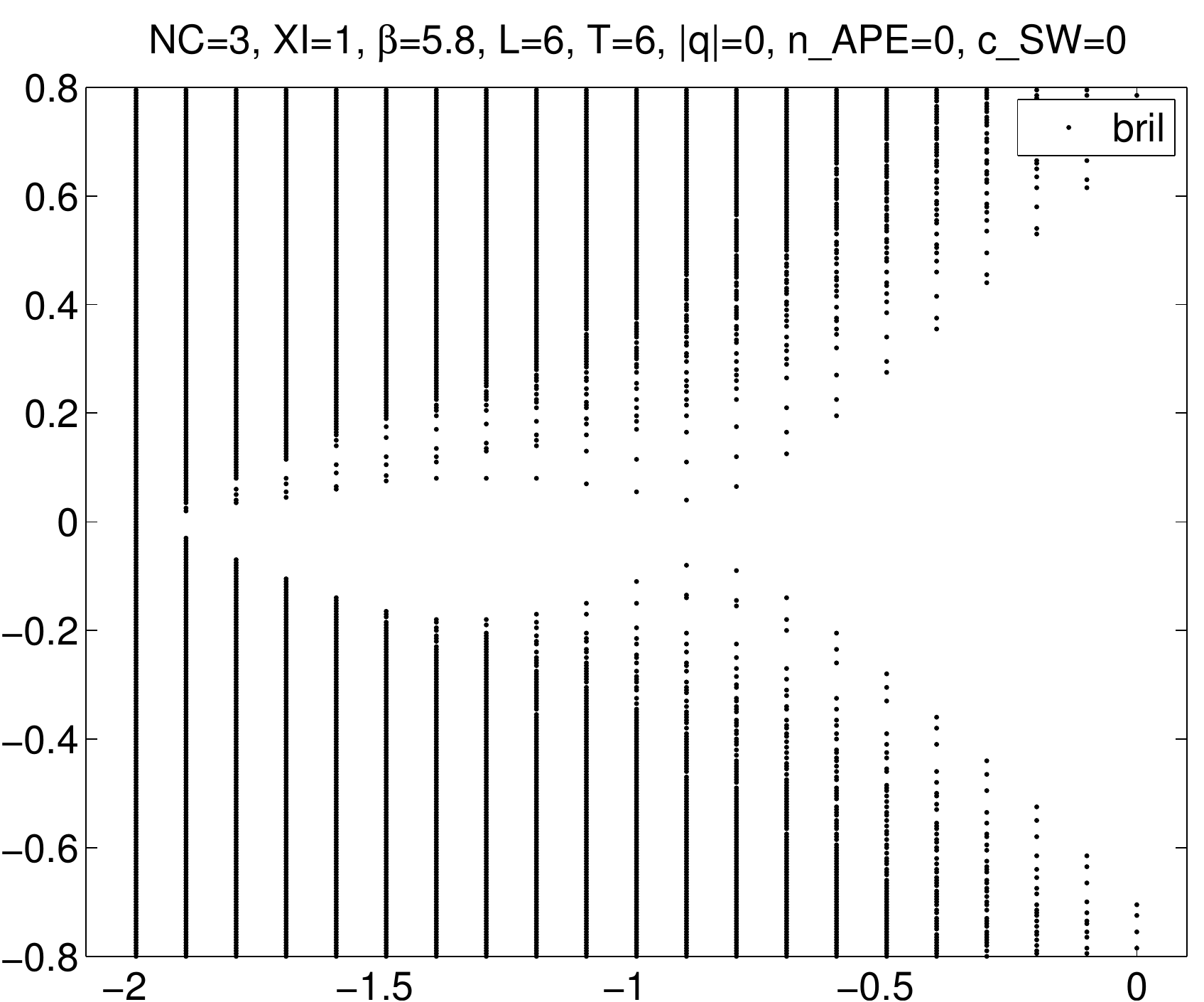}%
\includegraphics[width=8.4cm]{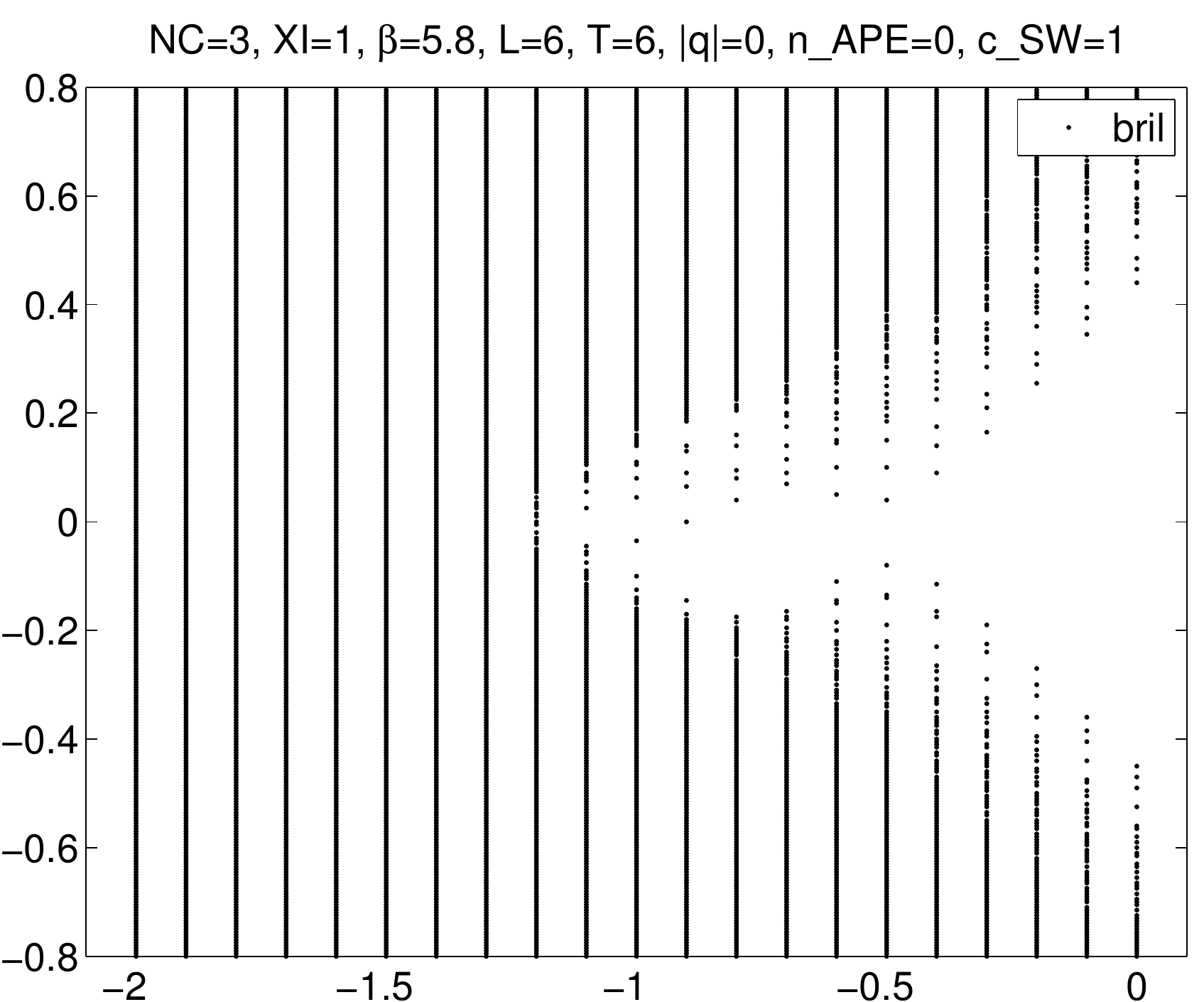}%
\\
\includegraphics[width=8.4cm]{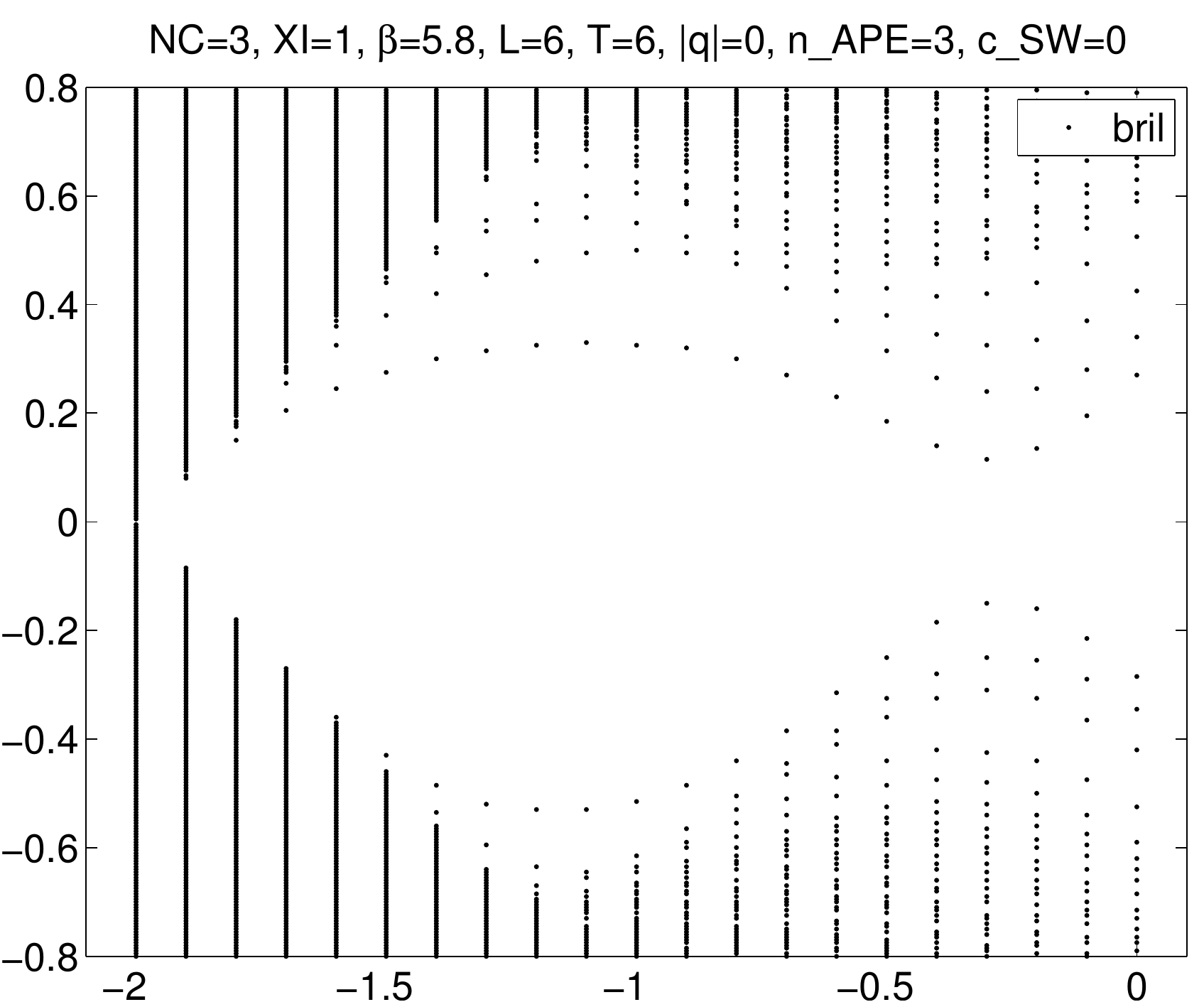}%
\includegraphics[width=8.4cm]{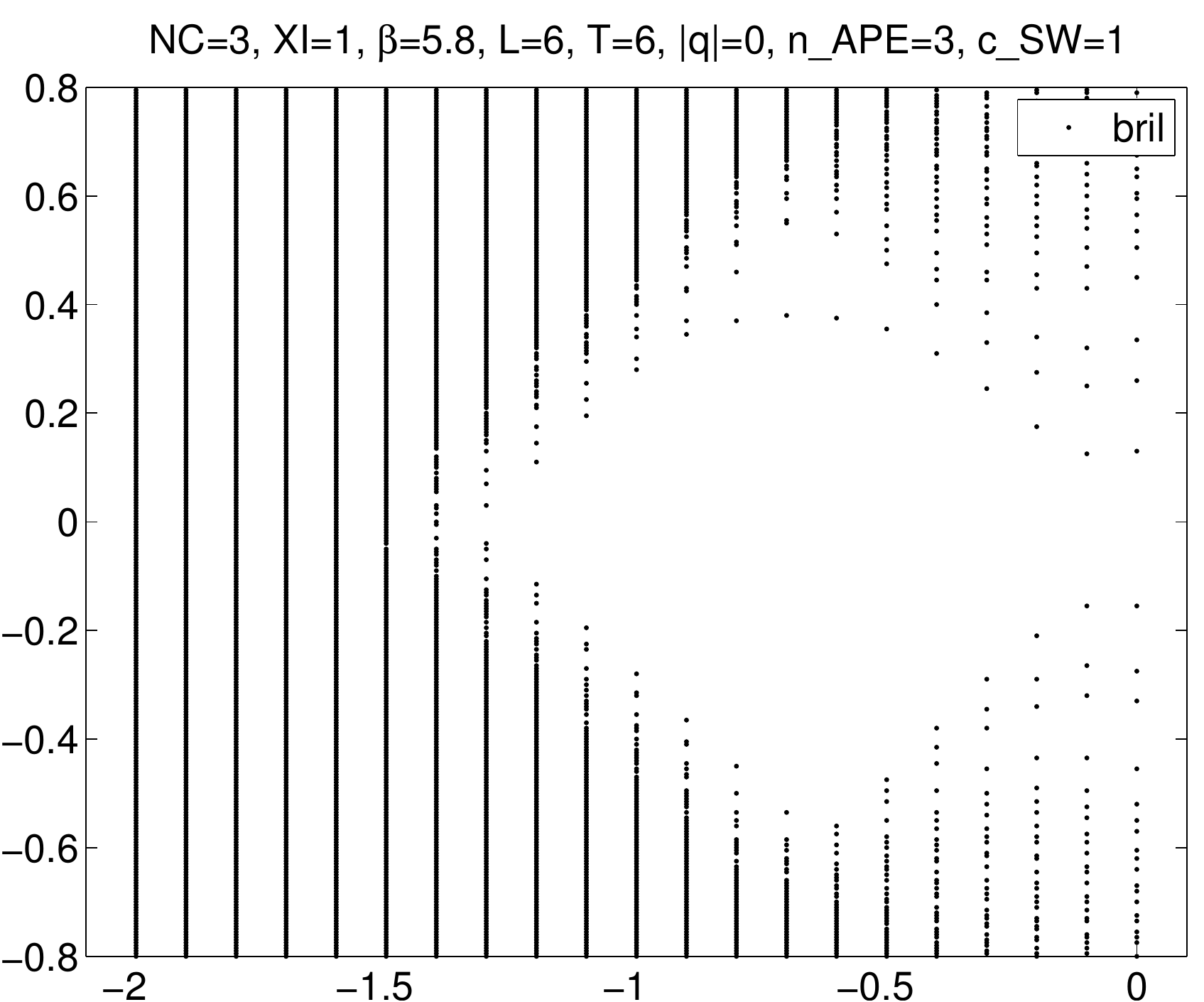}%
\caption{\label{fig:belly_bril}\sl
Same as Fig.\,\ref{fig:belly_wils}, but now for $\gaf D_{-\rh}^\mr{bril}$.
For the title details see the caption of Fig.\,\ref{fig:eigenvalues}.}
\end{figure}

\begin{figure}[!tb]
\centering
\includegraphics[width=8.4cm]{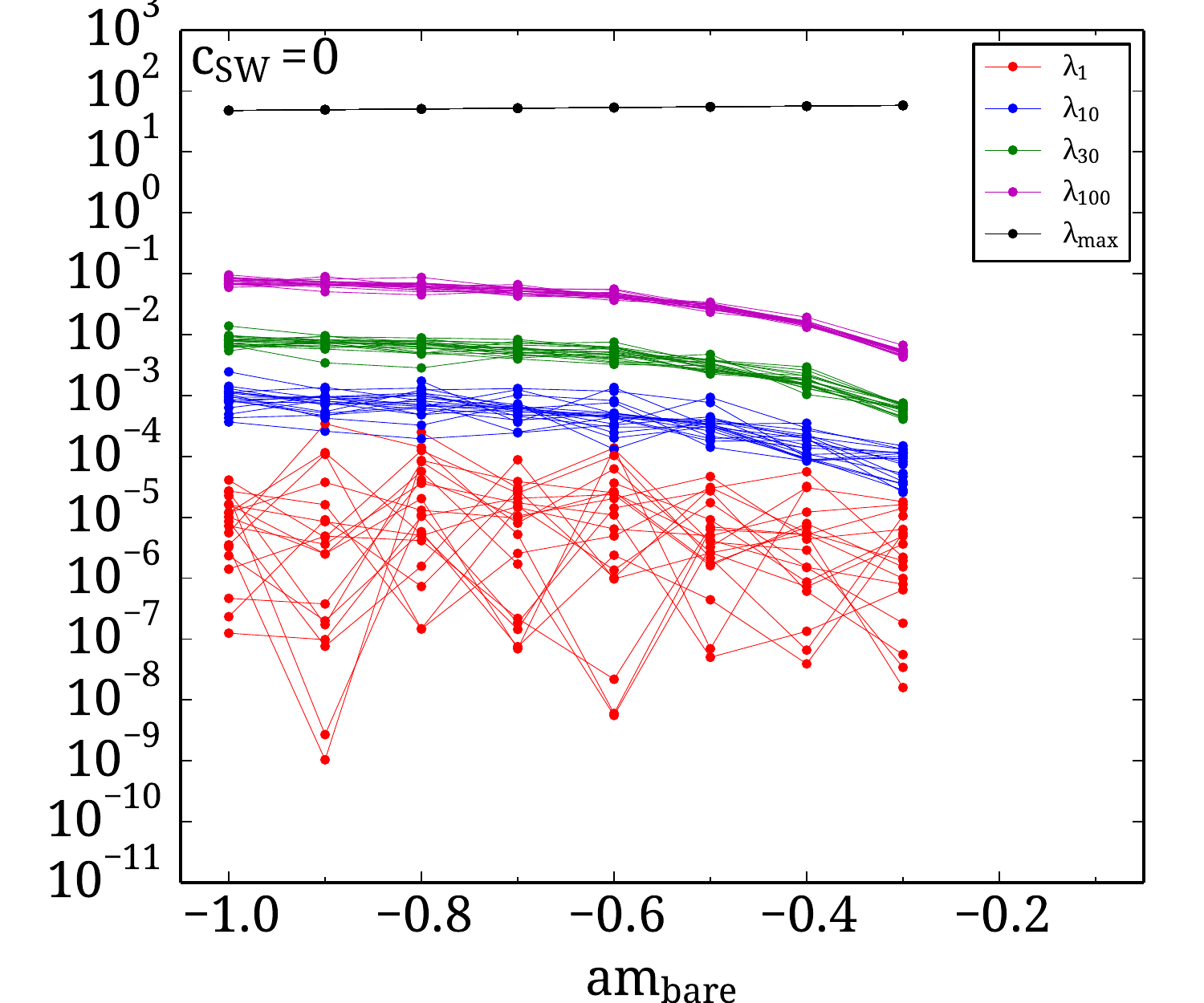}%
\includegraphics[width=8.4cm]{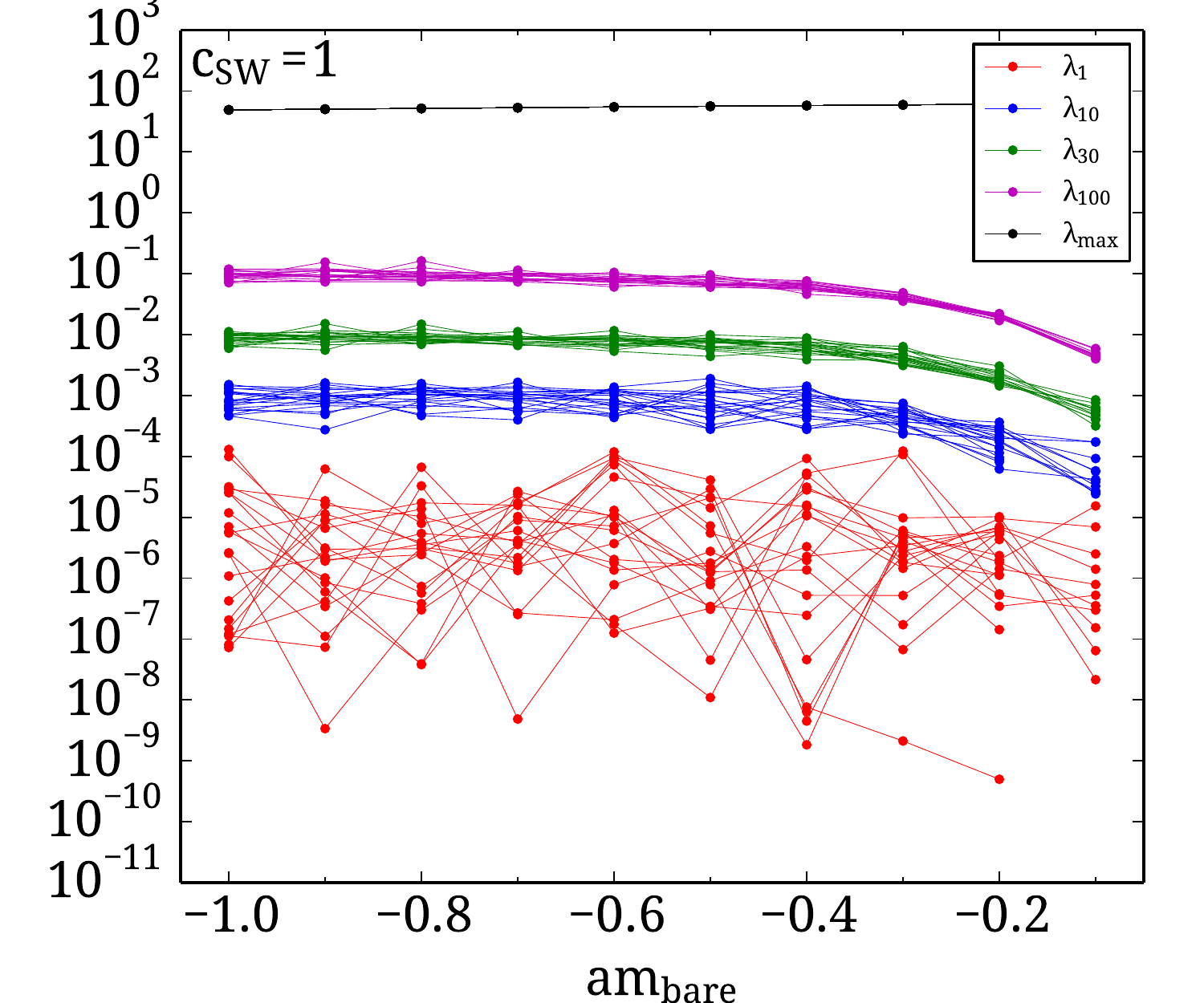}%
\\
\includegraphics[width=8.4cm]{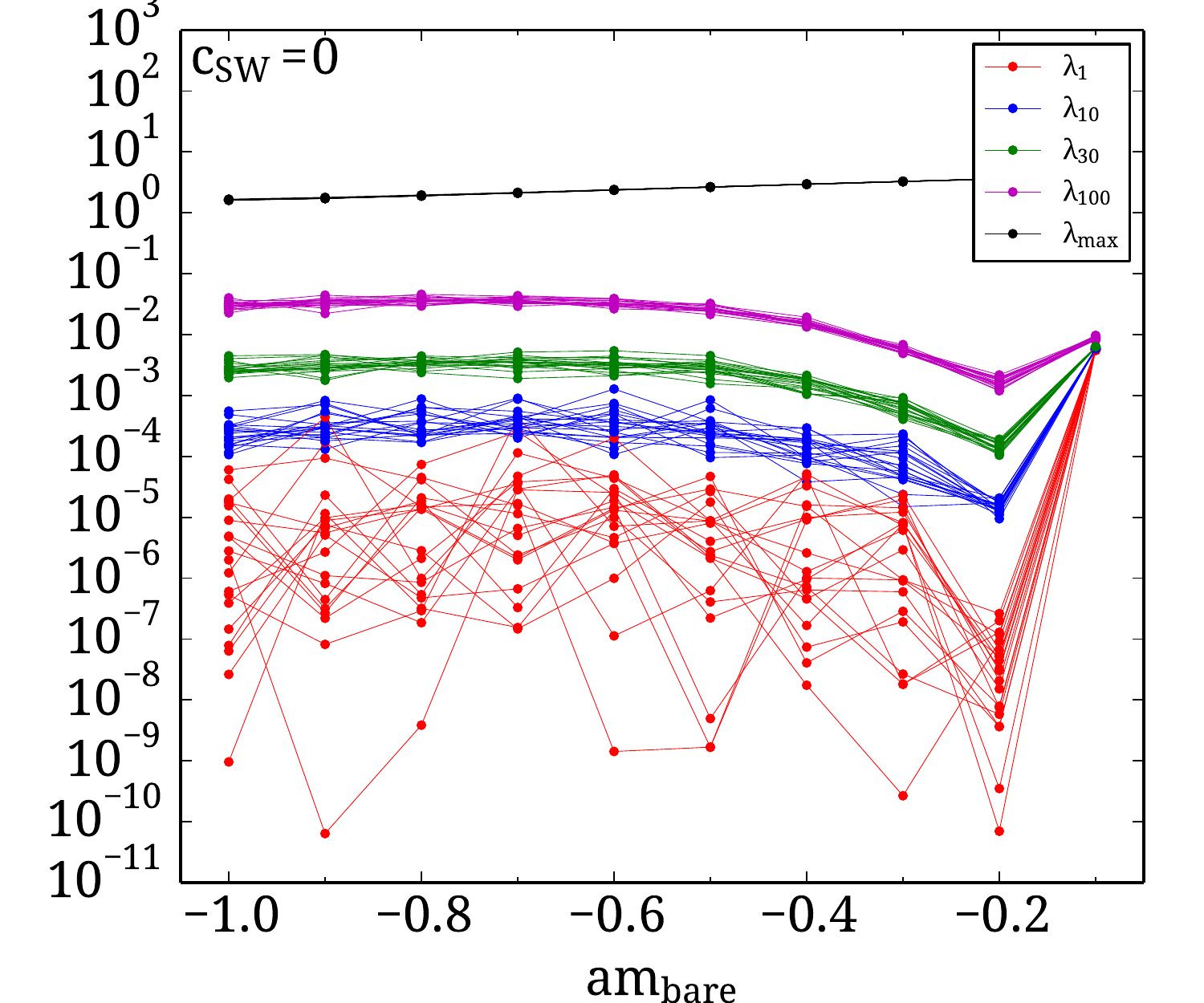}%
\includegraphics[width=8.4cm]{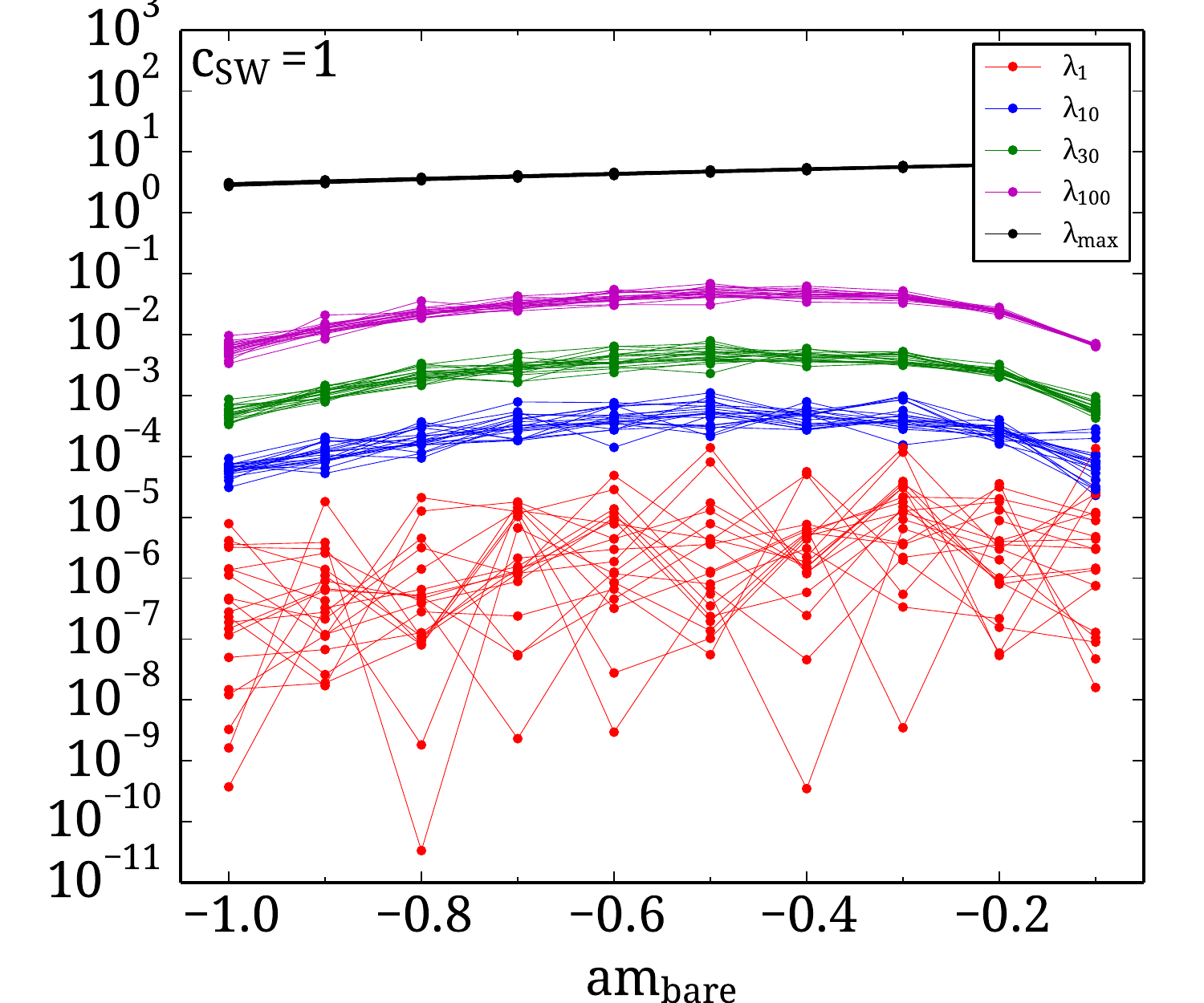}%
\caption{\label{fig:belly_qcdsf}\sl
Spectrum of $\Hrh^2=\Drh\dag\Drh^{}$ versus $am_\mr{bare}=-\rh$ on the QCDSF lattices, using the Wilson (top) and Brillouin (bottom) kernel with $c_\mr{SW}=0$ (left) and $c_\mr{SW}=1$ (right).
We show the first, tenth, thirtieth and 100th eigenvalues ($\la_{1,10,30,100}$), as well as the largest one ($\la_\mr{max}$).}
\end{figure}


\section{Spectral flow with Wilson and Brillouin kernel \label{sec:flow}}


Both kernels considered (Wilson or Brillouin) are $\gaf$-hermitean, but neither one is normal (i.e.\ $[\Dke,\Dke{}\dag]\neq0$ for either $\Dke\!=\!\Dwi$ or $\Dke\!=\!\Dbr$).
Accordingly, the spectral properties of $\Dke$ and $\gaf\Dke$ cannot be deduced from each other.

Fig.\,\ref{fig:belly_wils} shows the eigenvalues of $\gaf\Dwi$ on one gauge configuration for a scan of $m_0$ in the range from $-2$ to $0$, with and without link smearing, as well as with and without a clover term.
Ideally, one wants to choose all tunable parameters such that the ``eye'' in the hermitean eigenvalue spectrum (the leftmost ``bay'' in Fig.\,\ref{fig:belly_wils}; the ``open sea'' to the right of the ``straights'' is not shown) is wide open.
Clearly, link smearing helps a lot in this respect.
In comparison, the choice $c_\mr{SW}=0$ versus $c_\mr{SW}=1$ seems less important.
Still, what speaks in favor of a clover term in the Wilson kernel is that the ``magic'' value $\rh\simeq0.634$ of Sec.\,\ref{sec:DR} then fares reasonably, while without the clover term the opening of the ``eye'' is far from optimal at this value of $\rh$.

Fig.\,\ref{fig:belly_bril} shows the eigenvalues of $\gaf\Dbr$ on one gauge configuration for a scan of $m_0$ in the range from $-2$ to $0$, with and without link smearing, as well as with and without a clover term.
Again, link smearing is found to have a very beneficial effect on the opening of the ``eye''.
Also with the Brillouin kernel the choice $c_\mr{SW}=0$ versus $c_\mr{SW}=1$ seems insignificant regarding the maximum width of the ``eye'', but it affects the position (i.e.\ the value of $-m_0=\rh$) at which the maximum is realized.
Interestingly, with $c_\mr{SW}=1$ the ``magic'' value $\rh\simeq0.634$ of Sec.\,\ref{sec:DR} more-or-less coincides with the choice of $\rh$ which maximizes the opening of the ``eye''.

In Fig.\,\ref{fig:belly_qcdsf} we show similar eigenvalues (in fact eigenvalues of $\gaf\Dke\gaf\Dke$, i.e.\ without the sign information) on much larger lattices (the $40^3\times64$ lattices by QCDSF that will be discussed in Sec.\,\ref{sec:tests}).
In view of the lesson just learned, we restrict ourselves to the version with link smearing.
The spacing in $m_0=-\rh$ is too wide to allow for individual eigenvalue tracking.
Still, it is evident that the main difference between the Wilson and the Brillouin kernel is the \emph{upper} end of the eigenvalue spectrum; with the Brillouin kernel it is at least an order of magnitude lower.
What matters for the CPU time spent in large-scale computations is the effective condition number $\la_\mr{max}/\la_{n}$ after $n-1$ low modes are projected away (we show the situation for $n=1,10,30,100$).
It seems on such big lattices the difference between the Wilson and the Brillouin kernel is less pronounced than it appeared on the small lattices.
Still, it is encouraging to see that with the Brillouin kernel the ``magic'' choice $\rh\simeq0.634$ fares well, both for $c_\mr{SW}=0$ and $c_\mr{SW}=1$.


\section{Numerical tests with Brillouin and Wilson kernel \label{sec:tests}}


The massless overlap operator $\Dov$ as defined in (\ref{def_over}) differs from the kernel $\Dke$ in several ways:
($i$) $\Dov$ is normal, i.e.\ it commutes with $\Dov{}\dag$,
($ii$) $\Dov$ satisfies the GW relation (\ref{ginswils_one}),
($iii$) $\Dov$ is not ultralocal but just exponentially localized (with the fall-off pattern being a measure of the quality of the resulting operator).
Here we verify these properties numerically on matched quenched lattices (i.e.\ with a fixed physical box size $L$), using clover improved kernels ($c_\mr{SW}=1$) and 1 or 3 steps of $\al=0.72$ APE smearing.
In addition, we explore the inversion cost of the fixed-order Kenney-Laub overlap operator on large $\Nf=2$ lattices generated by QCDSF.


\subsection{Operator normality}

\begin{figure}[!tb]
\includegraphics[width=0.49\textwidth]{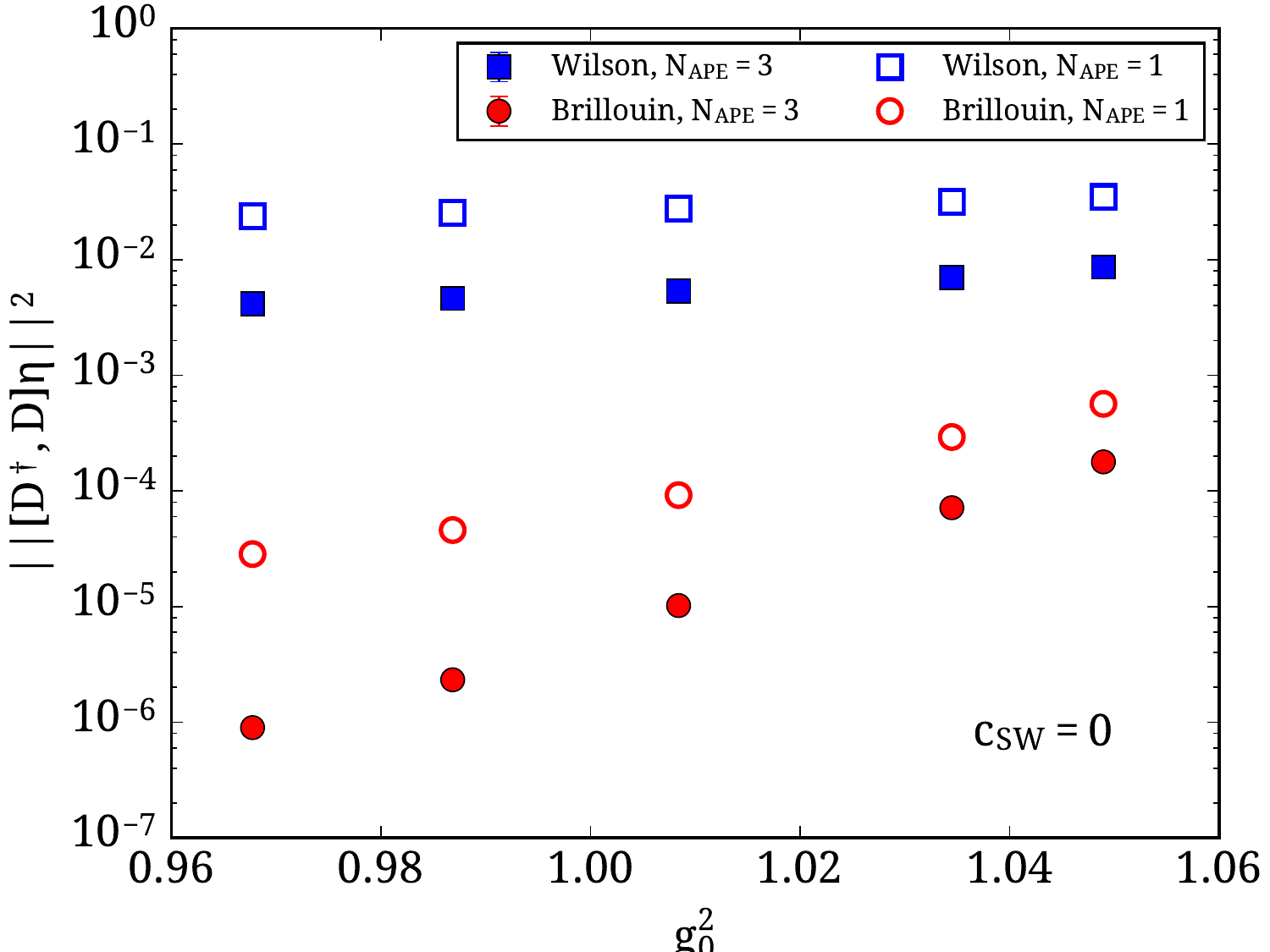}%
\includegraphics[width=0.49\textwidth]{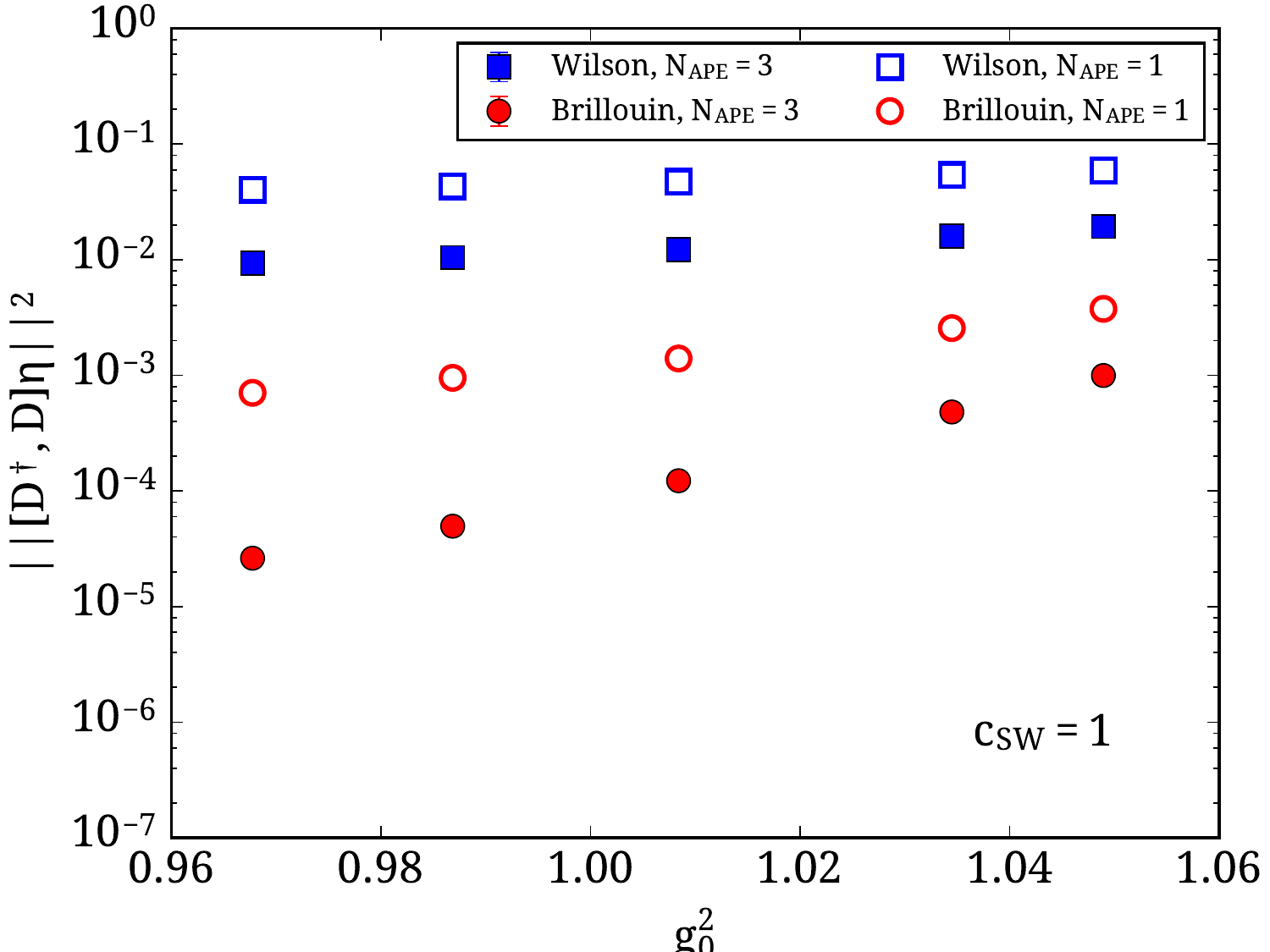}%
\caption{\sl\label{fig:KL11_normality}
Normality of the iterate $D^{(1)}_{-1}+1$ with the Kenney-Laub mapping $f_{1,1}$, and $D^{(0)}_{-1}$ the Wilson or Brillouin kernel with 1 or 3 APE smearings, and $c_\mr{SW}=0$ (left) or $c_\mr{SW}=1$ (right). Results on volume-matched ensembles of 40 quenched lattices each are plotted versus $6/\be$.}
\end{figure}

We select the fixed rational approximation to the sign function implied by the Kenney-Laub iterate $D=D^{(1)}_{-1}\!+\!1$ with $D^{(1)}_{-1}=f_{1,1}(D^{(0)}_{-1})$ and $D^{(0)}$ being the Wilson or Brillouin kernel.
We measure $||(DD\dag-D\dag D)\et||$ for a few dozen normalized Gaussian random vectors $\et$ on 40 configs for each $\be$ used in Ref.\,\cite{Durr:2010ch}, and Fig.\,\ref{fig:KL11_normality} shows the result.
Both in the Wilson and in the Brillouin case, the operator with 3 steps of link smearing in the kernel exhibits smaller deviations from normality than the one with 1 step of smearing in the kernel.
The main lesson to be learned is that both operators with Brillouin kernel have a smaller violation of normality than the two operators with Wilson kernel.
Evidently, in order to reach a fixed level of normality violation, e.g.\ $||(DD\dag-D\dag D)\et||<10^{-12}$, the order of the rational approximation must be enhanced most drastically for the unsmeared Wilson kernel and least so for the smeared Brillouin kernel.


\subsection{Ginsparg-Wilson relation}

\begin{figure}[!tb]
\includegraphics[width=0.49\textwidth]{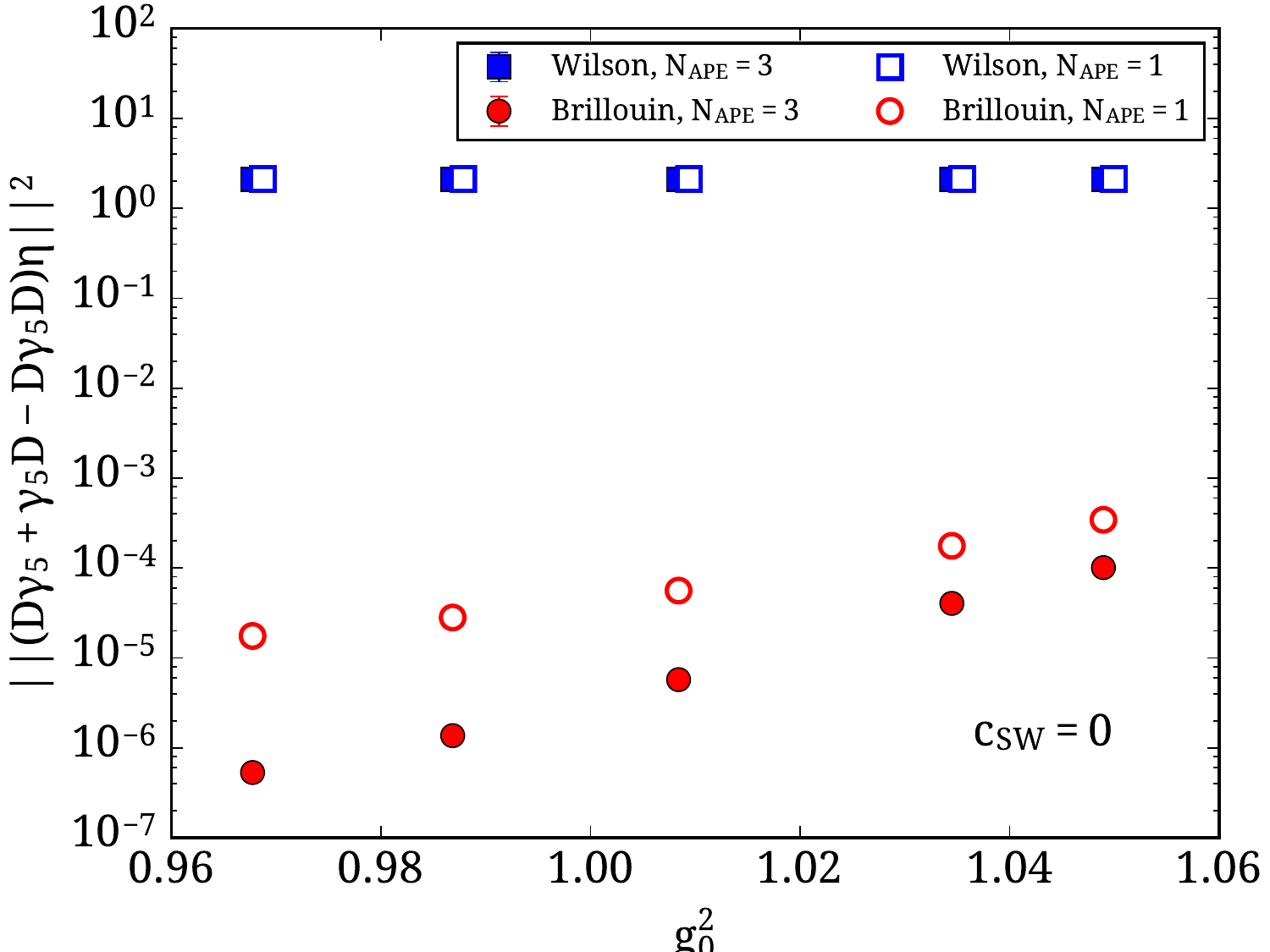}%
\includegraphics[width=0.49\textwidth]{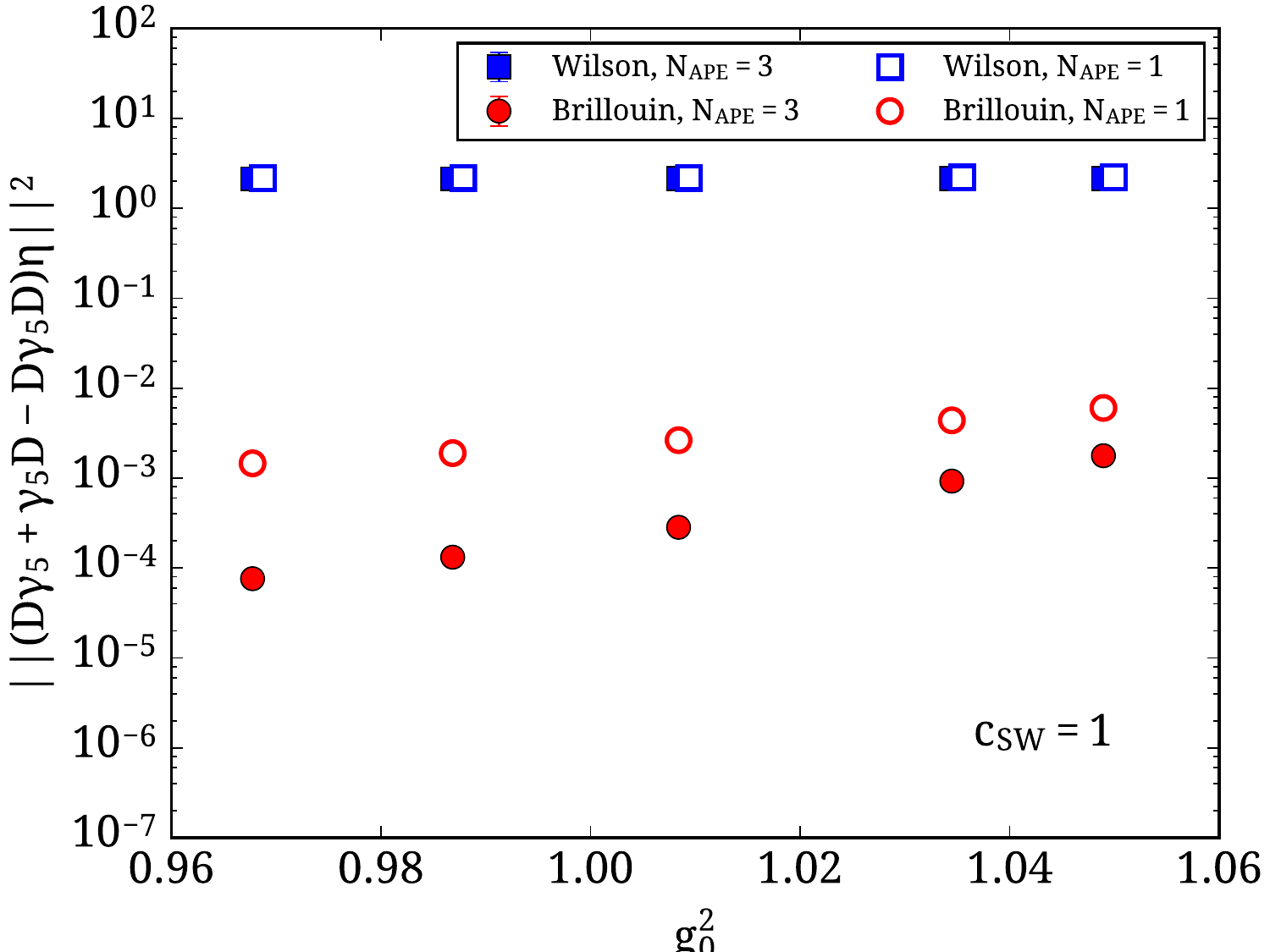}%
\caption{\sl\label{fig:KL11_GW}
GW defect of the iterate $D^{(1)}_{-1}+1$ with the Kenney-Laub function $f_{1,1}$, and $D^{(0)}_{-1}$ the Wilson or Brillouin kernel with 1 or 3 APE smearings, and $c_\mr{SW}=0$ (left) or $c_\mr{SW}=1$ (right). Results on volume-matched ensembles of 40 quenched lattices each are plotted versus $6/\be$.}
\end{figure}

We use the same (low-order) rational approximation to the sign function and the same pure gauge ensembles as in the previous subsection.
We measure $||(D\gaf+\gaf D-D\gaf D)\et||$, which we will refer to as the ``GW defect'', for a few dozen normalized Gaussian random vectors $\et$ on 40 configurations of each ensemble, see Fig.\,\ref{fig:KL11_GW}.
The GW defect with the Wilson kernel is several orders of magnitude larger than with the Brillouin kernel.
With the Wilson kernel the difference between the two smearing levels is barely visible, while with the Brillouin kernel increasing $N_\mr{APE}$ from $1$ to $3$ significantly reduces the GW defect.
Moreover, in the Brillouin case pushing to the continuum (i.e.\ to smaller $g_0^2$) has a beneficial effect, too, while no such effect is visible with the Wilson kernel.
Evidently, in order to achieve a fixed level of GW violation, say $||(D\gaf+\gaf D-D\gaf D)\et||<10^{-12}$, the order of the rational approximation needs to be increased much more drastically for the Wilson kernel than for the Brillouin kernel.


\subsection{Exponential operator localization}

The locality of the overlap action with the Wilson kernel was first studied in Ref.\,\cite{Hernandez:1998et}.
Ref.\,\cite{Bietenholz:2002ks} demonstrated that a more extended (but still ultralocal) kernel can significantly improve the coordinate-space locality of the resulting overlap action.
In Refs.\,\cite{DeGrand:2000tf,Kovacs:2002nz,Durr:2005an} it was shown that even a slight modification through some link-smearing can lead to a considerable improvement.
Therefore, one may hope that trading the Wilson kernel for the Brillouin kernel leads to a noticeable improvement of the locality of the overlap operator.
Note that all of this holds up to some gauge coupling $g_0^\mr{max}$, since Refs.\,\cite{Golterman:2003qe,Golterman:2004cy} pointed out that, once the gauge background becomes too rough, eigenmodes of the underlying shifted kernel $\Dke_{-\rh}$ delocalize, and mix into a band, with the effect that the overlap operator may cease to be exponentially localized.

\begin{figure}[!tb]
\includegraphics[width=0.99\textwidth]{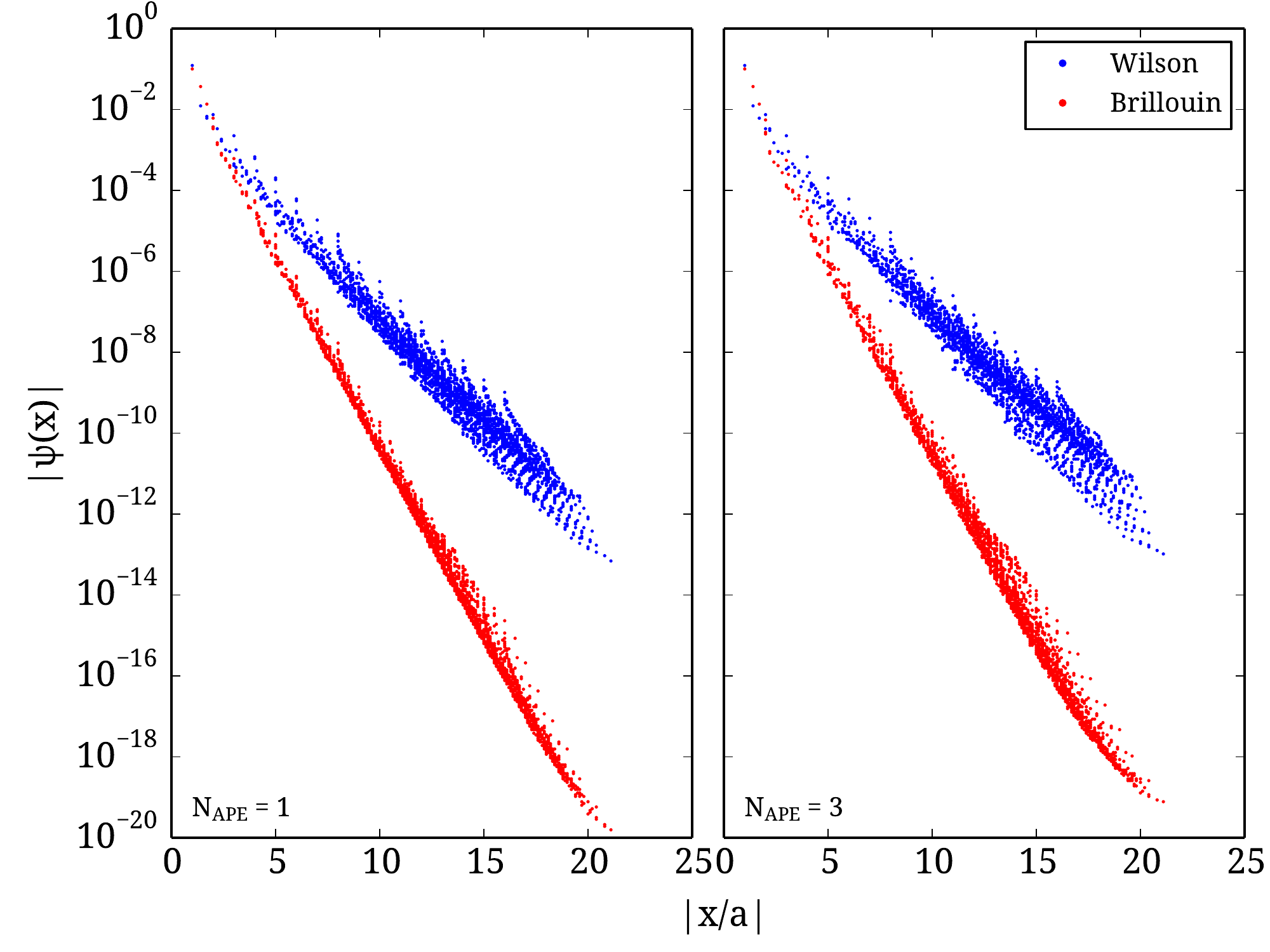}%
\caption{\sl\label{fig:locality_scatter}
Localization of the approximate overlap action defined through $f_{1,1}$ with Wilson or Brillouin kernel, 1 or 3 APE smearings and
$c_\mr{SW}=1$ prior to averaging over various directions with a fixed value
of $|x|$. Quenched lattices at $\be=5.95$ and $L/a=16,T/a=32$ are used.}
\end{figure}

\begin{figure}[!tb]
\includegraphics[width=0.49\textwidth]{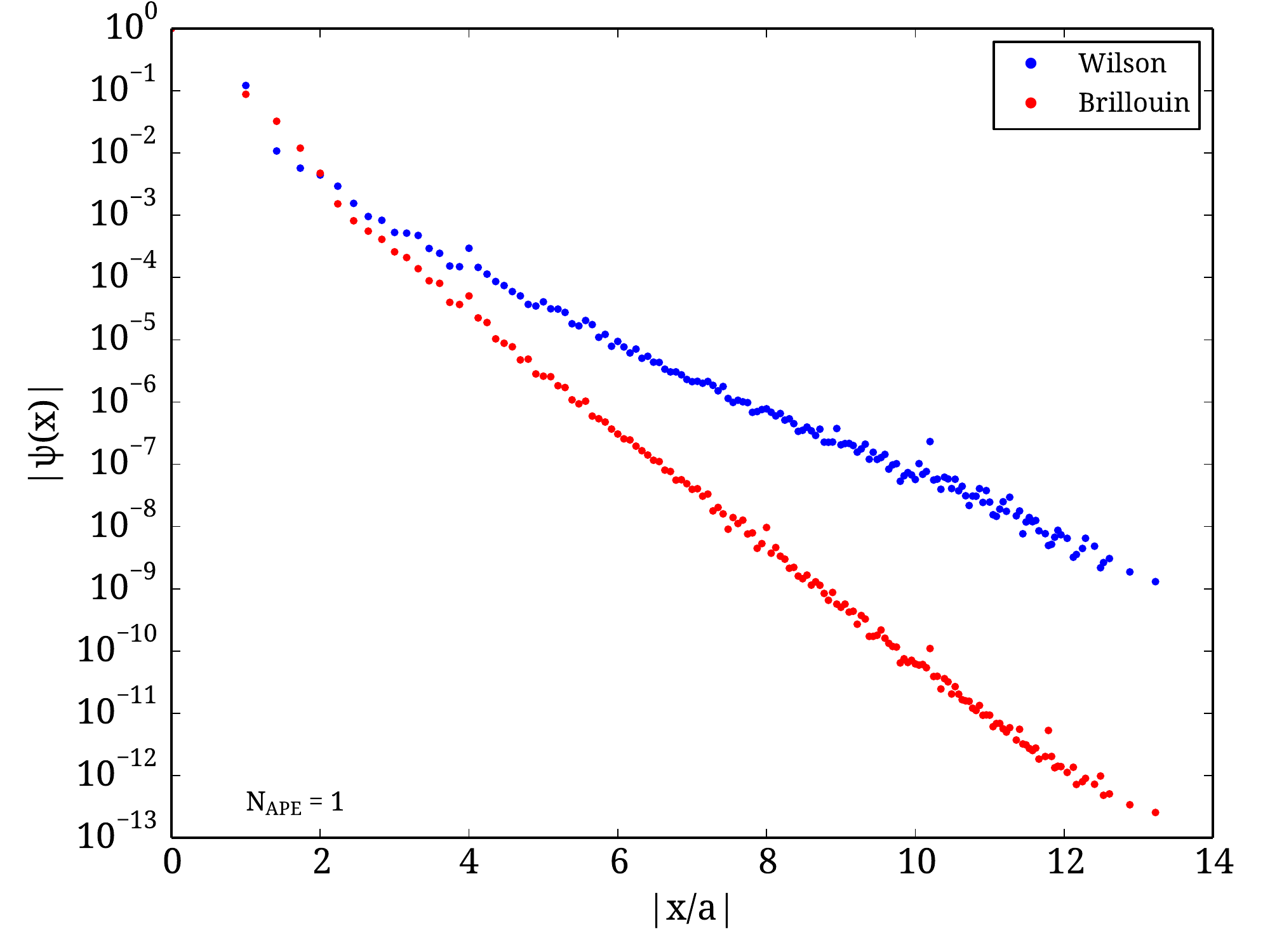}%
\includegraphics[width=0.49\textwidth]{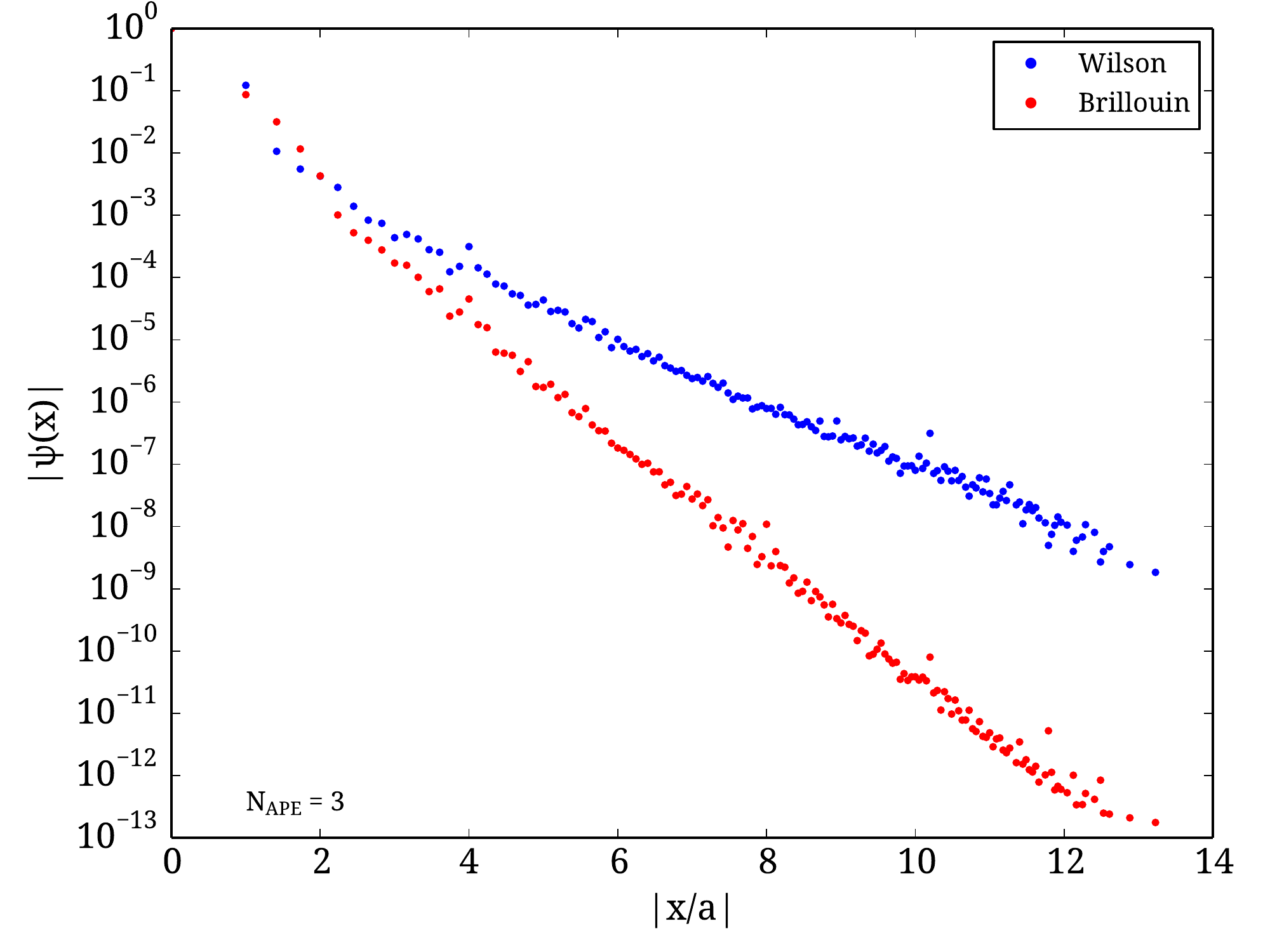}%
\\
\includegraphics[width=0.49\textwidth]{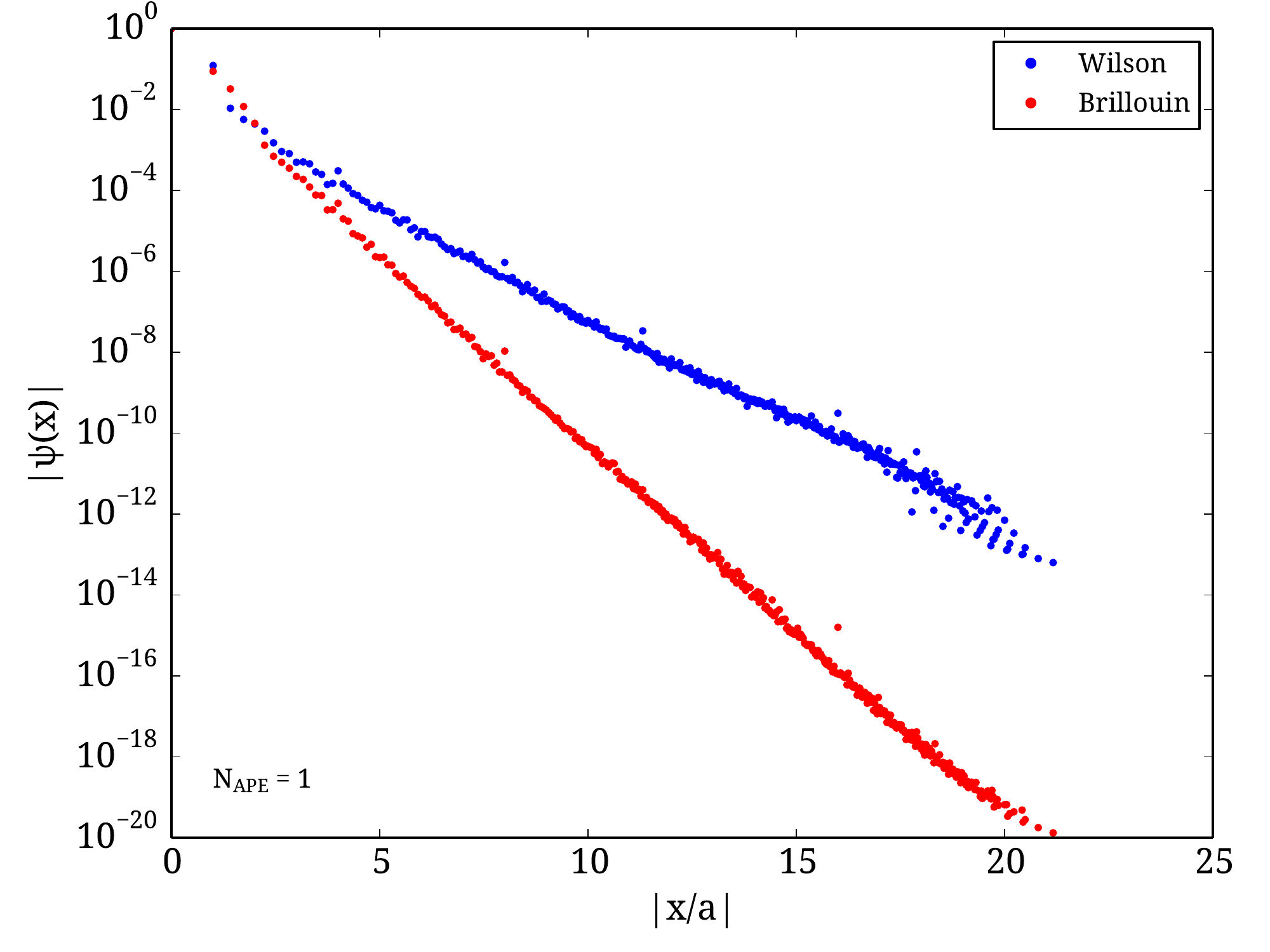}%
\includegraphics[width=0.49\textwidth]{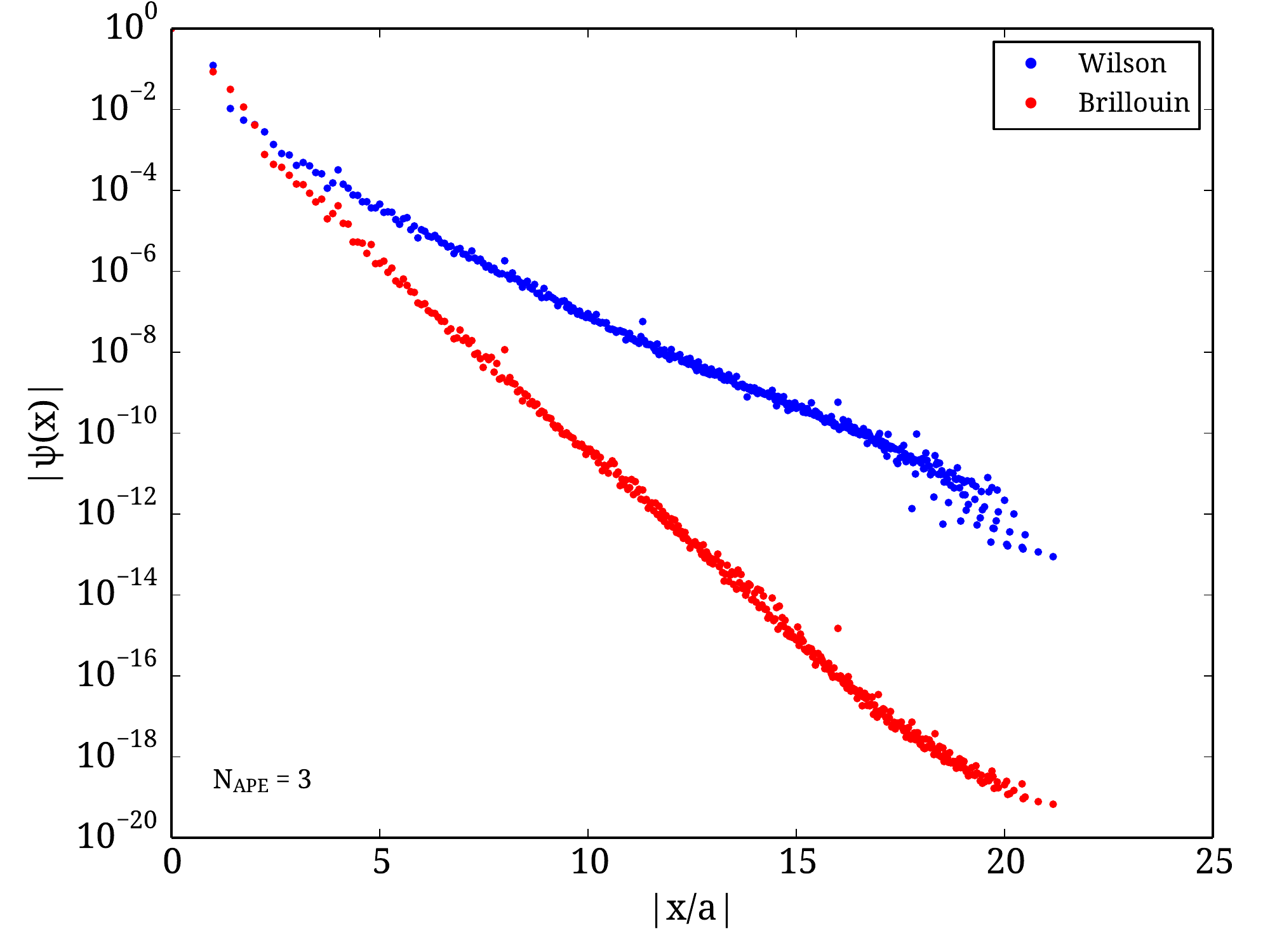}%
\\
\includegraphics[width=0.49\textwidth]{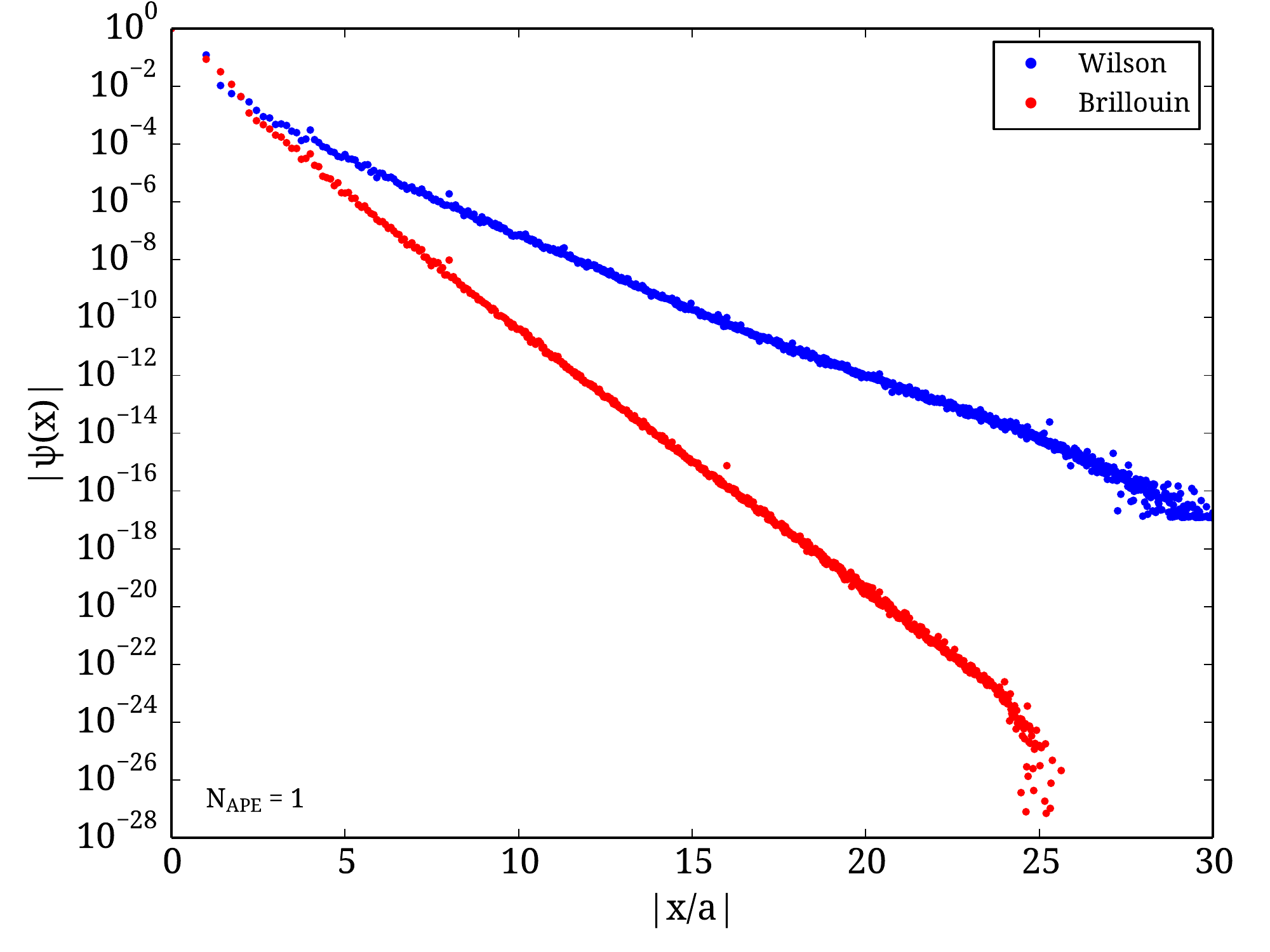}%
\includegraphics[width=0.49\textwidth]{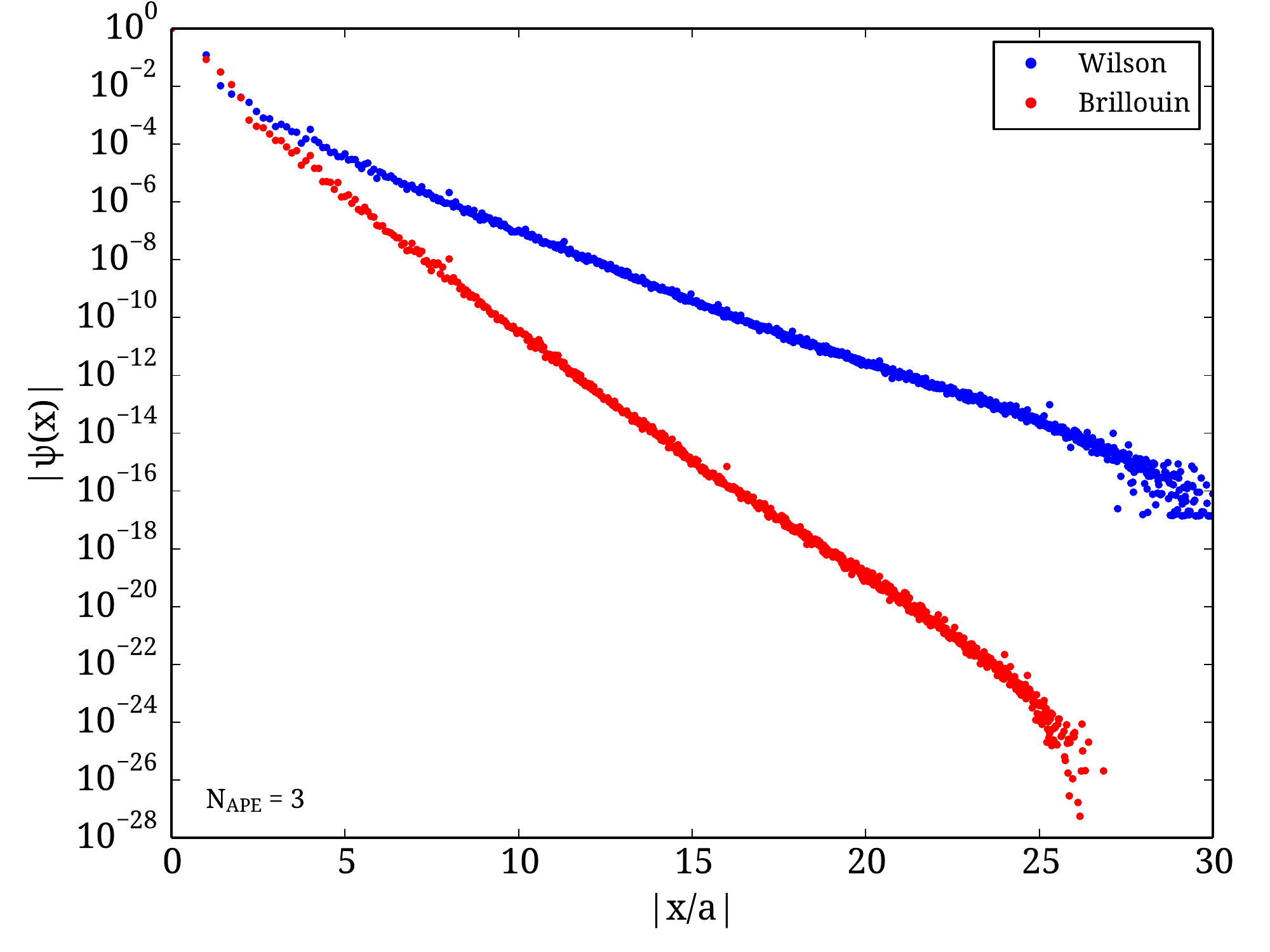}%
\caption{\sl\label{fig:locality_average}
Same as Fig.\,\ref{fig:locality_scatter} but after averaging over various directions with a common $|x|$.
The panel rows feature $\be=5.72,5.95,6.20$ and $L/a=10,16,24$ with $T=2L$ (from top to bottom).}
\end{figure}

We measure the fall-off of $|\ze(x)|$, with $\ze=D\et$ and $\et$ a normalized Gaussian random vector (in spinor/color space) with support at the single site $0$, for about a dozen $\et$ per config on 20 quenched configs per $\be$.
Fig.\,\ref{fig:locality_scatter} shows the result at the lattice spacing $a\simeq0.1\fm$ as a function of the Euclidean distance $|x|$.
The norm falls off exponentially with distance, but there are signs of rotational symmetry breaking -- different directions with a common $|x|$ do not lie on top of each other.
Clearly, this rotational symmetry breaking is more pronounced with the Wilson kernel, and a higher smearing level does not help.
The other marked difference is that the fall-off rate with the Brillouin kernel is better than with the Wilson kernel.
Fig.\,\ref{fig:locality_average} shows the locality after averaging over directions $x$ with the same $|x|$.
The fall-off pattern looks even more exponential than previously, and the better localization (roughly by a factor 2 at fixed $\rh=1$) of the Brillouin version is found to be virtually independent of the lattice spacing.
Note that the first row of Fig.\,\ref{fig:locality_average} demonstrates that the combination of $c_\mr{SW}=1$ and some link smearing ensures that either overlap action is exponentially localized on lattices with $a\simeq0.16\fm$.

\begin{figure}[!tb]
\includegraphics[width=0.49\textwidth]{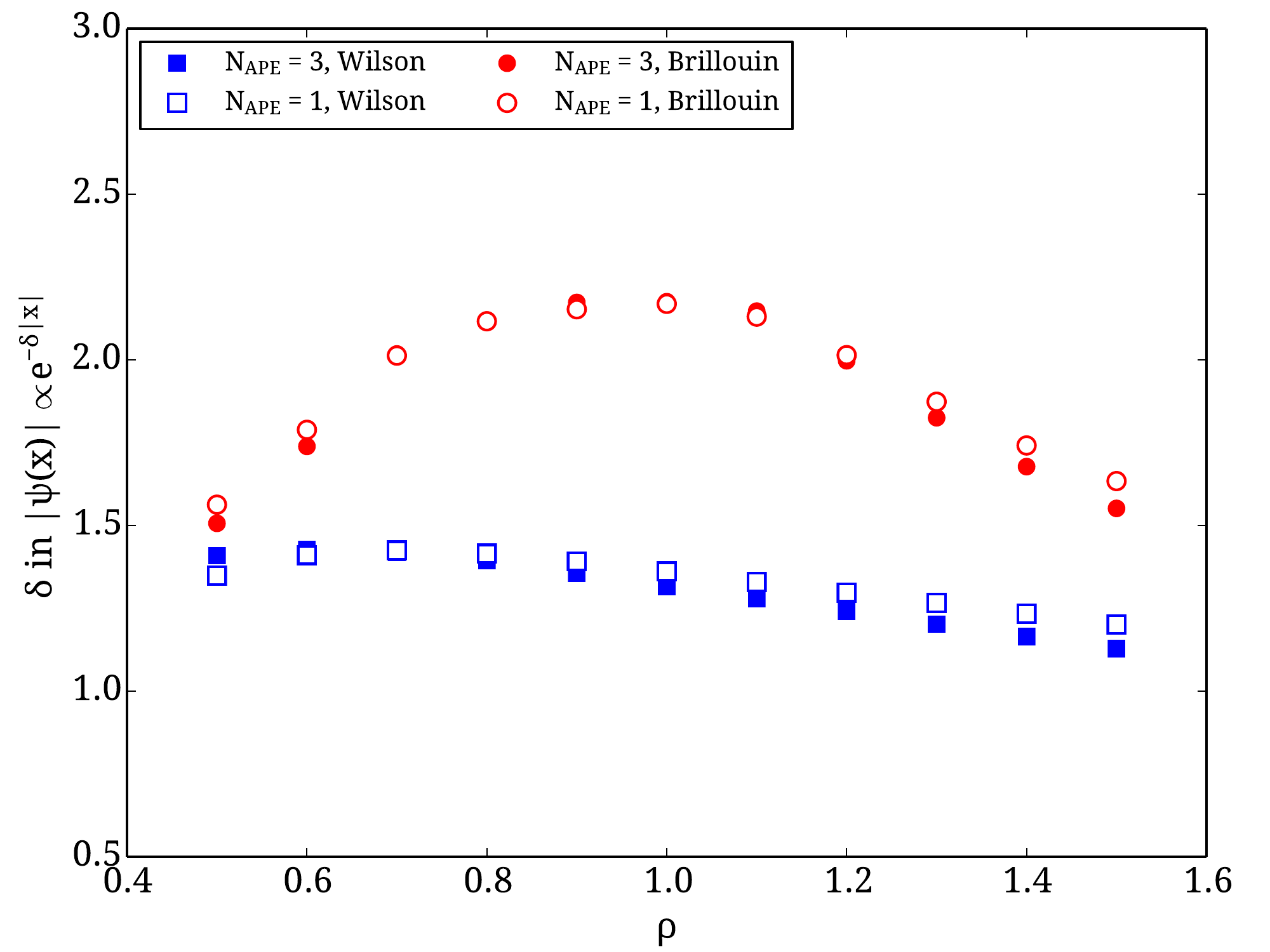}%
\includegraphics[width=0.49\textwidth]{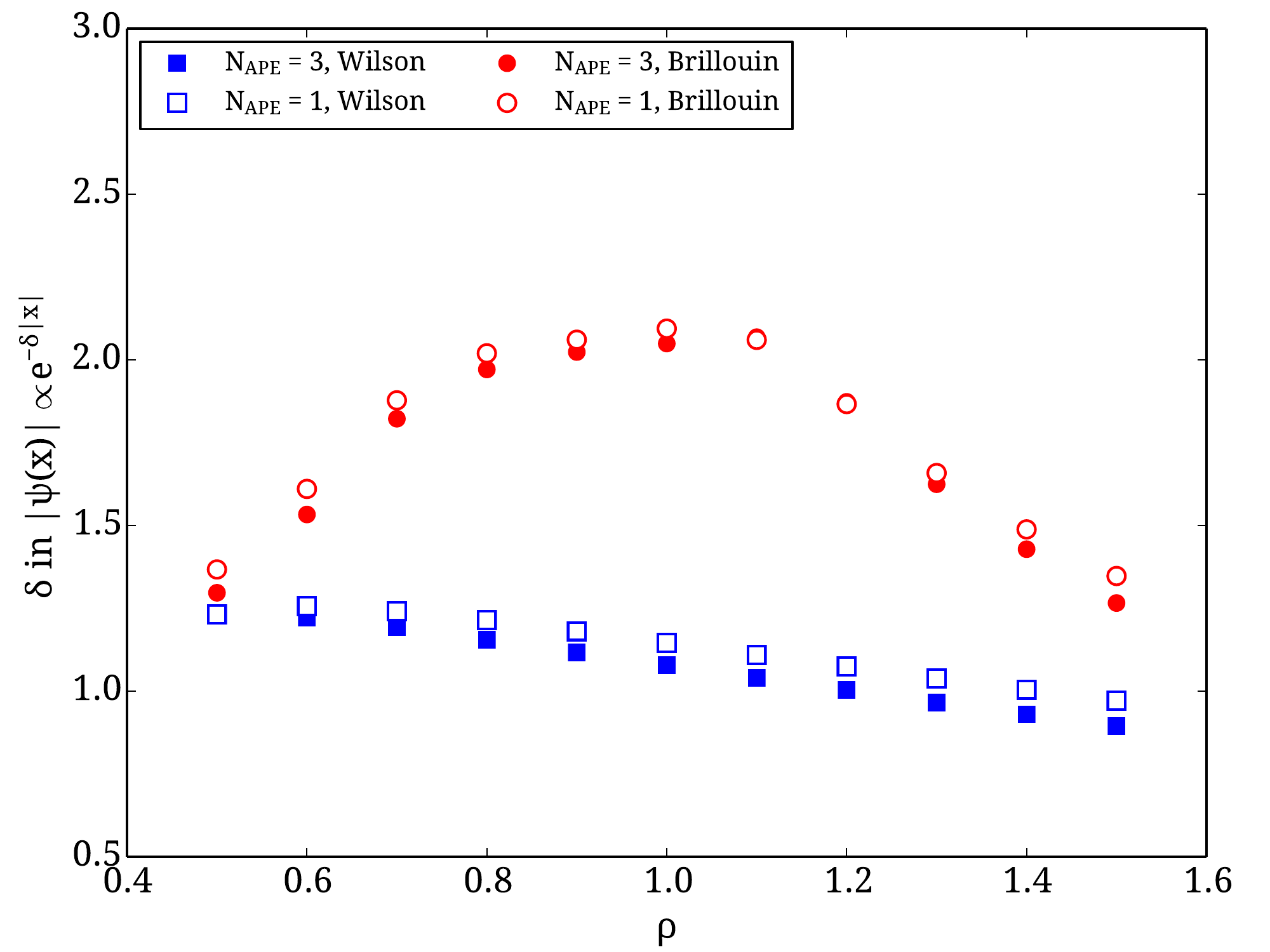}%
\caption{\sl\label{fig:locality_rhoscan}
Inverse localization $\de$ for the $f_{1,1}$ overlap actions with Wilson and Brillouin kernel, using 1-3 APE steps and $c_\mr{SW}\!=\!1$, as a function of $\rh$.
We use the coarsest ($a^{-1}\!=\!1.236\GeV$, $10^3\!\times\!20$ grid, left) and finest ($a^{-1}\!=\!2.964\GeV$, $24^3\!\times\!48$ grid, right) lattices of Ref.\,\cite{Durr:2010ch}.}
\end{figure}

How this localization, i.e.\ the ``effective mass'' $\de$ in $|\ps(x)|\propto\exp(-\de|x|)$, varies as a function of $\rh$ is shown in Fig.\,\ref{fig:locality_rhoscan}.
With an unsmeared and unimproved Wilson kernel frequently a value $\rh\simeq1.4$ is chosen to optimize locality on coarse lattices \cite{Hernandez:1998et}.
This, however, creates a clash with the free-field behavior where optimum locality is reached for $\rh\simeq0.6$ \cite{Durr:2005an}.
Our figure shows that even for the Wilson kernel this clash is resolved by some link smearing and putting $c_\mr{SW}=1$; then the optimum is assumed at $\rh\simeq0.6$.
Similarly, the Brillouin kernel with link smearing and $c_\mr{SW}=1$ has an optimum locality which (for accessible lattice spacings) is at $\rh\simeq1$, but also the ``magic'' value $\rh\simeq0.634$ of Sec.\,\ref{sec:DR} fares quite well.

In conclusion we find that the Brillouin kernel diminishes the anisotropy effects and results in an overlap operator that falls off significantly faster than the one with the Wilson kernel.
This may turn out to be relevant for QCD studies of bulk thermodynamic properties \cite{Hegde:2008nx}.


\subsection{Exploration of inversion cost and residual mass}

To further assess the suitability of the Brillouin operator as a kernel to the overlap procedure we conduct a pilot study of overlap inversions with a given source vector, as is typical in spectroscopy calculations.
The overall setup is standard \cite{Neuberger:1998my,Edwards:1998yw,vandenEshof:2002ms,Kennedy:2006ax}; we use a BiCGstab (``outer'') solver and a Kenney-Laub $f_{n,n}$ (``inner'') approximation to the matrix sign function.

\begin{figure}[!tb]
\centering
\includegraphics[width=0.75\textwidth]{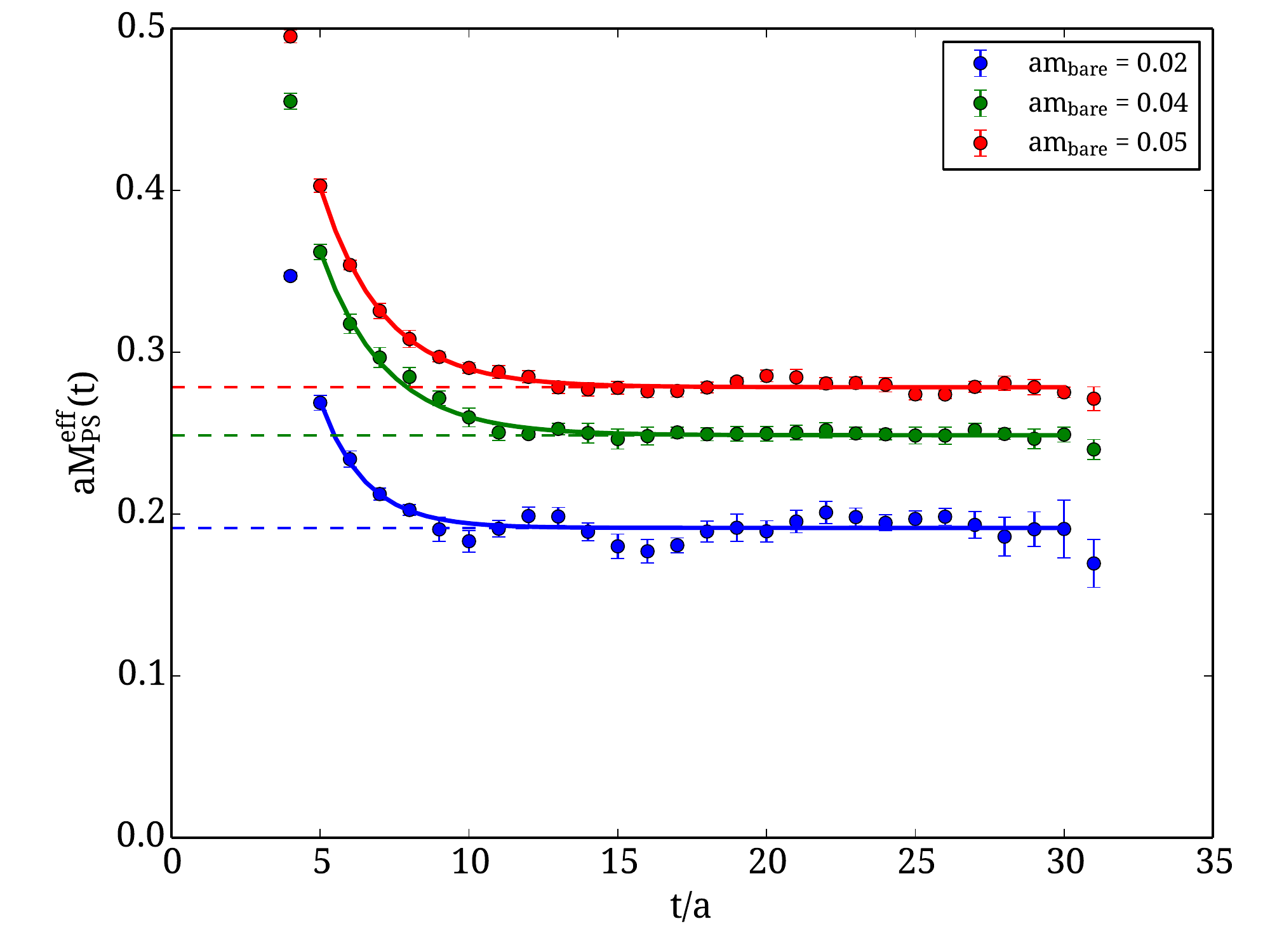}%
\vspace*{-5mm}
\caption{\sl\label{fig:meff}
Effective masses of pions built from two degenerate $f_{1,1}$ overlap fermions based on the Brillouin kernel with 3 APE steps and $c_\mr{SW}=1$.
The bare quark masses $am=0.02,0.04,0.05$ correspond to pion masses of $520\MeV,670\MeV,750\MeV$, respectively.}
\end{figure}

We use a freely available $\Nf=2$ ensemble by QCDSF, with geometry $40^3\times64$, sea pion mass $\Mpi\simeq280\MeV$ and lattice spacing $a\simeq0.0728(05)(19)\fm$ deduced from $a^{-1}=2.71(2)(7)\GeV$ at $\be=5.29$ \cite{Gockeler:2006jt,Bali:2012rs}.
Given the results in the previous subsections, we focus on the overlap operator with 3 smearings (at $\al_\mr{APE}=0.72$ each) and no ($c_\mr{SW}=0$) or tree-level ($c_\mr{SW}=1$) clover improvement in the kernel.
The shift parameter is pinned to the canonical value $\rh=1$ to avoid any tuning overhead; using the ``magic'' value $\rh\simeq0.634$ is not expected to bring any significant change.
The lattices are sufficiently long in Euclidean time such that we can identify clear effective mass plateaus for all bare quark masses studied.
A selection of such plateaus is shown in Fig.\,\ref{fig:meff}.
With fixed statistics, the statistical errors grow at small quark masses, but it is always evident that excited states contributions disappear at large $t/a$.

\begin{figure}[!tb]
\includegraphics[width=0.5\textwidth]{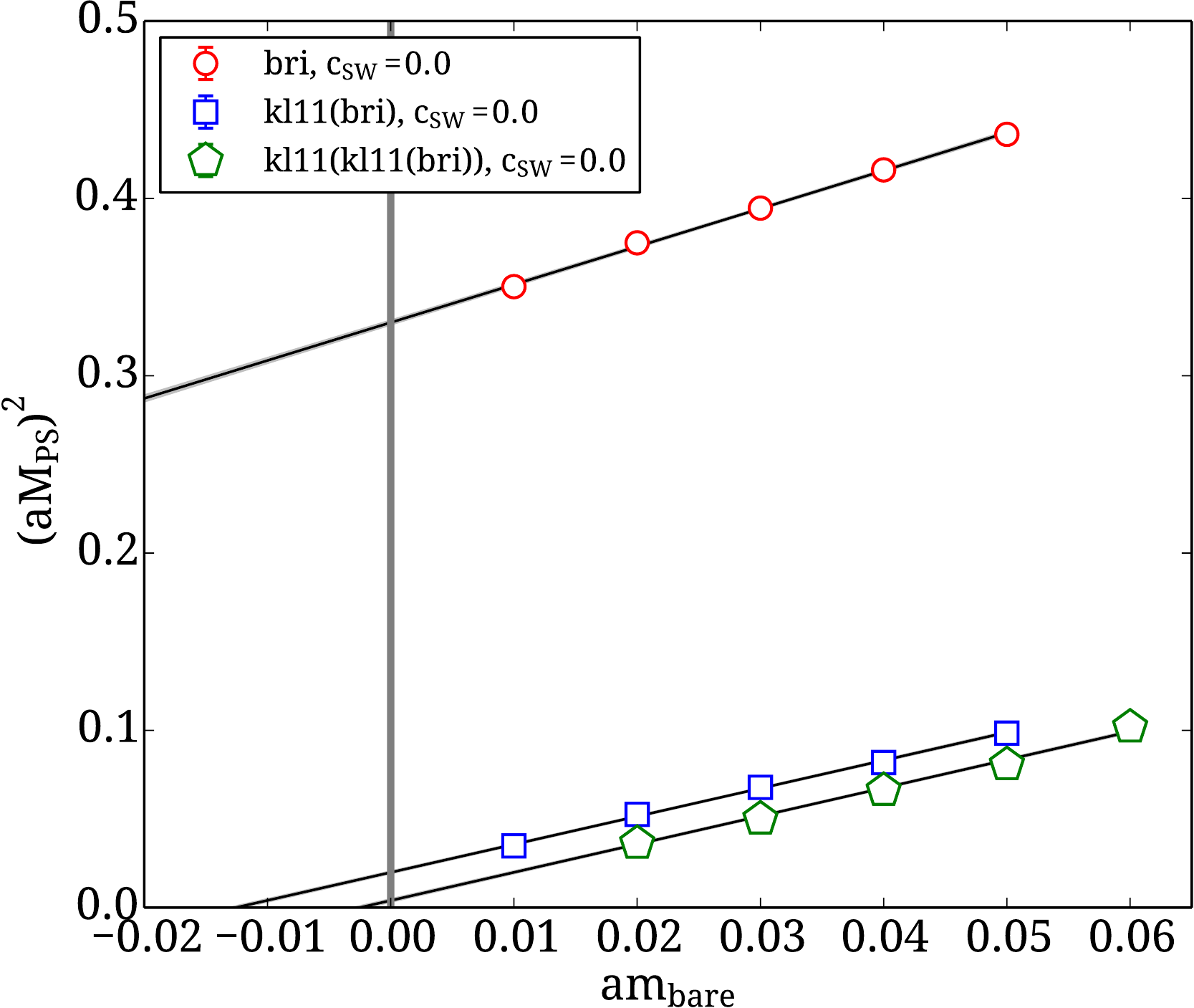}%
\includegraphics[width=0.5\textwidth]{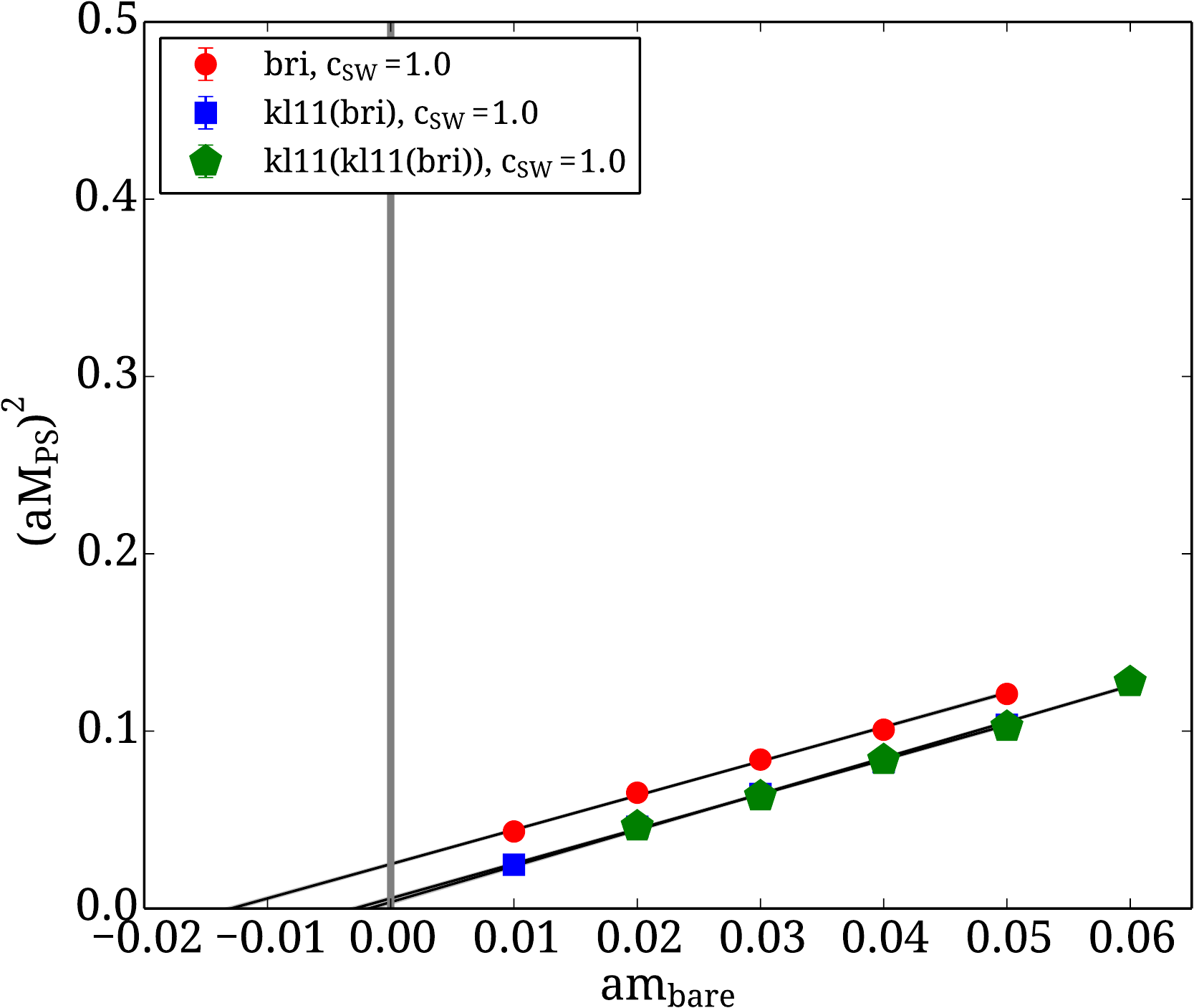}%
\caption{\sl\label{fig:mcrit}
$(a\Mpi)^2$ versus $am$ for $f_{1,1}$ and $f_{4,4}=f_{1,1}^{(2)}$ overlap fermions based on the
Brillouin kernel with $c_\mr{SW}=0$ (left) and $c_\mr{SW}=1$ (right).
Throughout 3 levels of APE smearing are used.}
\end{figure}

\begin{table}[!tb]
\centering
\begin{tabular}{|cccccc|}
\hline
 $am$ & $a\Mpi$ & $\Mpi$\,[MeV] & $n_\mr{iter}$ & time [sec] & $n_\mr{nodes}$ \\
\hline
 0.004 & 0.162(1) & 430 &  3433.9 &  865.2 & 320 \\
 0.010 & 0.192(1) & 520 &  2314.3 &  592.9 & 320 \\
 0.020 & 0.227(2) & 620 &  1311.9 &  320.9 & 320 \\
 0.035 & 0.271(2) & 730 &   878.7 &  215.6 & 320 \\
 0.050 & 0.313(1) & 850 &   652.8 &  178.6 & 320 \\
\hline
  0.01 & 0.155(2) & 420 & 13175.2 & 3469.8 & 320 \\
  0.02 & 0.191(2) & 520 &  6996.2 & 3858.8 & 200 \\
  0.03 & 0.222(2) & 600 &  3828.0 & 1570.6 & 160 \\
  0.04 & 0.249(2) & 670 &  2524.0 & 1656.6 & 160 \\
  0.05 & 0.278(2) & 750 &  2166.9 & 1463.3 & 160 \\
  0.06 & 0.307(2) & 830 &  1549.2 &  661.7 & 160 \\
\hline
\end{tabular}
\caption{\sl\label{tab:mcrit}
Overview of the pion mass $a\Mpi$ as a function of the quark mass $am$ for the $f_{1,1}$ Brillouin overlap fermion
with $c_\mr{SW}=0$ (top part) and $c_\mr{SW}=1$ (bottom part) in the kernel. In addition, we give the average number
of BiCGstab iterations and the average time per right-hand-side on $n_\mr{nodes}$ of the commodity cluster JUROPA.}
\end{table}

\begin{figure}[!tb]
\includegraphics[width=0.9\textwidth]{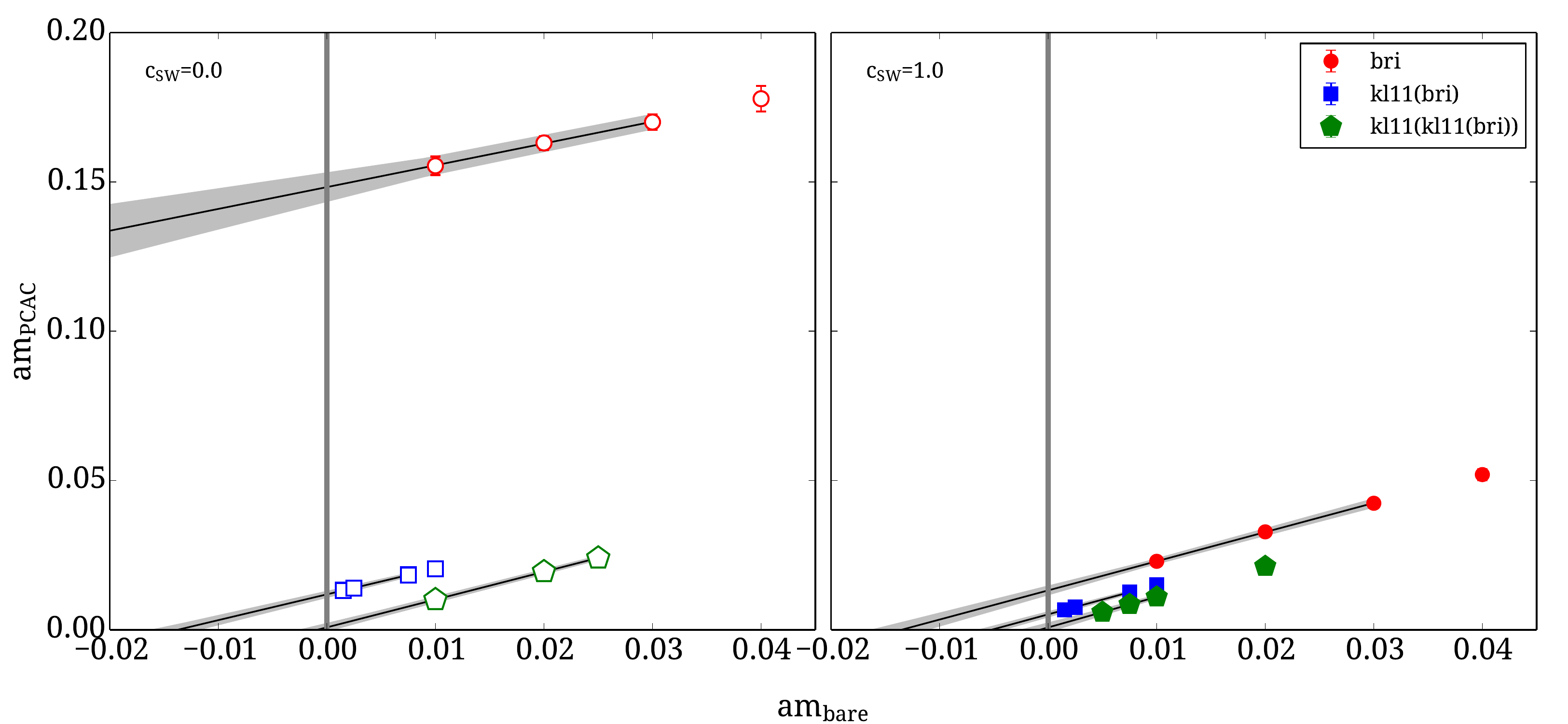}%
\caption{\sl\label{fig:mpcac}
PCAC versus bare quark mass for the Brillouin operator, its first and second KL11
iterates, with $c_\mr{SW}=0$ (left) and $c_\mr{SW}=1$ (right) on the $40^3\times64$
ensemble by QCDSF.}
\end{figure}

We measure $a\Mpi$ and monitor the number $N_\mr{iter}$ of outer iterations (i.e.\ of BiCGstab) for a selection of $f_{n,n}$ overlap masses $am$.
Results with $c_\mr{SW}=0$ and $c_\mr{SW}=1$ in the Brillouin kernel are shown in Fig.\,\ref{fig:mcrit} and presented in Tab.\,\ref{tab:mcrit}.
Since these are approximate overlap fermions, the additive mass shift is non-zero.
With the $c_\mr{SW}=0$ kernel, using $f_{1,1}$ brings more than an order of magnitude reduction, compared to the bare Brillouin action, and using $f_{4,4}$ makes it consistent with zero within our statistical precision.
On the other hand, with the $c_\mr{SW}=1$ kernel, the bare Brillouin action has a comparatively small mass shift, using $f_{1,1}$ makes it consistent with zero within $\sim2\si$, while using $f_{4,4}$ makes it consistent with zero within $\sim1\si$.

In the literature on approximate overlap fermions it is common practice to determine a ``residual mass'', i.e.\ an effective fermion mass evaluated at $am=0$.
In case of domain-wall fermions typically a version is used which explicitly refers to 5 dimensions \cite{Blum:2014tka}.
Since this is not an option for us, we choose the PCAC quark mass, employing the definition which is standard for Wilson fermions \cite{Bhattacharya:2005rb}.
The result is shown in Fig.\,\ref{fig:mpcac}, where the intercepts of the gray bands with the $y$-axis represent our residual quark masses.
The overall picture looks similar to the one in Fig.\,\ref{fig:mcrit}, except that this time also the $f_{1,1}$ overlap action with $c_\mr{SW}=1$ kernel shows a residual mass which is clearly non-zero.
But with $f_{4,4}$ the intercept is zero within errors, regardless of the value of $c_\mr{SW}$ in the kernel.


\section{Cascaded preconditioning \label{sec:precondition}}


In our opinion a dedicated research effort is needed to identify good preconditioners for the repeated inversions of the type $(\Dke_{-\rh}{}\dag\Dke_{-\rh}+\si)x=b$ which occur in the ``inner'' (CG-type) solver employed in the evaluation of partial fraction representation (\ref{KLnn_parfrac}).
For $n=1$ the shift is $\si=1/3$, for $n=4$ the smallest shift is $\si\simeq0.0311$; see App.\,\ref{app:kenneylaub} for details.

There is, however, a simple preconditioning strategy for the ``outer'' (BiCGstab-type) solver which is particularly convenient with a Brillouin kernel.
It builds on the relative proximity of $\Dke-1$ on one hand and various Kenney-Laub iterates of this combination on the other hand, see Fig.\,\ref{fig:eigenvalues}.
Suppose we wish to invert the operator defined by $f_{1,1}^{(3)}=f_{13,13}$.
One may then use
\beq
\Dke_{-1}+(1+\frac{\til{m}}{\rh})=\Dke+\frac{\til{m}}{\rh}
\eeq
as a preconditioner to the operator (with $A_{-\rh}^{(k)}=D_{-\rh}^{(k)}{}\dag D_{-\rh}^{(k)}$ and $D_{-\rh}^{(0)}=\Dke_{-\rh}$)
\beq
D^{(1)}_{-\rh}+(1+\frac{\til{m}}{\rh})\equiv
\Dke_{-\rh}\Big(\frac{1}{3}+\frac{8/9}{\Ake_{-\rh}+1/3}\Big)
+(1+\frac{\til{m}}{\rh})
\label{cascaded_one}
\eeq
which in turn is used as a preconditioner to the operator defined by $f_{1,1}^{(2)}=f_{4,4}$
\beq
D^{(2)}_{-\rh}+(1+\frac{\til{m}}{\rh})\equiv
D^{(1)}_{-\rh}\Big(\frac{1}{3}+\frac{8/9}{A^{(1)}_{-\rh}+1/3}\Big)
+(1+\frac{\til{m}}{\rh})
\label{cascaded_two}
\eeq
\bdm
=\Dke_{-\rh}\Big(\frac{1}{9}+\frac{0.229...}{\Ake_{-\rh}+0.0311...}+\frac{0.296...}{\Ake_{-\rh}+0.333...}+\frac{0.538...}{\Ake_{-\rh}+1.42...}+\frac{1.90...}{\Ake_{-\rh}+7.55...}\Big)
+(1+\frac{\til{m}}{\rh})
\edm
where the representation in terms of $\Ake$ uses the coefficients in the partial fraction expansion given in App.\,\ref{app:kenneylaub}.
The latter operator is used as a preconditioner to solve the equation
\bdm
\Big[D^{(3)}_{-\rh}+(1+\frac{\til{m}}{\rh})\Big]x=
\frac{\til{b}}{\rh}
\quad\mbox{where}\quad
D^{(3)}_{-\rh}\equiv
\Dke_{-\rh}\Big(
\frac{1}{27}+\frac{0.0743...}{\Ake_{-\rh}+0.00339...}+...+\frac{5.50...}{\Ake_{-\rh}+73.2...}
\Big)
\edm
for $x$, where $\til{m},\til{b}$ are given in (\ref{upshot_complete}), and the full set of coefficients is again found in App.\,\ref{app:kenneylaub}.

\begin{figure}[!tb]
\includegraphics[width=0.5\textwidth]{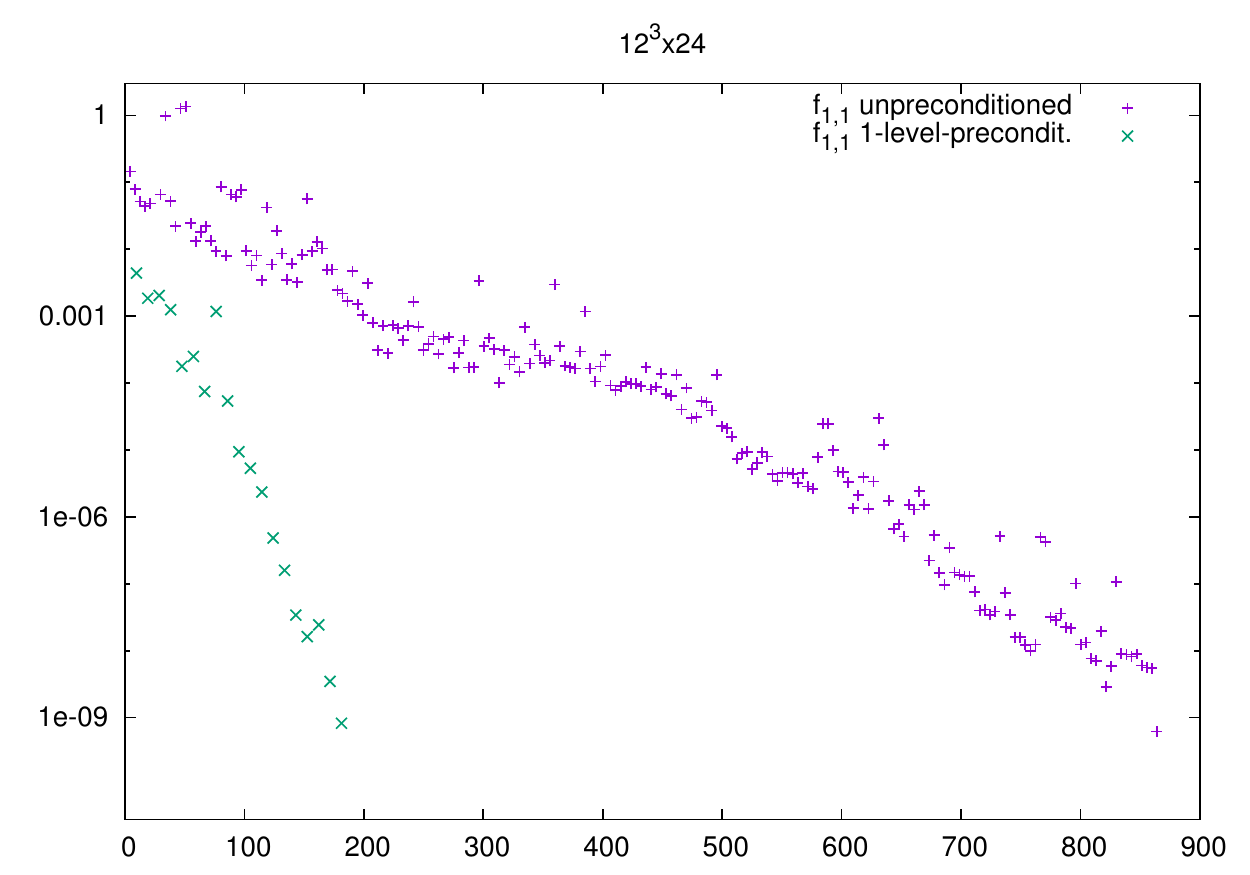}%
\includegraphics[width=0.5\textwidth]{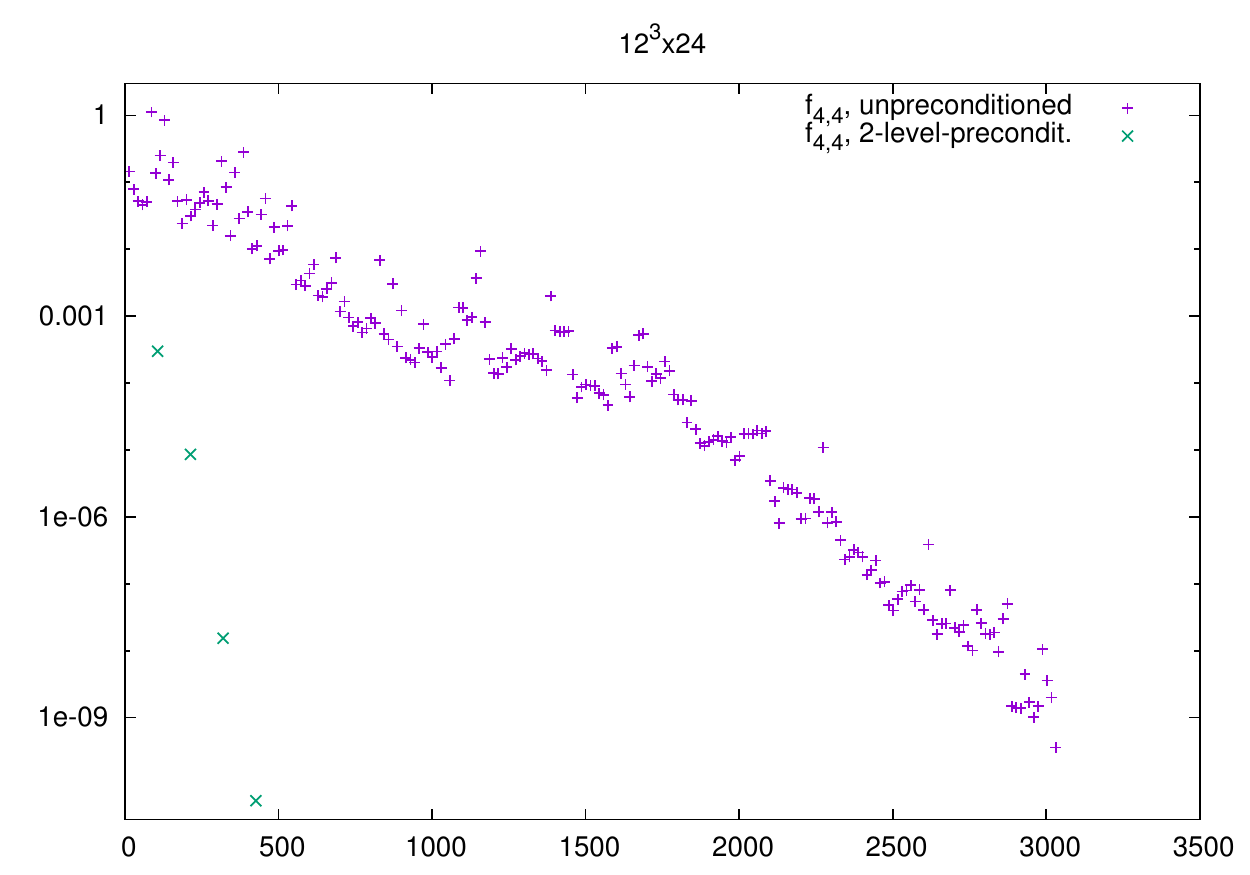}%
\caption{\sl\label{fig:cascaded}
Convergence of the BiCGstab solver for the Brillouin overlap with $\rh=1$, $c_\mr{SW}=0$, $n_\mr{smear}=7$ and $am=0.03$ on a pure glue $12^3\times24$ lattice at $\be=6.0$.
We use (\ref{cascaded_one}) as outer operator and no or one level of preconditioning (left) or (\ref{cascaded_two}) as outer operator and no or two levels of preconditioning (right).
We handle 12 right-hand sides simultaneously (cf.\ App.\,\ref{app:flopcount}), and the worst relative residual is shown.
The $x$-axis denotes wall-clock time in seconds.}
\end{figure}

\begin{table}[!tb]
\centering
\begin{tabular}{|cccrr|}
\hline
Volume       & $f_{n,n}$ & precond. & $n_\mr{iter}$ & time [sec]\\
\hline
$12^3\!\times\!24$ & 1,1 & none     & 204 &    864 \\
                   & 1,1 & 1-level  &  19 &    181 \\
                   & 4,4 & none     & 212 &   3031 \\
                   & 4,4 & 2-level  &   4 &    426 \\
\hline
$16^3\!\times\!32$ & 1,1 & none     & 249 &   4265 \\
                   & 1,1 & 1-level  &  22 &    732 \\
                   & 4,4 & none     & 285 &  16706 \\
                   & 4,4 & 2-level  &  10 &   3397 \\ 
\hline
$24^3\!\times\!48$ & 1,1 & none     & 264 &  27490 \\
                   & 1,1 & 1-level  &  25 &   5040 \\
                   & 4,4 & none     & 292 & 132975 \\
                   & 4,4 & 2-level  &   9 &  19835 \\
\hline
\end{tabular}
\caption{\label{tab:cascaded}\sl
Timings of the BiCGstab solver for the Brillouin overlap action with $\rh=1$, $c_\mr{SW}=0$, $n_\mr{smear}=7$ and $am=0.03$ on pure glue lattices at $\be=6.0$.
We use no or 1-level preconditioning for the operator defined through $f_{1,1}$, and no or 2-level preconditioning with $f_{4,4}$.
The exit criterion was set to $||r||/||b||\leq10^{-9}$.
The number of iterations is for the outer solver, the timings are per right-hand-side on one (4-core) node.
Details for $12^3\!\times\!24$ are shown in Fig.\,\ref{fig:cascaded}.}
\end{table}

In Fig.\,\ref{fig:cascaded} and Tab.\,\ref{tab:cascaded} we show that this concept works over one and two steps (i.e.\ for $D^{(1)}$ and $D^{(2)}$ implementing one and two iterations of $f_{1,1}$, respectively).
It seems that significant savings can be achieved even if none of the preconditioner masses is tuned (we use the same bare mass in all operators, even in the Brillouin action).
Without preconditioning the $f_{4,4}$-based approximant $D^{(2)}$ to the overlap action is significantly more expensive than the $f_{1,1}$-based operator $D^{(1)}$, but with cascaded preconditioning the extra cost of the better approximation becomes more tolerable.

To the best of our knowledge, preconditioning of an overlap operator by its kernel was first tried in Ref.\,\cite{Cundy:2004pza}, and a more elaborate version of this (with tuning of the preconditioner mass) was presented in Ref.\,\cite{Brannick:2014vda}.
Both of these references use the Wilson action as a preconditioner to the Wilson overlap.
Our Fig.\,\ref{fig:cascaded} demonstrates that the same concept works for Brillouin fermions, too, but obviously there is much room for optimization, still, on our side.


\section{Outlook on dynamical Brillouin overlap simulations \label{sec:outlook}}


We close with a brief outlook on how our findings fit into the perspective of carrying out dynamical overlap simulations based on the Brillouin kernel and the Hybrid Monte Carlo (HMC) algorithm \cite{Clark:2006wq}.
For dynamical overlap simulations with a Wilson kernel see e.g.\ Refs.\,\cite{Fodor:2003bh,DeGrand:2004nq,Cundy:2008zc,Allton:2008pn,Noaki:2008iy,Borsanyi:2012xf}.

The HMC algorithm is governed by the molecular dynamics time evolution which, in turn, builds on the HMC force.
Let $D_m$ be an undoubled fermion operator, which implicitly depends on the ``thin'' gauge links $U$.
The pseudo-fermion action for $\Nf$ degenerate fermions is
\beq
S_\mr{pf}=\<\ph|\,A_m^{-\Nf/2}\,|\ph\>
=\int\ph\dag(x)\,A_m^{-\Nf/2}(x,y)\,\ph(y)\;d^4\!x\,d^4\!y
\;,\quad
A_m=D_m\dag D_m^{}>0
\eeq
where $\ph$ denotes a boson field with the spinor/color components of a standard Dirac fermion.
The HMC force is defined as minus the derivative of $S_\mr{pf}$ with respect to the thin gauge links \cite{Clark:2006wq}.
For $\Nf=1$ one exploits the fact that the $p$-th order diagonal rational approximation of $x^{-1/2}$ over the relevant spectral range admits a partial fraction representation \cite{Clark:2006wq}
\beq
x^{-1/2}\simeq\al_0+\sum_{k=1}^p\frac{\al_k}{x+\be_k}
\eeq
with $\al_k>0$ for $1 \leq k \leq p$ and $0<\be_1<...<\be_p$.
For $\Nf=1$ the pseudo-fermion force is thus
\beq
F_\mr{pf}=-S_\mr{pf}'=\sum_{k=1}^p\al_k\,
\<\ph|(A_m+\be_k)^{-1}\,A_m'\,(A_m+\be_k)^{-1}|\ph\>
\eeq
where the prime denotes the derivative with respect to the gauge field element $A_\mu^a(x+\hat\mu/2)$, defined as a Gell-Mann component of $\log(U_\mu(x))$.
For $\Nf\in 2\mb{N}$ no rational representation is needed (though it still might be favorable for efficiency reasons \cite{Clark:2006wq}), and one gets away with products of powers of $A_m^{-1}$ and factor $A_m'$.
The bottom line is that in both cases -- even or odd $\Nf$ -- the derivative $A_m'$ with respect to the thin gauge field is required to work out the pseudo-fermion contribution to the molecular dynamics force.

This $A_m'$ is straightforward to write down in case $D_m$ is an ultralocal operator.
On the other hand, $A_m'$ is more involved with an overlap action.
The attentive reader will have noticed that we advocate using a fixed-order rational approximation to the matrix sign function (see Sec.\,\ref{sec:KL}), regardless of the spectral properties of $H^\mr{ke}_{-\rh}=\gaf\Dke_{-\rh}$ on the current gauge background $U$ (which is common practice with the domain-wall setup \cite{Allton:2008pn,Blum:2014tka}).
This is convenient, since the order of the rational approximation to the sign function does not increase as the belly-mode of the underlying hermitean kernel changes sign (in an attempt of the HMC algorithm to increase/decrease the global topological charge by one unit).
However, with a fixed-order approximation to the matrix sign function in the definition of $D_m$, it is still fairly easy to work out the inner derivative $A_m'$.
From the definition (\ref{def_gell}) we obtain
\beq
g_\ell'(x)=
2\ell\,\frac{(1+x)^\ell (1-x)^{\ell-1}+(1+x)^{\ell-1} (1-x)^\ell}{[(1+x)^\ell+(1-x)^\ell]^2}=
4\ell\,\frac{(1-x^2)^{\ell-1}}{[(1+x)^\ell+(1-x)^\ell]^2}
\eeq
which implies $g_\ell'(-x)=g_\ell'(x)$ for any $\ell\in\mb{N}$, along with $g_\ell'(x)\geq0$ for odd $\ell$.
In other words, the diagonal Kenney-Laub approximations $f_{n,n}$ to the matrix sign function grow monotonically on $]0,\infty[$ and strictly monotonically on the intervals $]0,1[$ and $]1,\infty[$.
This is in marked distinction to the situation with optimal rational approximations which show, within the accessible interval $]\la_\mr{min},\la_\mr{max}[$, many ``wiggles'', i.e.\ small-scale oscillations, in particular close to the endpoints.
It remains to be seen whether this peculiar property of the diagonal Kenney-Laub approximants has any impact on e.g.\ the topological tunneling rate at a fixed lattice spacing $a$.

In the end the computation of various inverses of $D_m$ and $A_m$ is required, and for this purpose the cascaded preconditioning technique as discussed in Sec.\,\ref{sec:precondition} will prove useful.
However, in the HMC algorithm there is still more room for optimization.
In principle, the pseudo-fermion force can be calculated with any fermion operator, as long as one includes the difference between the used and the desired pseudo-fermion action in the final accept-reject step.
For instance, in Ref.\,\cite{Durr:2004as} it was proposed to use the staggered action as a driving engine in the molecular dynamics evolution for HMC simulations of the Wilson overlap action.
Of course, the further the two actions involved and the longer the trajectory length the lower the acceptance rate.
However, in two space-time dimensions such games have been played successfully \cite{Bietenholz:2011ey}, and we feel optimistic that the relative proximity of the Brillouin kernel to the
Brillouin overlap action will allow for significant savings in four space-time dimensions, too.


\section{Summary \label{sec:summary}}


We summarize the main results of our investigation as follows:
\begin{enumerate}
\item
The free-field dispersion relation of the Brillouin overlap action with generic $\rh$ deviates from the continuum relation $(aE)^2-(a\mb{p})^2=(am)^2$ through a term proportional to $(am)^4$.
With the ``magic'' value $\rh=(3-\sqrt{3})/2$ the leading discretization error is lifted to order $(am)^6$.
We hope that this feature proves useful to compute properties of systems with charm quarks, perhaps in a further perspective even with bottom quarks.
\item
We advocate using any of the diagonal Kenney-Laub approximants $g_{2n+1}=f_{n,n}$ to the matrix sign function at fixed order $n$ in the definition of the overlap action.
This is close in spirit to what is done in the domain-wall setup, except that then a five-dimensional framework is used, and the effective four-dimensional operator corresponds to an element $f_{n-1,n}=g_{2n}$ of the Kenney-Laub family of matrix iterations.
In both cases the partial-fraction expansion involved in the definition of the overlap action does not depend on the gauge field details and can be constructed beforehand.
This, in turn, makes it easier to demonstrate good strong-scaling properties on massively parallel architectures.
\item
We advocate defining the massive overlap action $D_m$ through (\ref{def_mass_complete}) rather than (\ref{def_mass_traditional}).
Effectively this has been done in the past through the ``extra prescription'' of dressing external currents or densities with a ``chiral symmetry ensuring factor'' $(1-aD/[2\rh])$, where $D$ is the zero-mass overlap operator.
Hence our proposal adds to the ``piece of mind'', since the danger of missing an important ingredient in a later step of the calculation is bypassed.
\item
We checked that the eigenvalue spectra of both the non-hermitean and the hermitean kernel operator show promising features for the extraction of the unique unitary part of $\Dke_{-\rh}$ through an application of $f_{n,n}=g_{2n+1}$ with sufficiently high $n$.
With the Wilson and the Brillouin kernel link smearing helps to open the ``eye'' in the eigenvalue spectrum and hence to reduce the appropriate value of $n$.
On the other hand a clover term in the kernel was neither found to bring a clear advantage nor a clear disadvantage.
\item
In terms of physics properties the second important advantage of the Brillouin kernel over the Wilson kernel is the improved locality of the resulting overlap operator.
Whenever locality is important and the CPU cost scale with a high power of $a$ (as is the case in the study of bulk thermodynamics properties), a complete study with a valid continuum limit may be cheaper with the Brillouin overlap than with the Wilson overlap action.
\item
The proximity of the non-chiral Brillouin kernel to the (exact) Brillouin overlap operator (and the diagonal Kenney-Laub approximants to the latter) makes the Brillouin overlap action particularly susceptible to cascaded preconditioning strategies.
Even without any tuning effort, significant savings were found for a BiCGstab solver of the Brillouin overlap operator with one and two levels of preconditioning.
\end{enumerate}

\bigskip\noindent{\bf Acknowledgments}:\newline
The authors gratefully acknowledge the computing time granted by Forschungszentrum J\"ulich GmbH and provided on the supercomputer JUROPA at J\"ulich Supercomputing Centre (JSC).
This work was in parts supported by DFG through the program SFB-TR-55.


\appendix


\section{Details of quark-level dispersion relations \label{app:dispersion}}


In this appendix we shall use the boson momentum $\hat{p}$ as well as the std-fermion and iso-fermion momenta $\bar{p}$ and $\til{p}$ which are defined such that $\nab_\mu^\mr{std}=\ri\bar{p}_\mu$ and $\nab_\mu^\mr{iso}\,=\ri\til{p}_\mu$ \cite{Durr:2010ch}, viz.\
\beq
\hat{p}_\mu=\frac{2}{a}\sin(\frac{ap_\mu}{2}),\qquad
\bar{p}_\mu=\frac{1}{a}\sin(ap_\mu),\qquad
\til{p}_\mu=\frac{1}{27a}\sin(ap_\mu)\prod_{\nu\neq\mu}\{\cos(ap_\nu)\!+\!2\}
\;.
\eeq
Furthermore, we need the momentum space representations of the Laplacians \cite{Durr:2010ch}
\bea
\lap^\mr{std}&=&-\frac{4}{a^2}\sum_\mu\sin^2(\frac{ap_\mu}{2})=
\frac{2}{a^2}\sum_\mu\cos(ap_\mu)-\frac{8}{a^2}=-\sum_\mu\hat{p}_\mu^2\equiv-\hat{p}^2
\\
\lap^\mr{bri}&=&\frac{4}{a^2}\prod_\mu\cos^2(\frac{ap_\mu}{2})-\frac{4}{a^2}=
\frac{1}{4a^2}\prod_\mu\{\cos(ap_\mu)+1\}-\frac{4}{a^2}\,\equiv-\check{p}^2
\eea
and we caution that (unlike $\hat{p}^2$) the quantity $\check{p}^2$ is not a sum of squares.


\subsection{Dispersion relation for Wilson operator}

The Green's function of the Wilson operator at mass $am$ and $r=1$ follows as
\bea
D_{\mr{W},m}&=&\textstyle
\nab_\mu^\mr{std}\ga_\mu-\frac{a}{2}\lap^\mr{std}+m\:\:=\:\:
\ri\bar{p}_\mu\ga_\mu+\frac{a}{2}\hat{p}^2+m
\nonumber
\\
G_{\mr{W},m}&=&
\frac
{-\ri\bar{p}_\si\ga_\si+\frac{a}{2}\hat{p}^2+m}
{(\ri\bar{p}_\mu\ga_\mu+\frac{a}{2}\hat{p}^2+m)(-\ri\bar{p}_\nu\ga_\nu+\frac{a}{2}\hat{p}^2+m)}
\:\:=\:\:
\frac
{-\ri\bar{p}_\si\ga_\si+\frac{a}{2}\hat{p}^2+m}
{\bar{p}^2+(\frac{a}{2}\hat{p}^2+m)^2}
\;.
\eea
Searching for a zero of the denominator with
$\frac{a}{2}\hat{p}^2=-\frac{1}{a}\sum_\mu\cos(ap_\mu)+\frac{4}{a}$ and $p_4\to\ri E$ yields
\beq
\textstyle
\sinh^2(aE)-\sum_i\sin^2(ap_i)=\cosh^2(aE)+2\cosh(aE)\big[\sum_i\cos(ap_i)-4-am\big]+\big[...\big]^2
\label{DR_wils}
\eeq
and upon using $\cosh^2-\sinh^2=1$ this turns into a \emph{linear equation} in $\cosh(aE)$ which gives (\ref{free_wils}).
For the sake of a check we note that (\ref{DR_wils}) simplifies to $1+[1+am]^2=\cosh \cdot [2+2am]$ at $a\mb{p}=\mb{0}$.


\subsection{Dispersion relation for Brillouin operator}

The Green's function of the Brillouin operator at mass $am$ and $r=1$ follows as
\bea
D_{\mr{B},m}&=&\textstyle
\nab_\mu^\mr{iso}\ga_\mu-\frac{a}{2}\lap^\mr{bri}+m\:\:=\:\:
\ri\til{p}_\mu\ga_\mu+\frac{a}{2}\check{p}^2+m
\nonumber
\\
G_{\mr{B},m}&=&
\frac
{-\ri\til{p}_\si\ga_\si+\frac{a}{2}\check{p}^2+m}
 {( \ri\til{p}_\mu\ga_\mu+\frac{a}{2}\check{p}^2+m)
  (-\ri\til{p}_\nu\ga_\nu+\frac{a}{2}\check{p}^2+m)}
\:\:=\:\:
\frac
{-\ri\til{p}_\si\ga_\si+\frac{a}{2}\check{p}^2+m}
{\til{p}^2+(\frac{a}{2}\check{p}^2+m)^2}
\;.
\eea
Searching for a zero of the denominator with
$\frac{a}{2}\check{p}^2=\frac{2}{a}-\frac{1}{8a}\prod_\mu\{c_\mu+1\}$ yields
\bdm
\textstyle
a^2\sum_\mu\til{p}_\mu^2
+\frac{1}{64}\prod_\mu\{c_\mu+1\}^2
-\frac{1}{4}\prod_\mu\{c_\mu+1\}[2+am]
+[2+am]^2=0
\edm
with $a^2\til{p}^2=\frac{1}{729}\sum_\mu s_\mu^2\prod_{\nu\neq\mu}\{c_\nu+2\}^2$
and $c_\mu\equiv\cos(ap_\mu)$, $s_\mu\equiv\sin(ap_\mu)$.
Next, $p_4\to\ri E$ leads to
\bea
\textstyle
\frac{1}{729}\sum_i s_i^2 \prod_{j\neq i}\{c_j+2\}^2\{\cosh^2+4\cosh+4\}-
\frac{1}{729}\prod_{i}\{c_i+2\}^2\sinh^2&&
\nonumber
\\[2mm]
\textstyle
+\frac{1}{64}\prod_i\{c_i+1\}^2\{\cosh^2+2\cosh+1\}
-\frac{1}{4}\prod_i\{c_i+1\}\{\cosh+1\}[2+am]
+[2+am]^2&=&0
\qquad
\eea
and upon using $\cosh^2-\sinh^2=1$ this turns into a \emph{quadratic equation} in $\cosh(aE)$.
We note that it simplifies to $\{1-\cosh^2\}+\{\cosh^2+2\cosh+1\}-\{\cosh+1\}[4+2am]+[2+am]^2=0$ at $a\mb{p}=\mb{0}$, which agrees with the respective (linear) expression in the Wilson case, as it must be \cite{Cho:2015ffa}.
For $a\mb{p}\neq\mb{0}$ we have $A\cosh^2+B\cosh+C=0$ with
\bea
\frac{-B-\sqrt{B^2-4AC}}{2A}&=&
1+\frac{1}{2}(am)^2-\frac{1}{2}(am)^3+\frac{1}{2}(am)^4-\frac{1}{2}(am)^5
\nonumber
\\
&+&\Big[\frac{1}{2}+\frac{1}{12}(am)^2-\frac{1}{24}(am)^3\Big](a\mb{p})^2
\nonumber
\\
&+&\frac{1}{12}\sum_{i<j}a^4(p_i^2p_j^2) +\Big[\frac{1}{24}+\frac{1}{24}am\Big]\sum_{i}(ap_i)^4
+O(a^6)
\eea
and upon taking $\mr{arcosh}(.)$ one produces the dispersion relation (\ref{free_bril}).


\subsection{Dispersion relation for overlap operator with Wilson kernel}

To establish the dispersion relation of the overlap operator with Wilson kernel one starts from
\bdm
D_{\mr{W},-\rh}=\textstyle
\ri\bar{p}_\mu\ga_\mu+\frac{a}{2}\hat{p}^2-\frac{\rh}{a}
\edm
\bdm
D_{\mr{W},-\rh}\dag D_{\mr{W},-\rh}=\textstyle
(-\ri\bar{p}_\mu\ga_\mu+\frac{a}{2}\hat{p}^2-\frac{\rh}{a})
( \ri\bar{p}_\nu\ga_\nu+\frac{a}{2}\hat{p}^2-\frac{\rh}{a})
=
\bar{p}^2+(\frac{a}{2}\hat{p}^2-\frac{\rh}{a})^2
\edm
as this yields the the free-field form of the desired operator and of its inverse:
\bea
D_{\mr{NW},m}\!&\!=\!&\!\textstyle
\big(1-\frac{am}{2\rh}\big)D_\mr{NW}+m
\quad\mbox{with}\quad
D_\mr{NW}=\textstyle
\frac{\rh}{a}
\big\{1+D_{\mr{W},-\rh}[D_{\mr{W},-\rh}\dag D_{\mr{W},-\rh}]^{-1/2}\big\}
\nonumber
\\
D_{\mr{NW},m}\!&\!=\!&\!\textstyle
\underbrace{\textstyle\big(\frac{\rh}{a}+\frac{m}{2}\big)}_{\equiv c}+
\underbrace{\textstyle\big(\frac{\rh}{a}-\frac{m}{2}\big)}_{\equiv d}
\big(\ri\bar{p}_\mu\ga_\mu+\frac{a}{2}\hat{p}^2-\frac{\rh}{a}\big)
\big[\bar{p}^2+(\frac{a}{2}\hat{p}^2-\frac{\rh}{a})^2\big]^{-1/2}
\nonumber
\\
G_{\mr{NW},m}\!&\!=\!&\!\textstyle
\frac
{c+d[\bar{p}^2+(\frac{a}{2}\hat{p}^2-\frac{\rh}{a})^2]^{-1/2}(-\ri\bar{p}_\si\ga_\si+\frac{a}{2}\hat{p}^2-\frac{\rh}{a})}
{\big\{c+d[\bar{p}^2+(\frac{a}{2}\hat{p}^2-\frac{\rh}{a})^2]^{-1/2}(-\ri\bar{p}_\mu\ga_\mu+\frac{a}{2}\hat{p}^2-\frac{\rh}{a})\big\}
 \big\{c+d( \ri\bar{p}_\nu\ga_\nu+\frac{a}{2}\hat{p}^2-\frac{\rh}{a})[\bar{p}^2+(\frac{a}{2}\hat{p}^2-\frac{\rh}{a})^2]^{-1/2}\big\}}
\nonumber
\\
\!&\!=\!&\!\textstyle
\frac
{c+d[\bar{p}^2+(\frac{a}{2}\hat{p}^2-\frac{\rh}{a})^2]^{-1/2}(-\ri\bar{p}_\si\ga_\si+\frac{a}{2}\hat{p}^2-\frac{\rh}{a})}
{
c^2+
2cd(\frac{a}{2}\hat{p}^2-\frac{\rh}{a})[...]^{-1/2}
+d^2[...]^{-1/2}[...][...]^{-1/2}
}
\;.
\eea
Hence, one ends up searching for zero in
$c^2+2cd(\frac{a}{2}\hat{p}^2-\frac{\rh}{a})[\bar{p}^2+(\frac{a}{2}\hat{p}^2-\frac{\rh}{a})^2]^{-1/2}+d^2=0$
with
\bdm
\bar{p}^2=\textstyle\frac{1}{a^2}\sum_\mu s_\mu^2
\qquad\mbox{and}\qquad
\hat{p}^2=\textstyle\frac{8}{a^2}-\frac{2}{a^2}\sum_\mu c_\mu
\edm
or equivalently for a zero in $(c^2+d^2)[\bar{p}^2+(\frac{a}{2}\hat{p}^2-\frac{\rh}{a})^2]^{1/2}+2cd(\frac{a}{2}\hat{p}^2-\frac{\rh}{a})=0$, and the square root makes this a \emph{transcendental equation} in $\cosh(aE)$.
The result through $O(a^5)$ is given in (\ref{over_wils}).


\subsection{Dispersion relation for overlap operator with Brillouin kernel}

To establish the dispersion relation of the overlap operator with Brillouin kernel one starts from
\bdm
D_{\mr{B},-\rh}=\textstyle
\ri\til{p}_\mu\ga_\mu+\frac{a}{2}\check{p}^2-\frac{\rh}{a}
\edm
\bdm
D_{\mr{B},-\rh}\dag D_{\mr{B},-\rh}=\textstyle
(-\ri\til{p}_\mu\ga_\mu+\frac{a}{2}\check{p}^2-\frac{\rh}{a})
( \ri\til{p}_\nu\ga_\nu+\frac{a}{2}\check{p}^2-\frac{\rh}{a})
=
\til{p}^2+(\frac{a}{2}\check{p}^2-\frac{\rh}{a})^2
\edm
as this yields the the free-field form of the desired operator and of its inverse:
\bea
D_{\mr{NB},m}\!&\!=\!&\!\textstyle
\big(1-\frac{am}{2\rh}\big)D_\mr{NB}+m
\quad\mbox{with}\quad
D_\mr{NB}=\textstyle
\frac{\rh}{a}
\big\{1+D_{\mr{B},-\rh}[D_{\mr{B},-\rh}\dag D_{\mr{B},-\rh}]^{-1/2}\big\}
\nonumber
\\
D_{\mr{NB},m}\!&\!=\!&\!\textstyle
\underbrace{\textstyle\big(\frac{\rh}{a}+\frac{m}{2}\big)}_{\equiv c}+
\underbrace{\textstyle\big(\frac{\rh}{a}-\frac{m}{2}\big)}_{\equiv d}
\big(\ri\til{p}_\mu\ga_\mu+\frac{a}{2}\check{p}^2-\frac{\rh}{a}\big)
\big[\til{p}^2+(\frac{a}{2}\check{p}^2-\frac{\rh}{a})^2\big]^{-1/2}
\nonumber
\\
G_{\mr{NB},m}\!&\!=\!&\!\textstyle
\frac
{c+d[\til{p}^2+(\frac{a}{2}\check{p}^2-\frac{\rh}{a})^2]^{-1/2}(-\ri\til{p}_\si\ga_\si+\frac{a}{2}\check{p}^2-\frac{\rh}{a})}
{\big\{c+d[\til{p}^2+(\frac{a}{2}\check{p}^2-\frac{\rh}{a})^2]^{-1/2}(-\ri\til{p}_\mu\ga_\mu+\frac{a}{2}\check{p}^2-\frac{\rh}{a})\big\}
 \big\{c+d( \ri\til{p}_\nu\ga_\nu+\frac{a}{2}\check{p}^2-\frac{\rh}{a})[\til{p}^2+(\frac{a}{2}\check{p}^2-\frac{\rh}{a})^2]^{-1/2}\big\}}
\nonumber
\\
\!&\!=\!&\!\textstyle
\frac
{c+d[\til{p}^2+(\frac{a}{2}\check{p}^2-\frac{\rh}{a})^2]^{-1/2}(-\ri\til{p}_\si\ga_\si+\frac{a}{2}\check{p}^2-\frac{\rh}{a})}
{
c^2+
2cd(\frac{a}{2}\check{p}^2-\frac{\rh}{a})[...]^{-1/2}
+d^2[...]^{-1/2}[...][...]^{-1/2}
}
\eea
Hence, one ends up searching for zero in $c^2+2cd(\frac{a}{2}\check{p}^2-\frac{\rh}{a})[\til{p}^2+(\frac{a}{2}\check{p}^2-\frac{\rh}{a})^2]^{-1/2}+d^2=0$ with
\bdm
\til{p}^2=\textstyle\frac{1}{729a^2}\sum_\mu s_\mu^2\prod_{\nu\neq\mu}\{c_\nu+2\}^2
\qquad\mbox{and}\qquad
\check{p}^2=\textstyle\frac{4}{a^2}-\frac{1}{4a^2}\prod_\mu\{c_\mu+1\}
\edm
or equivalently for a zero in $(c^2+d^2)[\til{p}^2+(\frac{a}{2}\check{p}^2-\frac{\rh}{a})^2]^{1/2}+2cd(\frac{a}{2}\check{p}^2-\frac{\rh}{a})=0$, and the square root renders this a \emph{transcendental equation} in $\cosh(aE)$.
The result through $O(a^5)$ is given in (\ref{over_bril}).


\subsection{Dispersion relations with exponential quark mass trick}

\begin{figure}[!tb]
\centering
\includegraphics[width=0.5\textwidth]{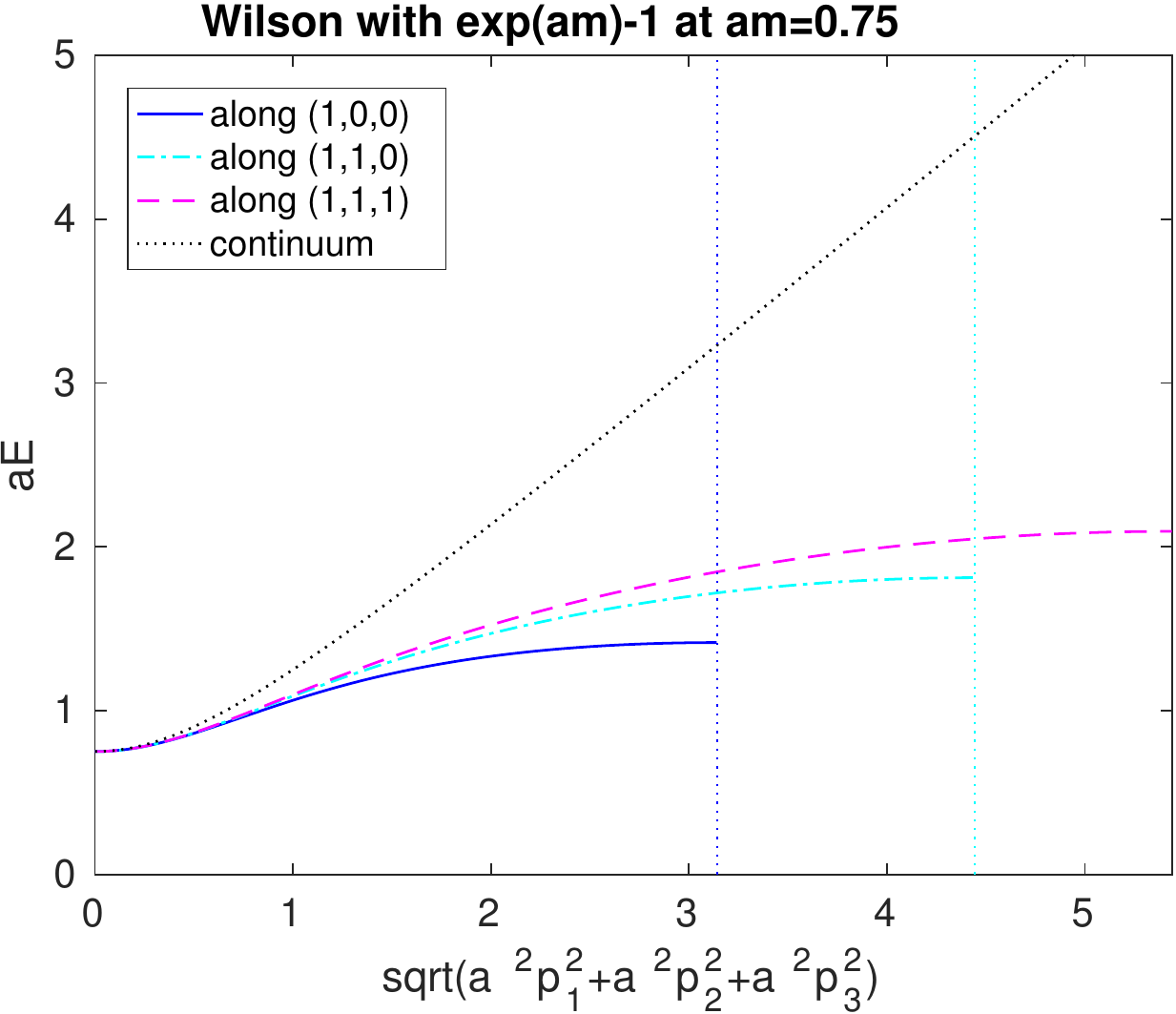}%
\includegraphics[width=0.5\textwidth]{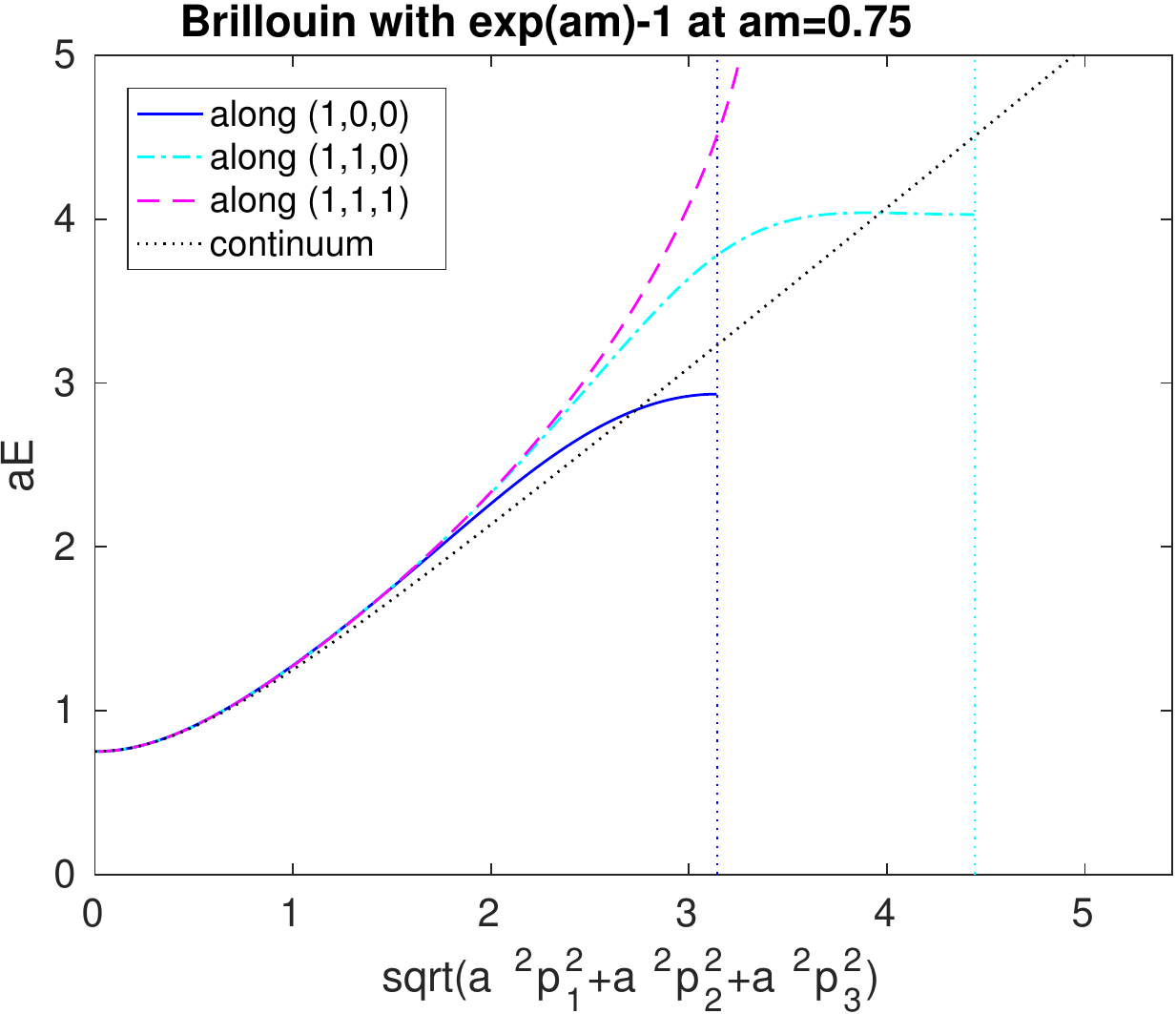}%
\caption{\label{fig:MDR_test}\sl
Same as the bottom part of Fig.\,\ref{fig:MDR_kern} but with the replacement $am\to\exp(am)-1$ in the bare Wilson (left) and Brillouin (right) actions.
Note that these dispersion relations extend over a larger portion of the Brillouin zone than the massive Wilson overlap and Brillouin overlap dispersion relations shown in the bottom part of Fig.\,\ref{fig:MDR_over}.}
\end{figure}

The observation that the first line in (\ref{free_wils}, \ref{free_bril}) is just an expansion of $\log^2(1+am)$ \cite{Cho:2015ffa} suggests that one might try the replacement $am\to\exp(am)-1$ in both the Wilson and the Brillouin actions (without the overlap procedure).
With this substitution we find
\bea
(aE)^2-(a\mb{p})^2
&=&(am)^2
\nonumber
\\
&+&\Big[-\frac{2}{3}(am)^2+\frac{1}{2}(am)^3\Big](a\mb{p})^2
\nonumber
\\
&+&\Big[-\frac{2}{3}+\frac{am}{2}\Big]\Big(\sum_{i<j}a^4p_i^2p_j^2+\sum_{i}(ap_i)^4\Big)+O(a^6)
\label{trick_wils}
\eea
for the (heavy-quark) Wilson operator and
\bea
(aE)^2-(a\mb{p})^2
&=&(am)^2
\nonumber
\\
&+&\Big[0+\frac{1}{12}(am)^3\Big](a\mb{p})^2
\nonumber
\\
&+&\Big[0+\frac{am}{12}\Big]\Big(\sum_{i<j}a^4p_i^2p_j^2+\sum_{i}(ap_i)^4\Big)+O(a^6)
\label{trick_bril}
\eea
for the (heavy-quark) Brillouin operator.
A comparison with (\ref{free_wils}, \ref{free_bril}) shows that indeed the first lines (the momentum independent parts) now match the continuum behavior, while the coefficients in the second and third lines are at most as large (in magnitude) as before.

A similar conclusion is suggested by plotting the dispersion relation of either the Wilson operator or the Brillouin operator with this substitution, as done in Fig.\,\ref{fig:MDR_test}.
The cut-off effects at $a\mb{p}=\mb{0}$ are gone (by construction), but even for non-zero momenta the free-field dispersion relation looks better than for the Wilson overlap and Brillouin overlap actions, respectively (see Fig.\,\ref{fig:MDR_over}).
Similarly to what was said about the Wilson overlap and Brillouin overlap actions, one might caution that such a behavior is only a necessary requirement for such a substitution to be useful in heavy-quark physics.
But this time the investigation was carried out long ago, since this observation was the starting point for the development of the Fermilab action \cite{ElKhadra:1996mp}.


\section{Details of diagonal Kenney-Laub iterates \label{app:kenneylaub}}


In this appendix we give details of the partial fraction expansion of some of the diagonal elements of Kenney-Laub mappings as defined in Tabs.\,\ref{tab1},\,\ref{tab2}.

\begin{table}[!p]
\centering\tt\small
\begin{tabular}{|c|rrr|}
\hline
KL(1,1)   & 1/3  & 0.8888888888888889 &   0.3333333333333333 \\
\hline
KL(2,2)   & 1/5  & 0.4422291236000336 &   0.1055728090000841 \\
          &      &  1.157770876399966 &    1.894427190999916 \\
\hline
KL(3,3)   & 1/7  & 0.3005985953147677 &  0.05209508360168703 \\
          &      & 0.4674182302787388 &   0.6359638059755859 \\
          &      &  1.517697460120779 &    4.311941110422727 \\
\hline
KL(4,4)   & 1/9  & 0.2291313786946141 &  0.03109120412576338 \\
          &      & 0.2962962962962963 &   0.3333333333333333 \\
          &      & 0.5378392501024903 &    1.420276625461206 \\
          &      &  1.899696037869562 &    7.548632170413030 \\
\hline
KL(5,5)   & 1/11 & 0.1855767632407464 &  0.02067219782410498 \\
          &      & 0.2197383478725930 &   0.2085609132992616 \\
          &      & 0.3183328723857196 &   0.7508307981214580 \\
          &      & 0.6220419130221986 &    2.421230521622092 \\
          &      &  2.290673739842379 &    11.59870556913308 \\
\hline
KL(6,6)   & 1/13 & 0.1561143535399426 &  0.01474329800962692 \\
          &      & 0.1759739298223624 &   0.1438305438453555 \\
          &      & 0.2271454286305450 &   0.4764452860985422 \\
          &      & 0.3498637189117062 &    1.274114172926091 \\
          &      & 0.7123574780333906 &    3.630323607217039 \\
          &      &  2.686237398754361 &    16.46054309190335 \\
\hline
KL(13,13) & 1/27 & 0.07432535480401804 & 0.003392289854243521 \\
          &      & 0.07637712623153803 &  0.03109120412576338 \\
          &      & 0.08071322905386415 &  0.08962859222716607 \\
          &      & 0.08785700982653639 &   0.1860696326582413 \\
          &      & 0.09876543209876543 &   0.3333333333333333 \\
          &      &  0.1151288280773304 &   0.5542391790439610 \\
          &      &  0.1400074395304560 &   0.8901004336611559 \\
          &      &  0.1792797500341634 &    1.420276625461206 \\
          &      &  0.2453107874470617 &    2.311695630535332 \\
          &      &  0.3677579938327626 &    3.964732916742294 \\
          &      &  0.6332320126231874 &    7.548632170413030 \\
          &      &   1.392797081723151 &    17.80276060326253 \\
          &      &   5.496102275704820 &    73.19738072201507 \\
\hline
\end{tabular}
\rm\normalsize
\caption{\label{tab5}\sl
Partial fraction form of the diagonal functions $f_{n,n}(x)=x\,p_{n,n}(x^2)/q_{n,n}(x^2)$ for $n=1,...,6,13$ from the Kenney-Laub family (\ref{pade_def}) for the matrix sign function.
The second column gives the constant contribution; the third and fourth columns give the numerator and the shift in the rational contribution.
For instance $f_{2,2}(x)=x(\frac{1}{5}+\frac{0.4422291236000336}{x^2+0.1055728090000841}+\frac{1.157770876399966}{x^2+1.894427190999916})$.}
\centering\tt\small
\begin{tabular}{|c|rr|}
\hline
KL(0,1) &  2.000000000000000 &   1.000000000000000 \\
\hline
KL(1,2) & 0.5857864376269050 &  0.1715728752538099 \\
        &  3.414213562373095 &   5.828427124746190 \\
\hline
KL(2,3) & 0.3572655899081636 & 0.07179676972449083 \\
        & 0.6666666666666667 &   1.000000000000000 \\
        &  4.976067743425180 &   13.92820323027551 \\
\hline
\end{tabular}
\rm\normalsize
\caption{\label{tab6}\sl
Partial fraction form of the first upper diagonal functions $f_{n-1,n}(x)$ for $n=1,...,3$ from the Kenney-Laub family (\ref{pade_def}) for the matrix sign function.
The meaning of the columns is as in Tab.\,\ref{tab5}, except that this time there is no constant contribution.}
\end{table}

The diagonal elements $f_{n,n}$ for $n=1,...,8$ are given in partial fraction form in Tab.\,\ref{tab5}.
One notices that the smallest shift (fourth column) decreases with increasing $n$.
Furthermore, for any fixed $n$, the weight (third column) is a monotonic (and positive) function of the shift.
This means that the stopping criterion in the CG solver can be relaxed for smaller shifts.
In fact, given the hierarchy among the shifts for a fixed $f_{n,n}$, it is clear that the cost of the numerical inversion is dominated by the cost of the smallest shift (or the smallest few shifts).

The elements $f_{n-1,n}$ for $n=1,...,3$ of the first upper diagonal are given in partial fraction form in Tab.\,\ref{tab6}.
One notices that the majority of the features discussed in the previous paragraph persist, except that there is no constant contribution any more.
Nonetheless, some of the properties of the overall function are quite different (see Sec.\,\ref{sec:KL}).


\section{Flop count and memory traffic considerations \label{app:flopcount}}


For an efficient implementation of the Brillouin operator it is vital to precompute the off-axis links that are implicitly used in the covariant derivative $\nab^\mr{iso}$ and the covariant Laplacian $\lap^\mr{bri}$.
The underlying reason is that, in order to maintain $\gaf$-hermiticity, one must average, within any $k$-hop contribution, over the $k!$ shortest paths [with optional backprojection to $SU(\Nc)$ in a quenched setting, but we favor an average].
Moreover, each individual $k$-hop path requires $k-1$ matrix multiplications in color space.

\subsection{Brillouin operator flop count}

Let {\tt U} and {\tt V} be objects which hold the original and smeared gauge fields, respectively.
In Fortran-style languages (which use column-major memory layout) they may be defined as rank 7 arrays, e.g.\ {\tt V(1:Nc,1:Nc,1:4,1:Nx,1:Ny,1:Nz,1:Nt)}.
Here {\tt 1:4} in the third slot limits the values that the direction index $\mu$ may take, $\Nc$ is the number of colors, and the box size is $N_x \times N_y \times N_z \times N_t$.
Next, number the $81$ elements in the $[-1:1]^4$ hypercube around a given position $n$ such that directions $\nu\in\{1,...,81\}$ and $82-\nu$ are opposite; in particular $\nu=41$ corresponds to the $0$-hop movement.
Since $W_\nu(n)$ and $W_{82-\nu}(n+\hat{\nu})$ relate to each other through hermitean conjugation, it suffices to store the first $40$ off-axis links (constructed from $V$) in the rank 7 array {\tt W(1:Nc,1:Nc,1:40,1:Nx,1:Ny,1:Nz,1:Nt)}.
In C-style languages (which use row-major memory layout) the ordering must be reversed, such that {\tt W[[.]][[.]][[.]][[.]][[.]][[0:Nc-1]][0:Nc-1]]} represents a $\Nc\times\Nc$ matrix which occupies a contiguous space in memory.
In the following we assume that $W$ is ready for use, and we ignore this kind of set-up cost, since on the overall scale it is negligible.

We now discuss the structure of the matrix-times-vector routine which constructs, for a given source vector $x$, the target vector $y=\Dbr x$.
The source and sink vectors may be represented by rank 3 arrays, e.g.\ {\tt x(1:Nc,1:4,1:Nx*Ny*Nz*Nt)} in Fortran-style languages.
This routine consists of an outer loop (or set of four loops) which runs over the position $n$ of the target $y$, and an inner loop (or set of four loops) which runs over the 81 elements of the hypercube around $n$ and thus over the positions $m$ of the source $x$ which contribute to $y(n)$.
In 80 out of the 81 cases the $\Nc\times4$ matrix $x(:,:,m)$ must be parallel transported through a left-multiplication with $W(:,:,\nu,n)$ or $W(:,:,\nu,m)\dag$.
In addition, the result (which is still a $\Nc\times4$ matrix) must be right-multiplied with 0 ($\nu=41$) to 4 ($\nu$ pointing to any of the 16 edges of the hypercube) elements of the set $\{\ga_1^t,...,\ga_4^t\}$, where $t$ means transposition.
In the chiral representation any $\ga$-matrix contains one of the elements $\pm1,\pm\ri$ in each row and colum, and the right-multiplication amounts to a re-ordering of the columns of this $\Nc\times4$ matrix (times factors of $\pm\ri$ which again implies reorderings of real and imaginary parts).
Since such reorderings can be done on the fly, we assume that the right-multiplication is for free, and we take only the left-multiplication into account in our cost estimate.

With this input, the flop count of the Brillouin matrix-times-vector routine is as follows:
\begin{enumerate}
\itemsep-1pt
\item[(i)]
$SU(\Nc)$-multiply the $\Nc\times4$ block for each non-trivial direction.
A comlex-times-complex multiplication takes 6 flops, a complex-plus-complex addition takes 2 flops, there are $\Nc$ multiplications and $\Nc-1$ additions per site, and there are $80$ directions.
Overall, this takes $\Nc\cdot4\cdot(6\Nc+2\Nc-2)\cdot80$ flops; hence $21120$ flops for $SU(3)$.
\item[(ii)]
Multiply the resulting $\Nc\times4$ matrix with the correct weight factor as given by the isotropic derivative and the hypercubic Laplacian.
These weight factors are real, and for each $\nab_\mu^\mr{iso}$ non-zero only for $54$ out of the $81$ directions.
The mass term may be incorporated into the $0$-hop (i.e.\ $\nu=41$) contribution of the Laplacian.
Overall, this takes $\Nc\cdot8\cdot(4\cdot54+81)$ flops; hence $7128$ flops for $SU(3)$.
\item[(iii)]
Accumulate the 81 contributions to the out-spinor.
Overall, this takes $\Nc\cdot8\cdot80$ flops; hence $1920$ flops for $SU(3)$.
\end{enumerate}
All together we arrive at a grand total of $30168$ flops per site for $SU(3)$.

\subsection{Wilson operator flop count}

For reference, let us give a brief account how such a flop count looks for the Wilson operator.
Here, the main difference is that for each one of the $8$ directions $\pm\mu$ the $\Nc\times4$ block $x(:,:,m)$ is right-multiplied by $\frac{1}{2}(1\pm\ga_\mu)^t$, and the latter operator is a projector whose eigenvectors can be precomputed.
In consequence, the block is shrunk into $\Nc\times2$ format before the left-multiplication with $V_\mu(n)$ or $V_\mu(m)\dag$ takes place, and expanded afterwards.

With this input, the flop count of the Wilson matrix-times-vector routine is as follows:
\begin{enumerate}
\itemsep-1pt
\item[(i)]
Spin project (from 4 to 2 components) the $\Nc\times4$ matrix for each direction.
Overall, this takes $\Nc\cdot4\cdot8$ flops; hence $96$ flops for $SU(3)$.
\item[(ii)]
$SU(\Nc)$-multiply the $\Nc\times2$ block for each direction, and expand back to $\Nc\times 4$ format (for free).
Overall, this takes $\Nc\cdot2\cdot(6\Nc+2\Nc-2)\cdot8$ flops; hence $1056$ flops for $SU(3)$.
\item[(iii)]
Accumulate these 8 directions, as well as the $0$-hop contribution which uses the precomputed factor $(4+m)$.
Overall, this takes $\Nc\cdot8\cdot9$ flops; hence $216$ flops for $SU(3)$.
\end{enumerate}
All together we arrive at a grand total of $1368$ flops per site for $SU(3)$.

\subsection{Brillouin operator memory traffic}

The memory traffic of the Brillouin matrix-times-vector routine is as follows:
\begin{enumerate}
\itemsep-1pt
\item[(a)]
Read one color-spinor block for each direction.
Overall, this amounts to $\Nc\cdot8\cdot81$ floats; hence $1944$ floats for $SU(3)$.
\item[(b)]
Read one gauge link $W_\nu$ for each non-trivial direction.
Overall, this amounts to $\Nc^2\cdot2\cdot80$ floats; hence $1440$ floats for $SU(3)$.
\item[(c)]
Write one color-spinor block back into memory.
Overall, this amounts to $\Nc\cdot8$ floats; hence $24$ floats for $SU(3)$.
\end{enumerate}
All together we arrive at a grand total of $3408$ floats per site for $SU(3)$, i.e.\ $13632$ bytes if everything is in single-precision, and twice as much in double-precision.
Here we assume that everything is to be read afresh, i.e.\ nothing is in cache.
By handling $N_v$ vectors simultaneously, the contribution (b) per vector is reduced by a factor $N_v$.
For instance for $\Nc=3$ and $N_v=12$ the grand total is $1940+120+24=2088$ floats from/to memory per vector and site.

\subsection{Wilson operator memory traffic}

The memory traffic of the Wilson matrix-times-vector routine is as follows:
\begin{enumerate}
\itemsep-1pt
\item[(a)]
Read one color-spinor block for each direction.
Overall, this amounts to $\Nc\cdot8\cdot9$ floats; hence $216$ floats for $SU(3)$.
\item[(b)]
Read one gauge link $V_\nu$ for each direction.
Overall, this amounts to $\Nc^2\cdot2\cdot8$ floats; hence $144$ floats for $SU(3)$.
\item[(c)]
Write one color-spinor block back into memory.
Overall, this amounts to $\Nc\cdot8$ floats; hence $24$ floats for $SU(3)$.
\end{enumerate}
All together we arrive at a grand total of $384$ floats per site for $SU(3)$, i.e.\ $1536$ bytes if everything is in single-precision, and twice as much in double-precision.
Here we assume that everything is to be read afresh, i.e.\ nothing is in cache.
By handling $N_v$ vectors simultaneously, the contribution (b) per vector is reduced by a factor $N_v$.
For instance for $\Nc=3$ and $N_v=12$ the grand total is $216+12+24=252$ floats from/to memory per vector and site.

\subsection{Technical summary}

The Brillouin operator flop count exceeds the Wilson flop count by a factor $30168/1368\simeq22$ at $\Nc=3$.
In the large-$\Nc$ limit the Brillouin flop count scales as $\Nc^2\cdot2560$, while the Wilson flop count scales as $\Nc^2\cdot128$.
This means that in the large-$\Nc$ limit this ratio approaches $20$.

The Brillouin memory traffic exceeds the Wilson memory traffic by a factor $3408/384\simeq8.9$ at $\Nc=3$, if one right-hand-side is handled at a time.
In the large-$\Nc$ limit the Brillouin memory traffic scales as $\Nc^2\cdot160$, while the Wilson traffic scales as $\Nc^2\cdot16$.
This means that in the large-$\Nc$ limit this ratio approaches $10$.

At any $\Nc$ the memory traffic per site and right-hand-side can be reduced by handling $N_v$ vectors simultaneously.
Overall, this brings an extra factor $N_v$ under (a) and (c), but no change under (b), for either operator.
On a per-vector basis this means that the traffic under (b) is reduced by a factor $N_v$, while (a) and (c) remain constant.
In other words, whenever memory bandwith is the main bottleneck in an actual computation (which on highly parallel architectures is usually true) handling $N_v$ right-hand-sides simultaneously is an efficient means to speed up both the Wilson and the Brillouin matrix-times-vector performance.


\end{document}